\documentclass[sn-mathphys-num]{sn-jnl}

\usepackage{chemformula} 
\usepackage{amssymb}
\usepackage[T2A,T1]{fontenc}
\usepackage[OT2,OT1]{fontenc}
\usepackage[german,english]{babel}
\usepackage[normalem]{ulem}
\usepackage{siunitx}

\begin{document}
	
	\author*[1]{\fnm{Niklas} \sur{Manz}}\email{nmanz@wooster.edu}
	
	\author[2,3,4,5]{\fnm{Yurij} \sur{Holovatch}}\email{hol@icmp.lviv.ua}
	
	\author[6]{\fnm{John} \sur{Tyson}}\email{tyson@vt.edu}
	
	\affil*[1]{\orgdiv{Department of Physics}, \orgname{College of Wooster}, \orgaddress{
			\city{Wooster}, \postcode{44691}, \state{OH}, \country{USA}}}
	
	\affil[2]{\orgdiv{Institute for Condensed Matter Physics of the National}, \orgname{Academy of Sciences of Ukraine}, \orgaddress{
			\city{Lviv}, \postcode{79011}, 
			\country{Ukraine}}}
	
	\affil[3]{\orgname{$\mathbb{L}^4$ Collaboration and Doctoral College for the Statistical Physics of Complex Systems}, \orgaddress{\city{Lviv-Leipzig-Lorraine-Coventry}, \country{Europe}}}
	
	\affil[4]{\orgdiv{Centre for Fluid and Complex Systems}, \orgname{Coventry University}, \orgaddress{
		\city{Coventry}, \postcode{CV1 5FB}, 
		\country{UK}}}
	
	\affil[5]{\orgname{Complexity Science Hub Vienna}, \orgaddress{
		\city{Vienna}, \postcode{1080}, 
		\country{Austria}}}

	\affil[6]{\orgdiv{Department of Biological Sciences}, \orgname{Virginia Polytechnic Institute and State University}, \orgaddress{
		\city{Blacksburg}, \postcode{24061}, \state{VA}, 
		\country{USA}}}
	
\title{Julian Hirniak, an early proponent of periodic chemical reactions}








\abstract{
In this article we present and discuss the work and scientific legacy of \textsc{Julian Hirniak}, the Ukrainian chemist and physicist who published two articles in 1908 and 1911 about periodic chemical reactions. Over the last 110+ years, his theoretical work has often been cited favorably in connection with \textsc{Alfred Lotka}'s theoretical model of an oscillating reaction system. Other authors have pointed out thermodynamic problems in \textsc{Hirniak}'s reaction scheme. Based on English translations of his 1908 Ukrainian and 1911 German articles, we show that \textsc{Hirniak}'s claim (that a cycle of inter-conversions of three chemical isomers in a closed reaction vessel can show damped periodic behavior) violates the \textit{Principle of Detailed Balance} (i.e., the Second Law of Thermodynamics), and that \textsc{Hirniak} was aware of this Principle. We also discuss his results in relation to \textsc{Lotka}'s first model of damped oscillations in an open system of chemical reactions involving an auto-catalytic reaction operating far from equilibrium. Taking hints from both \textsc{Hirniak} and \textsc{Lotka}, we show that the mundane case of a kinase enzyme catalyzing the phosphorylation of a sugar can satisfy \textsc{Hirniak}'s conditions for damped oscillations to its steady state flux (i.e., the \textsc{Michaelis--Menten} rate law), but that the oscillations are so highly damped as to be unobservable. Finally, we examine historical and factual misunderstandings related to \textsc{Julian Hirniak} and his publications.
}

\maketitle
\section{Introduction}
\label{sec:Introduction}

\textsc{Julian Hirniak}'s legacy in the field of nonlinear chemical reactions is based on his 1911 article in the  \textit{\foreignlanguage{German}{Zeitschrift für Physikalische Chemie}} with the title ``\foreignlanguage{German}{Zur Frage der periodischen Reaktionen}" (``On the question of periodic reactions")~\cite{Hirniak1911}, which he wrote in response to \textsc{Alfred Lotka}'s famous 1910 article in the same journal with the title ``\foreignlanguage{German}{Zur Theorie der periodischen Reaktionen}" (``On the theory of periodic reactions")~\cite{Lotka1910}. \textsc{Hirniak}'s 1911 paper is mainly a shorter version, including a response to \textsc{Lotka}'s 1910 paper, of the article ``\foreignlanguage{Ukrainian}{Про перiодичнi хемiчнi реакциї}" (``About periodic chemical reactions")~\cite{Hirniak1908a} published three years earlier in Ukrainian in the journal of the Shevchenko Scientific Society in Lviv.

During that time, many heterogeneous systems exhibiting oscillatory behavior were known, but it would be another decade before \textsc{William Bray} publishes his seminal paper ``A Periodic Reaction in Homogeneous Solution and Its Relation to Catalysis"~\cite{Bray1921} in 1921. The earliest examples are electrochemical systems such as the `beating mercury heart,' which was first described around 1800 by \textsc{Henry}~\cite{Henry1800} and \textsc{Volta}~\cite{Volta1800}, and \textsc{Fechner}'s discovery of the oscillating current (polarity change) produced in an electrochemical cell consisting of an iron/silver electrode pair in a nitric acid solution of silver nitrate~\cite{Fechner1828}. Other noteworthy systems were \textsc{Munck{~}af{~}Rosensch\"{o}ld}'s observation of alternating illumination of phosphorus in a bottle in contact with air in 1834~\cite{Munck-af-Rosenschoeld1834} and \textsc{Landolt}'s iodine clock, the first periodic heterogeneous chemical system, involving the autocatalytic iodate-bisulphite reaction in 1886~\cite{Landolt1886}. 

Around 1900, other periodic nonlinear chemical phenomena were published, sparking a need for theoretical models. \textsc{Liesegang} presented, starting in 1889, alternating precipitation zones from diffusing and reacting pairs of inorganic salts~\cite{Liesegang1898_book}, 
\textsc{Ostwald} published a short article in 1899 (followed by two longer in 1900) about the periodically increasing and decreasing dissolution of chromium in acid~\cite{Ostwald1899}, \textsc{B\'enard} reported spontaneous symmetry breaking in fluid dynamics~\cite{Benard1900} in 1900, \textsc{Heathcote} described an `iron nerve' (the periodic dissolution of iron wire in nitric acid)~\cite{Heathcote1901} one year later, and \textsc{Luther} published his famous paper in 1906 on a chemical trigger wave in the permanganate-oxalate reaction, a homogeneous liquid-phase chemical system~\cite{Luther1906}. \\

In this article, we discuss \textsc{Hirniak}'s contribution and his controversial legacy in the field of nonlinear chemical reactions, which is  based on his proposed linear model. 

In Chapter \ref{sec:Julian Hirniak} (\textbf{\emph{Julian Hirniak}}), we provide some background information about \textsc{Julian Hirniak} as a person and as a scientist and explain the confusion about his first name in the English literature. We also provide information about the term ``ruthenische Sprache" (``Ruthenian language") \textsc{Hirniak} used himself in his German 1911 article when referring to the language of his 1908 article. 

In Chapter \ref{sec:Periodic chemical reactions} (\textbf{\emph{Periodic chemical reactions}}), we present a short overview about the chemistry and mathematics of oscillatory chemical reactions and discuss the significant differences between the work published by \textsc{Hirniak} in 1908~\cite{Hirniak1908a} and 1911~\cite{Hirniak1911} and the work by \textsc{Lotka} in 1910~\cite{Lotka1910,Lotka1910a} and 1912~\cite{Lotka1912}. We present our opinion of whether \textsc{Hirniak} is a ``\textit{forgotten pioneer}" or a ``\textit{misguided proponent}" of periodic chemical reactions. 

The the final Chapter \ref{sec:Citations of Hirniak's work} (\textbf{\emph{Citations of Hirniak's work}}) gives an overview of all known publications citing \textsc{Hirniak}'s work in the last 110+ years. We present the historical reason why \textsc{Hirniak}'s 1911 article is, most of the time, cited as published in 1910, and why many scientists until today have a favorable view of his work and cite his article on a par with \textsc{Lotka}. 

In the Appendix, we provide 
in App.~\ref{app:Julian Hirniak's publications} a complete list (to our knowledge) of \textsc{Hirniak}'s scientific publications, with original and translated bibliography entries; 
in App.~\ref{app:Favorable and critical citations of Hirniak's work} a detailed presentation of all citations of \textsc{Hirniak}'s work not by \textsc{Lotka} and \textsc{Bray} since 1912;
in App.~\ref{app:1908Translation} translation of \textsc{Julian Hirniak}'s 1908 article from
Ukrainian into English; 
in App.~\ref{app:1911Translation} a translation of \textsc{Julian Hirniak}'s 1911 article from German into English; and 
in App.~\ref{app:1901Translation} a translation of Section IV from \textsc{Rudolf Wegscheider}’s 1901 German article.

\section{Julian Hirniak}
\label{sec:Julian Hirniak}

\begin{figure}
    \centering
    \includegraphics[width=0.4\linewidth]{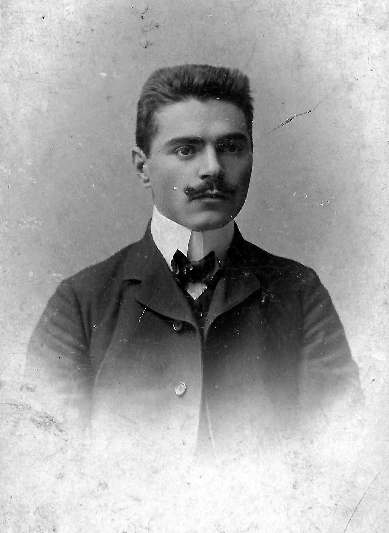}
    \caption{Julian Hirniak at the age of about 20 in 1901 (Reproduced with permission from the Joseph Hirniak collection).}
    \label{fig:JulianHirniak1901}
\end{figure}

\textsc{\foreignlanguage{Ukrainian}{Юліан Гірняк}}\footnote{Here and below, when appropriate, we give in parallel corresponding names in Ukrainian.}  (\textsc{Julian Hirniak}) was born on September 8, 1881 in the village of \foreignlanguage{Ukrainian}{Струсів} (Strusiv) which now belongs to the \foreignlanguage{Ukrainian}{Тернопільська область} (Ternopil province) in Ukraine, about \SI{150}{km} southeast of \foreignlanguage{Ukrainian}{Львів} (Lviv). Until 1918, this region was part of the Austro-Hungarian Empire, in the historical and geographic region of Galicia\footnote{Not to be confused with the  autonomous community (the first sub-national level) Galicia in the northwest of Spain.}. He died on June 5, 1970 in Passaic, New Jersey, USA. A picture of \textsc{Julian Hirniak} from 1901 is shown in Fig.~\ref{fig:JulianHirniak1901}. 

\textsc{Hirniak}'s work is often cited in connection with \textsc{Alfred Lotka}, the famous scientist who is credited with proposing the first theoretical model of damped oscillations in a homogeneous chemical reaction system in 1910. Curiously, there are two other geographical connections between the two: \textsc{Lotka} was born about 18 month earlier, on March 3, 1880 in Lviv, and died on December 5, 1949 in Red Bank, New Jersey, USA. 

\subsection{A short biography}
\label{sec:bio}

\textsc{Julian Hirniak}, born in 1881 in the Ukrainian family of \textsc{Josyph Hirniak} and \textsc{Sofia Ryzhevska}, was one of eight children (seven brothers and one sister), many of whom left a significant mark in the history of Ukraine. From 1893 to 1900, \textsc{Julian} attended a real school and gymnasium in \foreignlanguage{Ukrainian}{Івано-Франківськ} (Ivano-Frankivsk), as it is known nowadays after its renaming from the Ukrainian name Stanislaviv in 1962\footnote{Also known as Stanislau and Stanisławów in \textsc{Hirniak}'s time.}. In 1901, he became a student of the \emph{Lviv Polytechnic School} (now \emph{Lviv Polytechnic National University}) and at the same time studied chemistry and physics at the \emph{Lviv University} (now \emph{Ivan Franko National University of Lviv}). During the 1904/1905 winter semester, he joined the lab of Prof.\ \textsc{Ivan Puluj} at the \emph{Imperial and Royal German Technical High School} (now \emph{Czech Technical University}) in Prague (then also in the Austro-Hungarian Empire), studying the thermal conductivity of sugar in aqueous solution. For these results, he received a title of Doctor of Technical Sciences in Lviv in 1905 and published them in an article in 1907 (see Chapter~\ref{sec:Periodic chemical reactions}).

After receiving his doctorate degree, he started teaching at Ukrainian and Polish gymnasiums in Stanislaviv for about two years. In 1907, he received a scholarship from the Austrian Ministry of Education and conducted research at the \emph{\foreignlanguage{German}{Physikalisch-chemisches Institut}} in Leipzig, German Empire (now the \emph{\foreignlanguage{German}{Wilhelm-Ostwald-Institut für Physikalische und Theoretische Chemie der Universität Leipzig}}), one of the first institutions working in the field of non-equilibrium thermodynamics. Already during his studies in Lviv, he established contacts with the Shevchenko Scientific Society, the prominent Ukrainian scholarly society that played, in the beginning of the 20$^\text{th}$ century, the role of the Ukrainian Academy of Sciences. Similar to other contemporary academies, the society consisted of several sections. One of them was the Mathematical-Natural Sciences-Medical Section, which published a journal with a dual-language Ukrainian-German title ``\emph{\foreignlanguage{Ukrainian}{Збiрник Ма\-те\-матично-Природописно-Лї\-карсь\-кої Секциї Наукового Товарист\-ва імени Шевченка}}" and ``\emph{\foreignlanguage{German}{Sam\-mel\-schrift der Mathematisch-Naturwissenschaftlich-Ärztlichen Sektion der \v{S}ev\-\v{c}en\-ko-Ge\-sell\-schaft der Wissenschaften in Lemberg}}".\footnote{The English translation is \emph{Proceedings of the Mathematical-Natural Sciences-Medical Section of the Shevchenko Scientific Society in Lviv}.} Though the journal title and the table of content was bilingual, the actual articles were mainly in Ukrainian. The journal was established in 1897 and published in Lviv until 1939. \textsc{Hirniak} published most of his papers in this journal (see Appendix \ref{app:Julian Hirniak's publications} for a complete list of his publications).

On January 29, 1908, he was elected as a full member of the Shevchenko Scientific Society and three years later, he published his first book on theoretical studies of chemical reaction kinetics, published by the Shevchenko Scientific Society~\cite{Hirniak1911_book}.
Starting in 1912, he worked again as a teacher at the Ukrainian gymnasium in Lviv. 
After the \emph{Ukrainian State University} was established in \foreignlanguage{Ukrainian}{Кам'янець-Подільський} (Kamyanets'-Podilsky) in October 1918, \textsc{Hirniak} joined the faculty, but he returned to Lviv in 1920, when the university was closed after the defeat of the Ukrainian People's Republic.

Also in 1920, by the combined efforts of Ukrainian scientists and university students in Lviv, Ukrainian university courses started, hidden from the Polish authorities, which were in power after Poland won the Ukrainian-Polish war after World War I. At the end of July, the courses were transformed to the underground \emph{Ukrainian Higher University and Polytechnic} in Lviv \cite{Dudka2021_book-chapter}, and \textsc{Hirniak} started teaching  experimental physics and mathematics. Later, he also served as the rector of the 
underground  \emph{Ukrainian Higher Polytechnic School}, after the single institution  got separated into two.

In the 1930s, he was a member of the famous circle of Lviv mathematicians in the Scottish Caf\'e\footnote{The building is located at 27 Shevchenka Ave, Lviv, Ukraine}. One can read more about this period in the memoirs of \textsc{Stanislav Ulam} \cite{Ulam1991_book}, a prominent representative of the Lviv Mathematical School.

In 1939, he was appointed to the position of Associate Professor in the Department of Physical and Colloid Chemistry at \emph{Lviv Polytechnic Institute}\footnote{The used names and authorities of \emph{Lviv Polytechnic School} and \emph{Lviv University} where \textsc{Hirniak} studied and then worked changed several times between 1901 and 1939. They were under Austrian administration when \textsc{Hirniak} studied, became Polish institutions in 1918 and, after Germany and the Soviet Union invaded Poland on September 1 and September 17, 1939, respectively, Lviv was under the Soviet occupation. The new Soviet authorities continued the institutions after renaming them again.}. Five years later, in 1944, \textsc{Hirniak} emigrated first to Germany, and then in 1950 to the USA, where he died in 1970.

As a student, \textsc{Hirniak} made first attempts to promote the achievements of that time’s science and technology, publishing articles in some of Lviv's monthly publications for the youth, namely \foreignlanguage{Ukrainian}{Іскра} [Iskra] (The spark), \foreignlanguage{Ukrainian}{На Розсвіті} [Na rozsviti] (At dawn), as well as the journals \foreignlanguage{Ukrainian}{Ілюстрована Україна} [Iliustrovana Ukraina] (Illustrated Ukraine) and \foreignlanguage{Ukrainian}{Лі\-те\-ра\-тур\-но-науковий вістник} [Literaturno-naukovyy vistnyk] (Literary scientific herald). 
\textsc{Hirniak} also wrote many scientific and sociopolitical articles for `mass media' publications. \textsc{Kostyantyn Kurylyshyn}, for example, discovered about 30 articles between 1910 and 1928 in the newspaper \foreignlanguage{Ukrainian}{Діло} [Dilo]~\cite{Kurylyshyn_books}.

More information about \textsc{Julian Hirniak}, in Ukrainian, can be found in \textsc{Hryvnak}'s 2011 article ``\foreignlanguage{Ukrainian}{Юліан Гірняк: учений, педагог, громадський діяч, публіцист (до 130-річчя від народження)}" (``Julian Hirniak - scientist, teacher, public figure, columnist (On occasion of his 130th birthday)")~\cite{Hryvnak2011}, 
in \textsc{Shevchuk}'s 2018 article ``\foreignlanguage{Ukrainian}{Юліан Гірняк (1881 - 1970) - Відомий у світі український вчений}" (``Julian Hirniak (1881 - 1970) - A world-renowned Ukrainian scientist")~\cite{Shevchuk2018}, in a section of the 2018 article ``\foreignlanguage{Ukrainian}{Фізика і фізики в НТШ у Львові}" (``Physics and physicists in the Shevchenko Scientific Society in Lviv")~\cite{Holovatch2018}, and in two encyclopedia entries~\cite{Markus2009_encyclopedia,Kovaliv2016_encyclopedia}.

In the English literature, two book sections were dedicated to \textsc{Julian Hirniak}: 
the section ``Julian Hirniak" in \textsc{Alexander Pechenkin}'s 2018 book   ``\textit{The History of Research on Chemical Periodic Processes}~\cite{Pechenkin2018_book} (see a detailed discussion of this book chapter on page \pageref{Pechenkin2018book} in Section \ref{Pechenkin2018book}), and the chapter ``Physics and Physicists in the Shevchenko Scientific Society'' by \textsc{Holovatch}, \textsc{Honchar}, and \textsc{Krasnytska}~\cite{Holovatch2021_book-chapter} in the 2021 book \textit{Leopolis Scientifica: Exact Sciences in Lviv until the Middle of the 20$^\text{th}$ Century}. 

\subsection{Ruthenian Language}
\label{sec:Ruthenian Language}

When talking about \textsc{Hirniak}'s 1908 article~\cite{Hirniak1908a}, a question about language comes up. \textsc{Hirniak} himself used the term \textit{Ruthenian language} in his 1911 German article~\cite{Hirniak1911}, which is also used in the English literature. 
What he meant was the `contemporary Galician standard of Ukrainian', the language spoken and written by Ukrainians in the region Galicia (now western Ukraine and southeastern Poland) within the Austro-Hungarian empire\footnote{A more detailed discussion about Ruthenian as a contemporary Galician standard of Ukrainian  can be found in \textsc{Martin Rohde}'s book chapter (in German)~\cite{Rohde2021_book-chapter}.}. The Shevchenko Scientific Society in Lviv, established in Habsburg Galicia, adopted the `Ruthenian language of instruction', sanctioned by the Habsburg authorities as the language of education in schools. Within the time period we are discussing here, it was named \textit{\foreignlanguage{Ukrainian}{руська мова}} [\textit{rusjka mova}], whereas Ukrainians in the Russian empire called it \textit{\foreignlanguage{Ukrainian}{українська мова}} [\textit{ukrajinsjka mova}]~\footnote{\foreignlanguage{Ukrainian}{Мова} (language in Ukrainian).}
The corresponding \textit{\foreignlanguage{Ukrainian}{руська}} term in English or German is  \textit{Ruthenian}\footnote{Because \textit{Rusyn} / \textit{Ruthenian} are exonyms (name for a language that originate outside of that linguistic community), it seems reasonable to attribute that \textsc{Hirniak} used the term \textit{Ruthenian} at least in part to his German-speaking environment.}. Therefore, in the context of \textsc{Hirniak}'s early 20th-century papers, the Ruthenian language 
means Ukrainian language~\footnote{More information on this topic and an explanation of its complexity can be found in \textsc{Serhii Plokhy}'s book \textit{The origins of the Slavic nations: Premodern identities in Russia, Ukraine and Belarus}~\cite{Plokhy2006_book} or in \textsc{Michael Moser}'s book chapter \textit{Rusyn: A New-Old Language In-between Nations and States}~\cite{Moser2016_book-chapter}.}.

In general, references to the Ruthenian language  are confusing, as can be seen in \textsc{Cervellati} and \textsc{Greco}'s 2017 history of science article, mainly about \textsc{Lotka} and \textsc{Bray}~\cite{Cervellati2017}. In the section discussing \textsc{Hirniak}, they write
\begin{quote}
    ``On the other hand, Hirniak claimed the priority over Lotka’s work by citing his 1908 previous work written in the Ruthenian language,$^{32}$ \ldots"
\end{quote}
with the reference explaining the language as
\begin{quote}
    ``Ruthenian is a variant of the East Slavic language that was spoken in the Grand Duchy of Lithuania. It gave rise to the modern Byelorussian and the Ukrainian languages. The official language of the Grand Duchy was Ruthenian until 1700."
\end{quote}
It's correct that Belarusian and Ukrainian developed from Ruthenian, and all are East Slavic languages, but the reference to the Grand Duchy of Lithuania is misleading because its territorial sovereignty over the area in question ended about 200 years earlier.

Moreover, the term \emph{Ruthenian} is also  used for the ethnonym \emph{\foreignlanguage{Ukrainian}{русини}}, an East Slavic ethnic group also known as Carpatho-Rusyns. This double meaning may lead to misunderstandings and  errors when discussing \textsc{Hirniak}'s biography. Obviously, such a misinterpretation occurred when \textsc{Pechenkin} discussed the language of \textsc{Hirniak}'s 1908 article in his 2018 book~\cite{Pechenkin2018_book} (see also entry in App.~\ref{app:Favorable citations of Hirniak's work}).

\subsection{Romanization of Hirniak's name}
\label{sec:Romanization of Hirniak's name}

In Ukrainian, \textsc{Hirniak}'s mother tongue,  his name is written as \textit{\foreignlanguage{Ukrainian}{Юліан Гірняк}}. The Romanization (transliteration) of Cyrillic into Latin letters depends on the Cyrillic alphabet used by the initial language and on the final language it will be translated into. Furthermore, transliterations change over time and are subject to personal and organizational preferences. 
An example of the first factor can be found in \textsc{Hirniak}'s last name \foreignlanguage{Ukrainian}{Гірняк}. In the Bulgarian or Russian alphabet, for instance the letter `\foreignlanguage{Ukrainian}{Г}' would change to a `G', whereas the Romanization from the Ukrainian alphabet gives `H', the reason for his last name being \textit{Hirniak}.

Current transliteration into English would give his name as \textit{Hirnyak}, which can't be found in any publication. \textsc{Hirniak} himself used \textit{Hirniak} as the Romanization of his name in all non-Ukrainian publications. But one can also find \textit{Hirnjak} in a collection of German \textit{Meeting Minutes of the Mathematical-Natural Sciences-Medical Section of the Shevchenko Scientific Society in Lviv} published in 1924~\cite{Shevchenko1924_book} and 1925~\cite{Shevchenko1925_book}. \textsc{Martin Rohde} used the same last name, in connection with the first name \textit{Julijan} in his 2021 German book about the Shevchenko Society~\cite{Rohde2021_book}. He explained ``This is the German scholarly transliteration, which is based on the Czech alphabet [\ldots] which is the standard in German Slavic Studies and East European History".

Because of the second factor (the final language) his first name \textit{\foreignlanguage{Ukrainian}{Юліан}} could be written as \textit{Iulian},  \textit{Julian}, or \textit{Yulian}. The latter case had been used in the English abstract of \textsc{Hryvnak}'s 2011 Ukrainian article for \textsc{Hirniak}'s 130th birthday~\cite{Hryvnak2011}. 
But \textsc{Hirniak} himself used \textit{Juljan} in his lecture written in Polish~\cite{Hirniak1936_book} and \textit{Julius} in his German book~\cite{Hirniak1911_book}. Most importantly, he also used \textit{Julius} in his most cited work, the German article in the \textit{\foreignlanguage{German}{Zeitschrift für Physikalische Chemie}}~\cite{Hirniak1911}. The latter case is the reason that his name in English literature sometimes appears as \textit{Julius Hirniak}.

\section{Periodic chemical reactions}
\label{sec:Periodic chemical reactions}

Many scientific publications in the field of oscillating chemical reactions trace the origin of the idea to \textsc{Alfred Lotka}'s 1910 articles~\cite{Lotka1910,Lotka1910a} (the article \cite{Lotka1910a} is largely an English translation of his German article \cite{Lotka1910}), his 1912 article~\cite{Lotka1912}, and perhaps his 1920 articles~\cite{Lotka1920,Lotka1920b,Lotka1920c}. The first experimental demonstration of an oscillating reaction in homogeneous solution is usually attributed to \textsc{William Bray}'s 1921 article~\cite{Bray1921},
 although oscillatory behavior in heterogeneous chemical systems had been known since the 1830s (e.g., work by \textsc{Fechner}~\cite{Fechner1828}, \textsc{Herschel}~\cite{Herschel1833}, \textsc{Munck af Rosenschöld}~\cite{Munck-af-Rosenschoeld1834}, and \textsc{Schönbein}~\cite{Schoenbein1836a}).
 Often \textsc{Lotka}'s work is cited in conjunction with \textsc{Vito Volterra}'s two articles in 1926~\cite{Volterra1926} and 1928 \cite{Volterra1928}, or his 1931 book chapter~\cite{Volterra1931_book-chapter}, in the context of sustained oscillations in predator-prey systems, where it is known as the \textsc{Lotka-Volterra} model. 
Although 
\textsc{Lotka}'s work is perhaps best known in population dynamics, it should be remembered that he presented his model (as \textsc{Hirniak} did) as a hypothetical, periodic chemical reaction system. Only after \textsc{Volterra}'s theoretical publications on `fluctuating populations sizes of species living together' and \textsc{Gause}'s 1935 experimental demonstration of such oscillations in laboratory cultures~\cite{Gause1935}, did \textsc{Lotka}'s name become part of the eponymous \textsc{Lotka-Volterra} model.

Early books on periodic phenomena in chemistry are \textsc{Kremann}'s 1913 book ``\textit{\foreignlanguage{German}{Die periodischen Erscheinungen in der Chemie}}"~\cite{Kremann1913_book}, the 1926 book ``\textit{The Problem of Physico-Chemical Periodicity}"~\cite{Hedges1926_book} by \textsc{Hedges} and \textsc{Myers} (this book does not mention \textsc{Hirniak}), the two 1934 books by \textsc{Veil} ``\textit{\foreignlanguage{french}{Les phénomènes périodiques de la chimie I. Les  périodicités de structure}}"~\cite{Veil1934_book1} and ``\textit{\foreignlanguage{french}{Les phénomènes périodiques de la chimie II. Les  périodicités cinétiques}}" ~\cite{Veil1934_book2}, and the 1938 book ``\textit{\foreignlanguage{russian}{Физико-химические периодические процессы}}"~\cite{Shemiakin1938_book} by \textsc{Shemiakin} and \textsc{Mikhalev}. Furthermore, as previously mentioned, \textsc{William Bray} and his student \textsc{Asa Caulkins} had worked,  since at least 1916, on an experimental example of oscillations in  a homogeneous chemical system~\cite{Bray1921} (a nice review can  be found in reference~\cite{Cervellati2017}).  Nonetheless, many scientists, up to the 1960s, were not convinced that the concentrations of chemically reacting species could fluctuate periodically, even in a damped fashion. They pointed out that the Gibbs free energy of chemical reactions must decrease monotonically in a closed chemical system at constant temperature and pressure. Left alone, the system will  reach thermodynamic equilibrium in a monotonic fashion.

At the beginning of the $20^\text{th}$ century \textsc{Hirniak}'s publications were often cited in conjunction with \textsc{Lotka}'s work, and \textsc{Lotka} himself cited \textsc{Hirniak}. But over the years, \textsc{Hirniak}'s work fell out of view and is nowadays hardly known and cited. We will explore and explain this shift in scientific recognition based on \textsc{Hirniak}'s 1908 and 1911 papers in Section \textbf{\emph{Citations of Hirniak's work}}.

The number of citations are difficult to count for early $20^\text{th}$ century publications, but the total number over the last 110 years speak for itself. \textsc{Hirniak}'s 1908 article~\cite{Hirniak1908a} and 1911 article~\cite{Hirniak1911} have been cited 2 and 27 times, respectively (\textit{Web of Science}, 14 February 2024). But as can be seen in Table~\ref{tab:Citation table}, several publications citing \textsc{Hirniak}'s work are not listed in \textit{Web of Science}.

\textsc{Lotka}'s 1910 articles in \foreignlanguage{German}{\textit{Zeitschrift für Physikalische Chemie}}~\cite{Lotka1910} and \textit{Journal of Physical Chemistry}~\cite{Lotka1910a} have been cited 38 and 486 times, respectively (\textit{Web of Science}, 14 February 2024) and his most famous 1920 \textit{JACS} article~\cite{Lotka1920c} ``Undamped oscillations derived from the law of mass action'' has  607 citations (\textit{Web of Science}, 14 February 2024), although it is often cited in the context of predator-prey oscillations, where it is known as the \textsc{Lotka-Voterra} model, rather than as a model of an oscillating chemical reaction. In the context of chemical oscillations, most authors cite \textsc{Lotka}'s 1925 book \textit{Elements of physical biology}~\cite{Lotka1925_book} instead of his scientific articles, resulting in more than \num{3200} citations so far (\textit{Web of Science}, 14 February 2024).

\subsection{Analysis of Hirniak's work}
\label{sec:Analysis of Hirniak's work}

\begin{figure}
	\centering
	\includegraphics[width=0.5\linewidth]{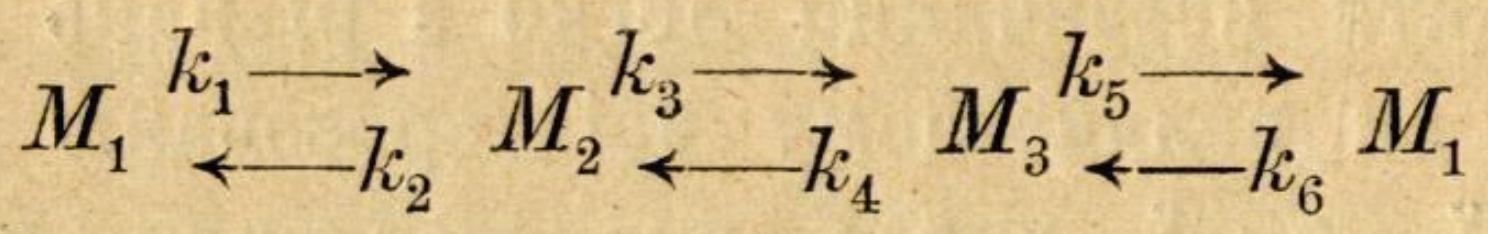}\\[2mm]
	\includegraphics[width=0.25\linewidth]{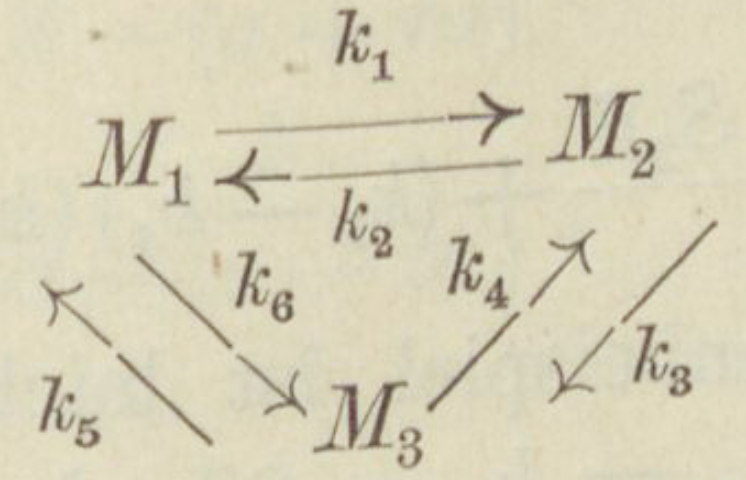}
	\caption{\textsc{Hirniak}'s reaction scheme using first-order kinetics between three isomers. Top from \textsc{Hirniak}'s 1908 article~\cite{Hirniak1908a}. Bottom from \textsc{Hirniak}'s 1911 article~\cite{Hirniak1911} (Public Domain. The original of the 1908 paper has been provided by the Stefanyk National Scientific Library in Lviv).}
    \label{fig:HirniakReactionSchemes}
\end{figure}

In 1908, \textsc{Hirniak} presented his analysis of the reaction dynamics of a system of three isomers (molecules with an identical molecular formula but different spatial arrangements of the atoms) inter-converting by first-order kinetics~\cite{Hirniak1908a}  as shown in Fig.~\ref{fig:HirniakReactionSchemes}, top (English translation of the original Ukrainian language in App.~\ref{app:1908Translation}). In his 1911 article~\cite{Hirniak1911} (English translation of the original German language in App.~\ref{app:1911Translation}, he discussed the same system but changed the sketch from a linear system to a cyclic system, shown in Fig.~\ref{fig:HirniakReactionSchemes}, bottom.

\textsc{Hirniak} clearly borrowed this reaction scheme  and copied much of the mathematical analysis directly from Section IV `\foreignlanguage{German}{Die gegenseitige Umwandlung von drei Isomeren}' (`The interconversion of three isomers') of \textsc{Rudolf Wegscheider} pioneering 1901 paper ``\textit{\foreignlanguage{German}{Über simultane Gleichgewichte und die Beziehungen zwischen Thermodynamik und Reactionskinetik homogener Systeme}" (``About simultaneous equilibria and the relationship between thermodynamics and reaction kinetics of homogeneous systems")}~\cite{Wegscheider1901}. This paper was identically republished in \textit{\foreignlanguage{German}{Zeitschrift für Physikalische Chemie}} in 1902~\cite{Wegscheider1902} and in the initial journal 10 years later again in 1911~\cite{Wegscheider1911}.

\textsc{Wegscheider} and \textsc{Hirniak} defined the concentrations of the isomeric species as $M_1$, $M_2$, and $M_3$, the forward rate constants as $k_1$, $k_3$, and $k_5$, and  the reverse rate constants as $k_2$, $k_4$, and  $k_6$. Under isothermal conditions, the time evolution of this closed system can  be described by
\begin{subequations}
\begin{align}
	\frac{\text{d}M_1}{\text{d}t} & = - \left( k_1 + k_6 \right)M_1 + k_2M_2 + k_5M_3\\
	\frac{\text{d}M_2}{\text{d}t} & = k_1M_1 - \left( k_2 + k_3 \right)M_2 + k_4M_3\\
	\frac{\text{d}M_3}{\text{d}t} &= k_6M_1 + k_3M_2 - \left( k_4 + k_5 \right)M_3~.
\end{align}
\label{eq:initial equation}
\end{subequations}
For a closed system, the total concentration of isomers, $M_\text{T} = \sum M_i$ must remain constant,
\begin{equation}
	M_\text{T} = M_1 + M_2 + M_3 = \text{constant}~,
	\label{eq:closed system}
\end{equation}
or
\begin{equation}
	\frac{\text{d}M_\text{T}}{\text{d}t} = 0~.
\end{equation}
Hence, one can compute the third isomer concentration as $M_3 = M_\text{T} - M_1 - M_2$ and rewrite Eqs.~\eqref{eq:initial equation} in matrix notation as
\begin{align}
	\frac{\text{d}}{\text{d}t}\begin{pmatrix}
	M_1 \\
	M_2 \\
	\end{pmatrix} = 
        &\begin{pmatrix}
	   {k_5M}_{\text{T}} \\
	   {k_4M}_{\text{T}} \\
	\end{pmatrix} + \nonumber\\
        & + 
        \begin{bmatrix}
	   - \left( k_1 + k_5 + k_6 \right) & k_2 - k_5 \\
	   k_1 - k_4 & - \left( k_2 + k_3 + k_4 \right) \\
	\end{bmatrix}
        \begin{pmatrix}
    	M_1 \\
    	M_2 \\
        \end{pmatrix}~.
	\label{eq:closed matrx notation}
\end{align}

The general solution of this linear ODE is 
\begin{equation}
	\begin{pmatrix}
		M_1(t) \\
		M_2(t) \\
	\end{pmatrix} = 
	\begin{pmatrix}
		M_1^{*} \\
		M_2^{*} \\
	\end{pmatrix} + c_1
	\begin{pmatrix}
		\xi_{11} \\
		\xi_{21} \\
	\end{pmatrix}
	\text{e}^{- \lambda_1t} + c_2\begin{pmatrix}
		\xi_{12} \\
		\xi_{22} \\
	\end{pmatrix}\text{e}^{- \lambda_2t}~,
 \label{eq:ODE solution}
\end{equation}
where $c_1$ and $c_2$ are arbitrary constants of integration, $\lambda_1$ and $\lambda_2$ are eigenvalues, $\xi_1$ and $\xi_2$ are eigenvectors of the $2\times2$ matrix of rate constants
\begin{equation}
	\textbf{K} = 
	\begin{bmatrix}
		- \left( k_1 + k_5 + k_6 \right) & k_2 - k_5 \\
		k_1 - k_4 & - \left( k_2 + k_3 + k_4 \right)\\
	\end{bmatrix}~,
\end{equation}
and \(\begin{pmatrix}
M_1^{*} \\
M_2^{*} \\
\end{pmatrix}\) is the `steady-state' solution of the system, which is given by
\begin{align}
	\begin{pmatrix}
		M_1^{*} \\
		M_2^{*} \\
	\end{pmatrix} &= - M_\text{T}\textbf{K}^{- 1}\begin{pmatrix}
		k_5 \\
		k_4 \\
	\end{pmatrix}\\
	& = \frac{M_\text{T}}{\text{det}(\textbf{K})}\begin{bmatrix}
		k_2 + k_3 + k_4 & k_2 - k_5 \\
		k_1 - k_4 & k_1 + k_5 + k_6 \\
	\end{bmatrix}
	\begin{pmatrix}
		k_5 \\
		k_4 \\
	\end{pmatrix}~. \label{eq:steady state solution}
\end{align}

The eigenvalues $\lambda_1$ and $\lambda_2$ are solutions of the characteristic equation
\begin{equation}
	\lambda^{2} - \text{tr}(\textbf{K}) \cdot \lambda + \text{det}(\mathbf{K}) = 0~,
\end{equation}
where the trace and determinant of \textbf{K} are
\begin{align}
	\text{tr}(\mathbf{K}) & = {- (k}_1 + k_2 + k_3 + k_4 + k_5 + k_6) \nonumber\\
        & < 0\\
	 \text{det}(\mathbf{K}) & = \left( k_1 + k_5 + k_6 \right)\left( k_2 + k_3 + k_4 \right) - \left( k_1 - k_4 \right)\left( k_2 - k_5 \right) \nonumber\\
	 & = k_1\left( k_3 + k_4 + k_5 \right) + k_2\left( k_4 + k_5 + k_6 \right) + k_3\left( k_5 + k_6 \right) + k_4k_6 \nonumber\\
	 &> 0~.
\end{align}
Because $\text{tr}(\textbf{K}) < 0$ and $\text{det}(\textbf{K}) > 0$, the eigenvalues have negative real parts and the
steady state is stable.

However, it is possible, in principle, that the discriminant of the qua\-dra\-\-tic equation,
\begin{equation}
	\text{Disc} = \text{tr}(\textbf{K})^2 - 4\cdot \text{det}(\mathbf{K}) <0~,
\end{equation}
is less than zero, and the system approaches the steady state by damped oscillations.

\textsc{Hirniak} reached this conclusion by a roundabout and difficult route; nonetheless, he saw the possibility of damped oscillations if
$\text{Disc} < 0$, and he looked for a set of rate constant values  $\left\{ k_1,k_2,k_3,k_4,k_5,k_6 \right\}$ that would fulfill the condition.
He also noted that the discriminant could be written as
\begin{equation}
	\text{Disc} = \left( S_1 - S_2 \right)^2 + 4\left( k_1 - k_4 \right)\left( k_2 - k_5 \right)~,
	\label{eq:disc}
\end{equation}
with the substitutions $S_1 = k_1 + k_5 + k_6$ and $S_2 = k_2 + k_3 + k_4$, a substitution \textsc{Wegscheider} already used in his articles. Clearly, it is 
possible to have $\text{Disc} < 0$ only if $k_4 > k_1$ and $k_2 > k_5$ or if $k_1 > k_4$ and $k_5 > k_2$ which are essentially the same conditions because of the symmetry of the reaction network.

It appears that, with these conditions in mind, \textsc{Hirniak} chose for the second term in Eq.~\eqref{eq:disc} 
$k_1 =1$, $k_4 = 30$ and $k_2 = 20$, $k_5 =1$ 
(see his first examples in App.~\ref{app:1908Translation} and App.~\ref{app:1911Translation}).
In this case, the second term in Eq.~\eqref{eq:disc} becomes
$4\cdot (-29)\cdot 19 = -2204$ and the substitutions become $S_1 = 2+ k_6$ and $S_2 = 50+ k_3$. Therefore,  he  had to choose $k_3$ and $k_6$ such that 
\begin{align}
\left((2+k_6)-(50+k_3)\right)^2 -2204 &< 0\\
|48 + k_3 - k_6| & < \sqrt{2204} = 46.95\\
1.05 < k_6 - k_3 & < 94.95 ~.
\end{align}
He chose $k_3 = 2$ and $k_6 = 20$, which gives 
\begin{subequations}
 \begin{align}
    \text{tr}(\textbf{K}) & = -74\\
    \text{det}(\textbf{K}) &= 1695~, 
\end{align}
\end{subequations}
so the eigenvalues of \textbf{K} are $-37 \pm 18i$, which corresponds to a damped oscillation with period = 0.348 and damping factor $\text{e}^{-37t}$. Over the course of one period, the amplitude of the oscillation will be damped by a factor of $\text{e}^{-37 \times 0.348} = \text{e}^{-12.9} = 2.5 \times 10^{-6}$; therefore, the ``oscillation" would be so severely damped as to be undetectable.

Is the problem just a poor choice of parameter values? It is not hard to show that, for \textsc{Hirniak}'s choice of $\left\{ k_1,k_2,k_3,k_4,k_5\right\}$, the damping factor over the course of one oscillation, as a function of $k_6$, is 
\begin{equation}
\text{e}^{-2 \pi (54+k_6) / \sqrt{(k_6-3)(97-k_6)}}~.
\label{eq:k6 damping factor}
\end{equation}
Using Eq.~\eqref{eq:k6 damping factor}, one can find that the damping factor is least extreme for $k_6 = 28.8$, when it has the value $\text{e}^{-12.4} = 4.1 \times 10^{-6}$.

\textsc{Hirniak} tried to find other parameter sets with less extreme damping. For a second example (see his second examples in App.~\ref{app:1908Translation} and App.~\ref{app:1911Translation}), he chose
\begin{equation}
	k_1=1, k_2=601, k_3=1, k_4=501, k_5=1, k_6=2196.426~.
\end{equation}
His odd choice of $k_6$ just barely satisfies the condition Disc $< 0$. In this case, the eigenvalues of \textbf{K} are $-1651 \pm 0.069i$, and the approach to the steady state is oscillatory with period = 91 and a damping factor, over the course of one oscillation, to be $\text{e}^{-1651\times 91} = \num{e-65237}$, which is an unimaginably small number. \textsc{Hirniak} himself realized this problem, writing
\begin{quote}
    ``{\small \foreignlanguage{German}{Der Vorgang spielt sich hier praktisch ganz aperiodisch ab, da von der ersten Exkursion an die Schwingung vollständig vernichtet wird. Trotz meiner ursprünglichen Absicht habe ich weitere Beispiele nicht durchgerechnet$^1$), \ldots}}"\\[2mm]    
    ``{\small The process here is practically completely aperiodic, because the oscillation is completely destroyed after the first oscillations. Despite my original intention, I did not calculate any further examples$^1$), \ldots.}"
\end{quote}
He further stated that he would expect `favorable' damped oscillations for some combinations of rate constant values but writes in  footnote $^1$)
\begin{quote}
    ``{\small \foreignlanguage{German}{Ich hatte ursprünglich die Absicht, systematisch die Rechnung für verschiedene typische Zahlenkombinationen durchzufhren, aber es erwies sich die Arbeit als so mühsam, dass ich davon abgesehen habe.}}"\\[2mm]
    ``{\small Originally, I intended to systematically carry out the calculation for various typical number combinations, but the work turned out to be so tedious that I decided against it.}"
\end{quote}
\textsc{Hirniak} gave up trying to find a `favorable' set of rate constants that would give damped oscillations that were not completely wiped out within the first period of oscillation. It seems that such a set is impossible. But searching for a `favorable' set is pointless, since \textsc{Hirniak}'s model has a much more severe problem.

The main problem with \textsc{Hirniak}'s approach is that \textsc{Wegscheider}  established already in 1901 the \emph{Principle of Detailed Balance}, which states that, for any circular loop of chemical reactions in a closed system governed by the \textit{Law of Mass Action}, the product of the forward rate constants must equal the product of the reverse rate constants.  This principle is a consequence of the reversibility of chemical reactions at thermodynamic equilibrium. This \textsc{Wegscheider} condition can be written as
\begin{equation}
	\prod_{i=1}^N k_i^+ = \prod_{i=1}^N k_i^-~,
	\label{eq:Wegscheider condition}
\end{equation}
with $k_i^+$ and $k_i^-$ as the rate constants in forward and reverse directions, respectively.
For \textsc{Hirniak}'s three isomer system we obtain
\begin{equation}
	k_1 \cdot k_3 \cdot k_5 = k_2 \cdot k_4 \cdot k_6~.
	\label{eq:PDB equation}
\end{equation}
A more detailed review of the \emph{Principle of Detailed Balance} for the three isomer interconversion can be found in Section \ref{sec:The Principle of Detailed Balance for the Interconversion of Three-Isomers}.

In \textsc{Hirniak}'s reaction network, `chemical equilibrium' is a special case of a `steady state' for
which the conditions
\begin{subequations}
	\begin{align}
		k_1 M_1^* & = k_2 M_2^*\\
		k_3 M_2^* & = k_4 M_3^*\\
		k_5 M_3^* & = k_6 M_1^*
	\end{align}
\end{subequations}
need to be fulfilled. Therefore,
\begin{equation}
	M_1^* = \frac{k_2}{k_1}M_2^* = \frac{k_2k_4}{k_1k_3}M_3^* = \frac{k_2k_4k_6}{k_1k_3k_5}M_1^*
\end{equation}
from which \textsc{Wegscheider}'s condition follows.
Insisting that, in this closed system, the total species concentration, defined as $M_\text{T} =M_1^* + M_2^* + M_3^*$ is constant and keeping detailed balance in mind, we can write expressions for the equilibrium concentrations in terms of $M_\text{T}$ and the rate constants $\left\{ k_1,k_2,k_3,k_4,k_5,k_6\right\}$.

In his 1908 paper, \textsc{Hirniak} cited \textsc{Wegscheider} 1901 paper and reproduced much of \textsc{Wegscheier}'s mathematical analysis of the three-isomer system. Therefore, \textsc{Hirniak} should have noticed that his choice of parameter values was in extreme violation of the \emph{Principle of Detailed Balance}
\begin{equation}
	2 = k_1k_3k_5 \neq k_2k_4k_6 = 12,000~.
	\label{eq:detailed balance violation}
\end{equation}
If he had taken a hint from \textsc{Wegscheider}'s principle, \textsc{Hirniak} might have gone on to show that it excludes damped oscillations to the equilibrium state of his closed reaction system.

For an algebraic proof of this claim in \textsc{Hirniak}'s specific case, let's suppose, quite generally, that $k_4 = \alpha\cdot k_1$ and $k_2 = \beta\cdot k_5$ with $\alpha >1$ and $\beta > 1$. In order to satisfy the \emph{Principle of Detailed Balance}, we must insist that $k_3 = \alpha\beta\cdot k_6$. To calculate the discriminant as in Eq.~\eqref{eq:disc}, we need
\begin{subequations}
	\begin{align}
		S_1 &= k_1 + k_5 + k_6,\\
		S_2 &= k_2 + k_3 + k_4 = \beta k_5 + \alpha\beta k_6 + \alpha k_1~,
	\end{align}
\end{subequations}
and 
\begin{equation}
		4\left( k_1 - k_4 \right)\left( k_2 - k_5 \right) = - 4k_1k_5(\alpha - 1)(\beta - 1)~,
\end{equation}
resulting in
\begin{align}
	\text{Disc} & =  [(\alpha - 1)k_1 + (\beta - 1)k_5 + (\alpha\beta - 1)k_6]^2\nonumber \\
	&\;\;\;\; - 4{(\alpha - 1)(\beta - 1)k}_1k_5\\
	& = [(\alpha - 1)k_1 + (\beta - 1)k_5]^2 + 2[(\alpha - 1)k_1 \nonumber\\
	&\;\;\;\; + (\beta - 1)k_5] (\alpha\beta - 1)k_6 \nonumber \\
	&\;\;\;\; + [(\alpha\beta - 1) k_6]^2 - 4(\alpha - 1)(\beta - 1)k_1 k_5\\
	& = [(\alpha - 1)k_1 - (\beta - 1)k_5]^2 + 2[(\alpha - 1)k_1  \nonumber\\
	&\;\;\;\; + (\beta - 1)k_5] (\alpha\beta - 1)k_6 + [(\alpha\beta - 1)k_6 ]^2 \\
	&> 0~.
\end{align}
Since $\text{Disc} > 0$, the \emph{Principle of Detailed Balance} precludes a damped oscillatory approach to equilibrium.

Between the 1940s and 1960s, \textsc{Jost}~\cite{Jost1947}, \textsc{Hearon}~\cite{Hearon1953}, and \textsc{Wei} and \textsc{Prater}~\cite{Wei1962} showed that a set of linear ODEs satisfying the \emph{Principle of Detailed Balance} cannot exhibit damped oscillations of the form 
\begin{equation}
	y(t)= \text{e}^{-kt}\sin(\omega t)~.
\end{equation}

\textsc{Wei} and \textsc{Prater} provided an elegant proof~\cite[pp.~364-371]{Wei1962}, based on the \emph{Principle of Detailed Balance} for a non-oscillatory approach to equilibrium in a closed chemical reaction system governed by
elementary, first-order kinetics $\text{d}\textbf{M}/\text{d}t = \mathbf{KM}$. Here \textbf{M} is an
\textit{n}-dimensional vector of chemical concentrations and \textbf{K} is
an $(n\times n)$-matrix of first-order rate constants given by (in \textsc{Hirniak}'s case)
\begin{equation}
\mathbf{K} = \begin{bmatrix}
	- \left( k_1 + k_6 \right) & k_2 & k_5 \\
	k_1 & - \left( k_2 + k_3 \right) & k_4 \\
	k_6 & k_3 & - \left( k_4 + k_5 \right) \\
\end{bmatrix}~.
\end{equation}

Notice that $\text{det}(\textbf{K}) = 0$, so $\lambda = 0$ is an eigenvalue of \textbf{K} corresponding to the eigenvector
$\textbf{M}^* = (M_1^*, M_2^*, M_3^*)^\text{T}$ of equilibrium concentrations. \textsc{Wei} and \textsc{Prater} introduced a similarity transformation,
$\textbf{X} = \textbf{D}^{-1/2}\textbf{M}$ and $\textbf{L} =
\textbf{D}^{-1/2}\textbf{KD}^{1/2}$, where
\textbf{D} is an $(n \times n)$ diagonal matrix whose elements are $D_{ii} = M_i^*$ = equilibrium concentration
of $\textit{M}_i$ and $D_{ij} = 0$  for $i \ne j$.

For \textsc{Hirniak}'s case  with $n = 3$ one finds
\begin{align}
\frac{\text{d}\mathbf{X}}{\text{d}t} & = \mathbf{D}^{- 1/2}\frac{\text{d}\mathbf{M}}{\text{d}t} \\
 & = \mathbf{D}^{- 1/2}\mathbf{KM} \\
 & = \mathbf{D}^{- 1/2}\mathbf{K}\mathbf{D}^{1/2}\mathbf{D}^{- 1/2}\mathbf{M} \\
 &= \mathbf{LX}~,
\end{align}
where \textbf{L} is the symmetric matrix
\begin{equation}
\mathbf{L} = \begin{bmatrix}
	- \left( k_1 + k_6 \right) & \sqrt{k_1k_2} & \sqrt{k_5k_6} \\
	\sqrt{k_1k_2} & - \left( k_2 + k_3 \right) & \sqrt{k_3k_4} \\
	\sqrt{k_5k_6} & \sqrt{k_3k_4} & - \left( k_4 + k_5 \right) \\
	\end{bmatrix}~.
\end{equation}

Because \textbf{L} is a symmetric matrix, all its eigenvalues are real; hence, no damped (or growing) oscillations are possible. To prove stability of the equilibrium state (i.e., all eigenvalues $\le 0$), \textsc{Wei} and \textsc{Prater} rely on a theorem that the eigenvalues of a symmetric matrix \textbf{L} are non-positive if the quadratic form $\mathbf{\gamma}^\text{T}\mathbf{L\gamma} \le 0$ for an arbitrary vector $\mathbf{\gamma}$. After some algebraic gymnastics, one finds that
\begin{align}
	\mathbf{\gamma}^\text{T} \mathbf{L}\mathbf{\gamma} 
	&= - \left( \gamma_1\sqrt{k_1} - \gamma_2\sqrt{k_2} \right)^2 - \left( \gamma_2\sqrt{k_3} - \gamma_3\sqrt{k_4} \right)^2 \nonumber \\
	&\;\;\;\; - \left( \gamma_3\sqrt{k_5} - \gamma_1\sqrt{k_6} \right)^2 \\
	& \leq 0~,
\end{align}
with equality at the equilibrium state
\begin{equation}
	\mathbf{\gamma} = \left( \sqrt{M_1^{*}},\sqrt{M_2^{*}},\sqrt{M_3^{*}} \right)^{\text{T}}~.
\end{equation}

\subsection{The Principle of Detailed Balance for the Interconversion of Three-Isomers}
\label{sec:The Principle of Detailed Balance for the Interconversion of Three-Isomers}

From Eq.~\eqref{eq:steady state solution} 
we derive
\begin{equation}
\begin{pmatrix}
M_1^{*} \\
M_2^{*} \\
M_3^{*} \\
\end{pmatrix} = \frac{M_\text{T}}{\text{det}(\mathbf{K})}\begin{pmatrix}
k_2k_4 + k_2k_5 + k_3k_5 \\
k_1k_4 + k_1k_5 + k_4k_6 \\
k_1k_3 + k_2k_6 + k_3k_6 \\
\end{pmatrix}
\end{equation}
as the `steady state' solution of the kinetic equations (\ref{eq:initial equation}) describing the interconversion of three isomers in general (i.e., with no restrictions on the rate constants). This steady state solution is found on page 863 of \textsc{Wegscheider}'s 1901 article~\cite{Wegscheider1901}. He went on to show that, at the steady state, there is a constant net flux $\Phi$ through
each reaction in
\begin{align}
    M_1 &\rightarrow M_2~,\nonumber\\
    M_2 &\rightarrow M_3~,\label{eq:Three reactions}\\ 
    M_3 &\rightarrow M_1~,\nonumber
\end{align}
%
given by
\begin{equation}
    \begin{pmatrix}
\Phi_1^{*} \\
\Phi_2^{*} \\
\Phi_3^{*} \\
\end{pmatrix} = \begin{pmatrix}
k_1M_1^{*} - k_2M_2^{*} \\
k_3M_2^{*} - k_4M_3^{*} \\
k_5M_3^{*} - k_6M_1^{*} \\
\end{pmatrix} = \frac{M_\text{T}}{\text{det}(\mathbf{K})}\left( k_1k_3k_5 - k_2k_4k_6 \right)\begin{pmatrix}
1 \\
1 \\
1 \\
\end{pmatrix}.
\end{equation}   
If $k_1 k_3 k_5 = k_2 k_4 k_6$, the net flux is zero, and each interconversion in Eq.~\eqref{eq:Three reactions} is at equilibrium: forward reaction rate = reverse reaction rate (\emph{Principle of Detailed Balance}). \\

If $k_1 k_3 k_5 > k_2 k_4 k_6$, the three reactions in Eq.~\eqref{eq:Three reactions} would be occurring spontaneously. In this case, according to the Second Law of Thermodynamics, which requires that the Gibbs free energy change $\Delta G$ is negative for any spontaneous reaction occurring at constant temperature and pressure,
\begin{align}
    \Delta G_{1 \rightarrow 2} & = G_2^{*} - G_1^{*} < 0~,\nonumber\\
    \Delta G_{2 \rightarrow 3} & = G_3^{*} - G_2^{*} < 0~,\\ 
    \Delta G_{3 \rightarrow 1} & = G_3^{*} - G_1^{*} < 0~.\nonumber
\end{align}
But these inequalities imply that
\begin{equation}
    G_1^* > G_2^* > G_3^* > G_1^*~.
\end{equation} 
which is impossible. Hence, the \emph{Principle of Detailed Balance} is a consequence of the Second Law of Thermodynamics.

\subsection{Comparison of the work by Hirniak and Lotka}
\label{sec:Comparison of Hirniak and Lotka}

\textsc{Lotka}'s 1910 model~\cite{Lotka1910,Lotka1910a} was very different from \textsc{Hirniak}'s 1908/1911 model. \textsc{Lotka}'s
reaction system
\begin{align}
	\text{H} &\rightarrow \text{A} \nonumber\\
	\text{A} + \text{B} &\rightarrow 2 \text{B} \\
	 \text{B} &\rightarrow  \text{C}~,\nonumber
\end{align}
is open ($\text{H} \rightarrow \text{C}$ via the intermediates A and B), irreversible (no loops that must satisfy detailed balance), and nonlinear (synthesis of B from A is autocatalytic). Assuming $[\text{H}]= \text{constant}$, the mass-action rate equations are
\begin{subequations}
	\begin{align}
		\frac{\text{d}A}{\text{d}t} &= k_1 H - k_2 A \cdot B\\
	 	\frac{\text{d}B}{\text{d}t} & = k_2 A \cdot B - k_3B~.
	\end{align}
\end{subequations}
Defining $a  = k_2 A/k_3$, $b = k_2 B/k_3$, $h = k_1k_2H/k_3^2$, and $\tau = k_3 t$ we find that
\begin{subequations}
	\begin{align}
		\frac{\text{d}a}{\text{d}\tau} &= h -ab\\
		\frac{\text{d}b}{\text{d}\tau} & = ab - b~,
	\end{align}
\end{subequations}
with steady state at $a^* = 1$ and $b^* = h$. Defining $x= a - 1$ and $y = b - h$, we obtain
\begin{subequations}
	\begin{align}
		\frac{\text{d}x}{\text{d}\tau} &=-hx -y -xy\\
		\frac{\text{d}y}{\text{d}\tau} & = hx+xy~.
	\end{align}
\end{subequations}

The stability of the steady state here ($x=0, y=0$) is determined by solutions of the linear approximation
\begin{equation}
	\frac{d}{dt}\begin{pmatrix}
		x \\
		y \\
	\end{pmatrix} = \begin{bmatrix}
		- h & - 1 \\
		h & 0 \\
	\end{bmatrix}\begin{pmatrix}
		x \\
		y \\
	\end{pmatrix}~.
\end{equation}
The eigenvalues of the coefficient matrix are
\begin{equation}
	\lambda = \frac{h}{2}\left( - 1 \pm \sqrt{1 - \frac{4}{h}} \right)~,
\end{equation}
which is a complex conjugate pair if $h<4$. For $0 < h << 1$, the damping factor of \textsc{Lotka}'s example is  $\text{e}^{-h \pi / \sqrt{h}} \rightarrow 1$, so the oscillation is negligibly damped for $h$ small.

From \textsc{Lotka}'s example, we learn that sufficient conditions for damped oscillations on the way to a chemical steady state is an open reaction system providing a flux of material and free energy, which then drive a set of spontaneous (irreversible) reactions governed by some nonlinear rate laws. In \textsc{Lotka}'s example, an autocatalytic nonlinearity generates a burst of component B, whose accumulation is limited by depletion of substrate A, which must be replaced from the store of H. If the replacement rate is small enough
\begin{equation}
 k_1 H <  4 \cdot \frac{k_3^2}{k_2}~,   
\end{equation}
the approach to steady state is oscillatory. Two very complete early review articles about oscillatory chemical reactions can be found in \cite{Nicolis1973,Noyes1974}.

For us, it seems unwarranted to maintain that \textsc{Hirniak} proposed a theoretical model for a damped oscillatory chemical reaction system. Furthermore, \textsc{Hirniak}'s claim of priority over \textsc{Lotka}'s 1910 papers~\cite{Lotka1910,Lotka1910a} is also unwarranted, because \textsc{Hirniak}'s `demonstration' of damped oscillations is thermodynamically impossible, and it misses the important conditions for damped oscillations identified by \textsc{Lotka}\footnote{A short discussion of this topic can also be found in \textsc{Cervellati} and \textsc{Greco}'s review of the work by \textsc{Bray} and \textsc{Lotka}.~\cite{Cervellati2017}}.

\textsc{Hirniak}  starts the conclusion in his 1911 paper with ``\foreignlanguage{German}{Zum Schluss habe ich noch folgendes zu bemerken''} (``Finally, I have to note the following") and then writes:
\begin{quote}
    ``{\small \foreignlanguage{German}{Mein Beispiel ist einfacher und zufälligerweise radikaler, als das von Herrn A.\ Lotka, und zwar deshalb, weil es sich auf eine periodische Reaktion in homogenem Systeme bezieht, und eine solche ist bis jetzt noch vollständig unbekannt.
    Der Fall, welchen Herr A.\ Lotka behandelt, ist vielleicht in dieser Beziehung aktueller, da wir einer Theorie der periodischen Reaktionen für heterogene Systeme, ich  möchte sagen, dringend benötigen.}}"\\[2mm]
    ``{\small My example is simpler and coincidentally more radical than that of Mr.\ A.\ Lotka, because it refers to a periodic reaction in a homogeneous system, and such a reaction is still completely unknown.
    The case that Mr.\ A.\ Lotka is dealing with is perhaps more relevant, since we urgently need, I would say, a theory of periodic reactions for heterogeneous systems.}"
\end{quote}
\textsc{Hirniak} is correct that a homogeneous periodic reaction hasn't been discovered in 1911, but heterogeneous systems have been published before. It will take 10 more years before \textsc{Bray} publishes his famous paper in 1921.

In his final paragraph \textsc{Hirniak} writes:
\begin{quote}
    ``{\small Auch wollen mir einige Vernachlässigungen in der Rechnung des Herrn Lotka nicht ganz einwandfrei erscheinen; da aber ihre Diskussion zu weit führen würde, will ich lieber darauf verzichten, bis weitere Untersuchung auf diesem Gebiete ihre Erörterung wichtig erscheinen lässt.}"\\[2mm]
    ``{\small Also, some of the simplifications in Mr.\ Lotka's calculation do not seem entirely correct to me; but since their discussion would be too much, I rather forego it until further research in this area appears to make their considerations important.}"
\end{quote}
In this paragraph, \textsc{Hiniak} seems to refer to some inconclusive calculations by \textsc{Lotka}, but he doesn't specify further. He promises further work in the future but, unfortunately, never published a follow-up article.

\textsc{Hirniak}'s own analysis of oscillating chemical reactions is 
chemically flawed, because it violates the \emph{Principle of Detailed Balance}. Furthermore, it gives the false impression that oscillations are possible in the very mundane case of a set of three first-order inter-conversions of isomers in a closed reaction vessel (what could be simpler?). \textsc{Lotka}'s example is simple and his analysis straightforward. It also gives insight into the peculiar requirements for oscillations: open, nonlinear, irreversible (i.e., far from equilibrium), and self-limited (i.e., negative feedback). 
\textsc{Lotka}'s examples, in 1910 of damped oscillations and in 1920 of sustained oscillations, rely on autocatalytic reactions to provide nonlinearity and 'substrate depletion' to provide negative feedback. Much later, it will become apparent that a nonlinear negative feedback loop alone (without autocatalysis) is sufficient to generate sustained (bio)chemical oscillations, by the work of \textsc{Goodwin}~\cite{Goodwin1965} and \textsc{Griffith} ~\cite{Griffith1968a,Griffith1968b}.

Nonetheless, \textsc{Hirniak}'s legacy could be seen as ``\ldots promoting the possibility of damped oscillations if the \textsc{Wegscheider} constraints are strongly violated." as \textsc{Fiedler} wrote in his 2021 article.~\cite{Fiedler2021}

\subsection{Damped Oscillations in a Cyclic Reaction Driven by Free-Energy Dissipation}
\label{sec:Damped Oscillations}

We showed earlier, that the interconversion of three isomers in a closed reaction vessel, as examined by \textsc{Wegscheider} and \textsc{Hirniak}, cannot exhibit damped oscillations on its way to equilibrium because of a constraint on the rate constants imposed by the  \emph{Principle of Detailed Balance}. Nonetheless, immediately after introducing the constraint ($k_1 k_3 k_5 = k_2 k_4 k_6$), \textsc{Wegscheider} left open the possibility that interesting things might happen if a reaction system were to approach ‘equilibrium’ (a modern physical chemist would say ‘steady state’) in the absence of opposing reactions (i.e., in a circular sequence of reactions). 

\begin{figure}
    \centering
    \includegraphics[width=0.4\linewidth]{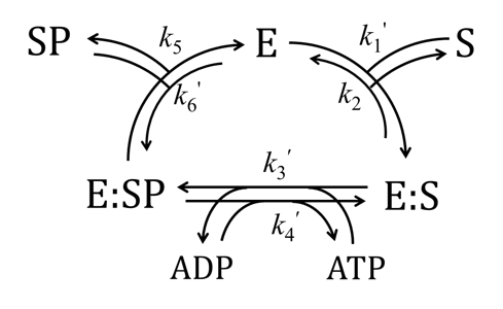}
    \caption{Enzyme E catalyzes the reaction $\text{S} + \text{ATP}  \rightarrow \text{SP} + \text{ADP}$ by a \textsc{Michaelis--Menten} mechanism. Because the standard Gibbs free energy change for this reaction is significantly negative, $\Delta G^\circ \approx -\SI{17}{kJ/mole}$, the equilibrium constant for the overall reaction is $K_\text{eq} \approx 10^3$, implying that the forward rate constants are larger than the reverse rate constants, as detailed in the text. The rate constants $k_2$ and $k_5$ are first-order and $k_1'$, \ldots are second-order. 
    }
    \label{fig:MM}
\end{figure}

Although \textsc{Wegscheider} did not give an example, we know from chemical thermodynamics that the reaction system must be open, with a flow of material and \textsc{Gibbs} free energy through the reactions. One simple possibility is an enzyme kinase E that phosphorylates a substrate S according to \textsc{Michaelis--Menten} kinetics, as illustrated in Fig.~\ref{fig:MM}. This reaction mechanism has the same topology as the three-isomer interconversion reactions, but because of the large free energy dissipation in the reaction, it proceeds irreversibly in a clockwise direction.

Let $E_1(t) = [\text{E}]$, $E_2(t) = [\text{E:S}]$, and $E_3(t) = [\text{E:SP}]$. Then

\begin{equation}
\frac{\text{d}}{\text{d}t}\begin{pmatrix}
E_1 \\
E_2 \\
E_3 \\
\end{pmatrix} = 
\begin{bmatrix}
 - \left( k_1^{'}\text{[S]+}k_6^{'}\text{[SP]} \right) & k_2 & k_5 \\
k_1^{'}\text{[S]} & - \left( k_2 + k_3^{'}\lbrack\text{ATP}\rbrack \right) & k_4^{'}\lbrack\text{ADP}\rbrack \\
k_6^{'}\text{[SP]} & k_3^{'}\lbrack\text{ATP}\rbrack & - (k_5 + k_4^{'}\text{[ADP]}) \\
\end{bmatrix}\begin{pmatrix}
E_1 \\
E_2 \\
E_3 \\
\end{pmatrix}    
\end{equation}

This is \textsc{Hirniak}'s system of equations with
$k_1 = k_1'[\text{S}]$, $k_3 = k_3'[\text{ATP}]$, $k_4 = k_4'[\text{ADP}]$, and $k_6 = k_6'[\text{SP}]$.
In Table \ref{tab:Parameter} we make some reasonable assumptions about the parameters. \\

\begin{table}
    \centering
    \caption{Parameter values for the \textsc{Michaelis--Menten} mechanism used in Fig.~\ref{fig:MM}.}
    \begin{tabular}{llll}\hline
         [S]   = \SI{10}{mM}	& 
         [ATP] = \SI{10}{mM}	& 
         [ADP] = \SI{2}{mM}	&
         [SP]  = \SI{0}{mM} \\
         $k_1' = \SI{0.1}{mM^{-1} min^{-1}}$ &
         $k_3' = \SI{0.1}{mM^{-1} min^{-1}}$ &
         $k_5 = \kappa~\si{min^{-1}}$ &
         $\kappa = \si{constant}$\\
         $k_2 = \SI{0.01}{min^{-1}}$ &
         $k_4' = \SI{e-3}{mM^{-1} min^{-1}}$ &
         $k_6' = \kappa~\si{mM^{-1} min^{-1}}$ &
         \\ \hline
    \end{tabular}
    \label{tab:Parameter}
\end{table}

First, let's check that our rate constants satisfy the \emph{Principle of Detailed Balance}. If the reactions are taking place in a closed
vessel at constant temperature and pressure, then they will reach an
equilibrium state where each reaction has a net flux of zero:
\begin{align}
    k_1'\text{[S]}_\text{eq}\text{[E]}_\text{eq} &= k_2\text{[E:S]}_\text{eq}\nonumber\\
    k_3'\text{[ATP]}_\text{eq}\text{[E:S]}_\text{eq} &= k_4'\text{[ADP]}_\text{eq}\text{[E:SP]}_\text{eq}\\
    k_5\text{[E:SP]}_\text{eq} &= k_6'\text{[SP]}_\text{eq}\text{[E]}_\text{eq}\nonumber
\end{align}
or 
\begin{align}
    K_1 &= \frac{\text{[E:S]}_\text{eq}}{\text{[S]}_\text{eq}\text{[E]}_\text{eq}} = \frac{k_1^{'}}{k_2} = \SI{10}{mM^{-1}}\nonumber\\
    K_2 &= \frac{{\text{[ATP]}_\text{eq}\text{[E:S]}}_\text{eq}}{\text{[ADP]}_\text{eq}\text{[E:SP]}_\text{eq}} = \frac{k_3^{'}}{k_4^{'}} = 10^{2}\\
    K_3 &= \frac{\text{[SP]}_\text{eq}\text{[E]}_\text{eq}}{\text{[E:SP]}_\text{eq}} = \frac{k_5}{k_6'} = \SI{1}{mM},\nonumber
\end{align}
from which we find
\begin{equation}
    K_1K_2K_3 = \frac{\text{[ADP]}_\text{eq} \text{[SP]}_\text{eq}}{\text{[ATP]}_\text{eq} \text{[S]}_\text{eq}} = \frac{k_1'k_3'k_5}{k_2k_4'k_6'} = 10^3~,
\end{equation}
which is the equilibrium constant for the overall reaction $\text{S} + \text{ATP} \rightarrow \text{SP} + \text{ADP}$. From the equilibrium constant, we can calculate
$\Delta G^\circ = -RT \cdot\ln K_\text{eq} \approx -\SI{17}{kJ/mole}$, which is quite reasonable for the phosphorylation of a sugar. 

Given the parameter values in Table~\ref{tab:Parameter}, we have, for the open system of reactions $k_1 = 1$, $k_2 = 0.01$, $k_3 = 1$, $k_4 = 0.002$, $k_5 = \kappa$, $k_ 6 = 0$.
In this case,
\begin{align}
    \text{tr}(\mathbf{K}) & \approx - 2 - \kappa\\
    \text{det}(\mathbf{K}) & \approx 1 + 2\kappa~,\
\end{align}
and
\begin{align}
    \text{Disc} & \approx \kappa(\kappa -4)~.
\end{align}
For the system to execute damped oscillations on the way to steady state, we must have $\kappa < 4$, in which case the eigenvalues of $\mathbf{K}$ are
\begin{equation}
    \lambda  = -\frac{1}{2} (2 + \kappa) \pm i \frac{1}{2}\sqrt{\kappa(4-\kappa)}~,
\end{equation}
which corresponds to damped oscillations with $\text{period} = 4\pi/\sqrt{\kappa(4 - \kappa)}$ and damping factor over the course of one oscillation of
\begin{equation}
    DF = \text{e}^{- 2\pi(2 + \kappa)/\sqrt{\kappa(4 - \kappa)}}~.
\end{equation}
The damping factor has a minimum at $\kappa = 1$, in which case $DF = e^{- 2\pi\sqrt{3}} \approx 2 \cdot 10^{- 5}$. 

Hence, under appropriate conditions, the \textsc{Michaelis--Menten} mechanism for
a phosphatase reaction approaches its steady-state rate of substrate dephosphorylation by damped oscillations (!), but the oscillation is so heavily damped that it would be unobservable. 

In a sense, \textsc{Hirniak}'s theoretical musings in 1908 may have quite a mundane realization, though not the one he thought (the interconversion of three isomers).

\section{Citations of Hirniak's work}
\label{sec:Citations of Hirniak's work}

\textsc{Julian Hirniak} published two articles which are of wider interest for scientists until today. His 1908 article~\cite{Hirniak1908a} was first cited in \textsc{Kremann}'s 1913 book \textit{\foreignlanguage{German}{Die periodischen Erscheinungen in der Chemie}}  (\textit{The periodic phenomena in chemistry})~\cite{Kremann1913_book} but it took more than 70 years, until \textsc{Krug} and \textsc{Kuhnert} cited it again in their 1985 German article~\cite{Krug1985}. \textsc{Hirniak}'s 1911 article~\cite{Hirniak1911} was immediately cited the following year in \textsc{Alfred Lotka}'s famous article in the \textit{\foreignlanguage{German}{Zeitschrift für Physikalische Chemie}}~\cite{Lotka1912}, in a nearly unknown article by \textsc{Max Trautz} in 1912~\cite{Trautz1912}, and in 1913 in \textsc{Robert Kremann}'s book~\cite{Kremann1913_book}.

From a science historical point of view it's interesting how \textsc{Hirniak}'s citations are distributed and, playing a game with words, one could call it oscillatory. In the first nearly 30 years, his contribution to the field of oscillating chemical have been seen very favorably, but the following 40 years resulted in only four citations, all pointing out the problems of his proposed system. Between 1981 and 1991, one can find seven favorable citations and a more critical reflection in \textsc{Anatol Zhabotinsky}'s 1991 `\emph{A History of Chemical Oscillations and Waves}' article~\cite{Zhabotinsky1991a}. This was followed by no citations in the next nearly quarter century before citations of \textsc{Hirniak}'s work started to come back in 2015.

In Section \ref{sec:Wrong publication year}, we explain why many citations of \textsc{Hirniak}'s 1911 article use the wrong publication year 1910. This is followed by Section \ref{sec:Discrepancy in evaluating Hirniak's work} in which we attempt to explain the discrepancy between more favorable and more critical views of \textsc{Hirniak}'s work. After those two, more general sections, we discuss how other scientists cited \textsc{Hirniak}'s 1908 and 1911 articles, using text quotes and translations if necessary. Because \textsc{Lotka} and \textsc{Bray} had the most influence on \textsc{Hirniak}'s citation, we present those in Section \ref{sec:Citations by Alfred Lotka and William Bray}. All other citations of \textsc{Hirniak}'s work can be found in Appendix \ref{app:Favorable and critical citations of Hirniak's work}, highlighting the controversial view of his idea until today.

\subsection{Wrong publication year in bibliographies}
\label{sec:Wrong publication year}

\textsc{Hirniak}'s most cited article from 1911 in the \textit{\foreignlanguage{German}{Zeitschrift für Physikalische Chemie}}~\cite{Hirniak1911} appears in many articles as published in 1910. This wrong year is based on \textsc{Alfred Lotka}'s citations in his 1912 article~\cite{Lotka1912} and his three 1920 articles~\cite{Lotka1920a,Lotka1920b,Lotka1920c}. It is likely that \textsc{Lotka} made the mistake in his articles because he was citing his own 1910 articles in the \textit{\foreignlanguage{German}{Zeitschrift für Physikalische Chemie}}~\cite{Lotka1910} and in \textit{The Journal of physical chemistry}~\cite{Lotka1910a} in conjunction with \textsc{Hirniak}'s work and simply used the same publication year.

\textsc{William Bray} continued using the wrong year in his seminal 1921 paper~\cite{Bray1921}. Together with \textsc{Lotka}'s four articles, all highly important early publications about oscillatory chemical reactions, except of \textsc{Lotka}'s two 1910 article, used the wrong year. Since then, other publications followed this mistake, adding up to 18, which is just two fewer than the citations we found using the correct year of 1911.

Table \ref{tab:Citation table} summarizes the important information about \textsc{Hirniak}'s citations. Besides the year, name of the first author, and the used language of the citing publication, we indicated if the authors had a more favorable (+) or a more critical (-) perception of \textsc{Hirniak}'s work. We indicated two articles with a neutral perception (o) because they discussed the problems in \textsc{Hirniak}'s work but then suggested how to implement other methods to the simple reaction scheme. The last two columns indicate the five publications citing \textsc{Hirniak}'s 1908 articles until now and the year used for \textsc{Hirniak}'s 1911 article. 

\begin{table}
    \centering
    \caption{All publications citing \textsc{Hirniak}'s 1908 and 1911 articles. Columns 1-3: publication year, the first author name, and the publication's language. Column 4: general perception 
    as more favorable (\textcolor{green}{\textbf{+}}), neutral (\textcolor{orange}{\textbf{o}}), or more critical (\textcolor{red}{\textbf{-}}). Columns 5-6: Used citation years. 
     }
   \vspace*{2mm}
    \begin{tabular}{|c|l|l|c|c|c|}
    \hline
        	\textbf{Year} & \multicolumn{1}{|c|}{\textbf{First author}} & \multicolumn{1}{|c|}{\textbf{Language}} & \textbf{Perception} & \multicolumn{2}{|c|}{\textbf{Citation}} \\ \hline
        1912 & Lotka (JPC) \cite{Lotka1912} & German & \textcolor{green}{\textbf{+}} & ~ & \textcolor{red}{1910}  \\ \hline
        1912 & Trautz \cite{Trautz1912} & German & \textcolor{green}{\textbf{+}} & 1908 & \textcolor{green}{1911}  \\ \hline
        1913 & Kremann \cite{Kremann1913_book} & German & \textcolor{green}{\textbf{+}} & ~ & \textcolor{green}{1911}  \\ \hline
        1920 & Lotka (PNAS) \cite{Lotka1920b} & English & \textcolor{green}{\textbf{+}} & ~ & \textcolor{red}{1910}  \\ \hline
        1920 & Lotka (AAAS) \cite{Lotka1920b} & English & \textcolor{green}{\textbf{+}} & ~ & \textcolor{red}{1910}  \\ \hline
        1920 & Lotka (JACS) \cite{Lotka1920c} & English & \textcolor{green}{\textbf{+}} & ~ & \textcolor{red}{1910}  \\ \hline
        1921 & Bray (JACS) \cite{Bray1921} & English & \textcolor{green}{\textbf{+}} & ~ & \textcolor{red}{1910}  \\ \hline
        1926 & Hedges \cite{Hedges1926_book} & English & \textcolor{green}{\textbf{+}} & ~ & \textcolor{green}{1911}  \\ \hline
       1929 & von Raschevsky \cite{Raschevsky1929} & German & \textcolor{green}{\textbf{+}} & ~ & \textcolor{red}{1910}  \\ \hline
        1930 & Skrabal \cite{Skrabal1930} & German & \textcolor{red}{\textbf{-}} & ~ & \textcolor{green}{1911}  \\ \hline
        1931 & Copisarow \cite{Copisarow1931} & German & \textcolor{green}{\textbf{+}} & ~ & \textcolor{green}{1911}  \\ \hline
        1934 & Veil \cite{Veil1934_book2} & French & \textcolor{green}{\textbf{+}} & ~ & \textcolor{green}{1911}  \\ \hline
        1938 & Shemiakin \cite{Shemiakin1938_book} & Russian & \textcolor{green}{\textbf{+}}  & ~ & \textcolor{green}{1911}  \\ \hline
        1939 & Arvanitaki \cite{Arvanitaki1939} & French & \textcolor{green}{\textbf{+}} & ~ & \textcolor{green}{1911}  \\ \hline
        1939 & Frank-Kamenetskii \cite{Frank-Kamenetskii1939} & Russian & \textcolor{green}{\textbf{+}} & ~ & \textcolor{red}{1910}  \\ \hline
        1947 & Jost \cite{Jost1947} & German & \textcolor{red}{\textbf{-}} & ~ & \textcolor{green}{1911}  \\ \hline
        1953 & Hearon \cite{Hearon1953} & English & \textcolor{red}{\textbf{-}} & ~ & \textcolor{green}{1911}  \\ \hline
        1963 & Hearon \cite{Hearon1963} & English & \textcolor{red}{\textbf{-}} & ~ & \textcolor{green}{1911}  \\ \hline
        1973 & Nitzan \cite{Nitzan1973} & English & \textcolor{orange}{\textbf{o}} & ~ & \textcolor{green}{1911}  \\ \hline
        1981 & Pohl \cite{Pohl1981a} & English & \textcolor{green}{\textbf{+}} & ~ & \textcolor{red}{1910}  \\ \hline
        1981 & Pohl \cite{Pohl1981b} & English & \textcolor{green}{\textbf{+}} & ~ & \textcolor{red}{1910}  \\ \hline
        1982 & Fong \cite{Fong1982} & English & \textcolor{green}{\textbf{+}} & ~ & \textcolor{green}{1911}  \\ \hline
        1985 & Krug \cite{Krug1985} & German & \textcolor{green}{\textbf{+}} & 1908 & \textcolor{red}{1910}  \\ \hline
        1987 & Rapp \cite{Rapp1987} & English & \textcolor{green}{\textbf{+}} & ~ & \textcolor{red}{1910}  \\ \hline
        1987 & Bruckner \cite{Bruckner1987} & German & \textcolor{green}{\textbf{+}} & ~ & \textcolor{red}{1910}  \\ \hline
        1991 & Zhabotinsky \cite{Zhabotinsky1991a} & English & \textcolor{red}{\textbf{-}} & ~ & \textcolor{red}{1910}  \\ \hline
        2015 & Gallas \cite{Gallas2015} & English & \textcolor{green}{\textbf{+}} & ~ & \textcolor{red}{1910}  \\ \hline
        2016 & Pechenkin \cite{Pechenkin2016} & Russian & \textcolor{green}{\textbf{+}} & ~ & \textcolor{red}{1910}  \\ \hline
        2016 & Pechenkin \cite{Pechenkin2016b} & English & \textcolor{green}{\textbf{+}} & ~ & \textcolor{green}{1911}  \\ \hline
        2017 & Freire \cite{Freire2017} & English & \textcolor{green}{\textbf{+}} & ~ & \textcolor{red}{1910}  \\ \hline
        2017 & Cervellati \cite{Cervellati2017} & English & \textcolor{red}{\textbf{-}} & 1908 & \textcolor{green}{1911}  \\ \hline
        2018 & Pechenkin \cite{Pechenkin2018_book} & English & \textcolor{green}{\textbf{+}} & ~ & \textcolor{red}{1910}, \textcolor{green}{1911}  \\ \hline
        2018 & Shevchuk \cite{Shevchuk2018} & Ukrainian & \textcolor{green}{\textbf{+}} & 1908 &  \textcolor{green}{1911}  \\ \hline
        2019 & Letellier \cite{Letellier2019_book} & English & \textcolor{green}{\textbf{+}} & ~ & \textcolor{green}{1911}  \\ \hline
        2021 & Fiedler \cite{Fiedler2021} & English & \textcolor{orange}{\textbf{o}} & ~ & \textcolor{green}{1911}  \\ \hline
        2022 & Cherniha \cite{Cherniha2022} & English & \textcolor{green}{\textbf{+}} & 1908 & \textcolor{green}{1911}  \\ \hline
        2022 & Gallas \cite{Gallas2022} & English & \textcolor{green}{\textbf{+}} & ~ & \textcolor{red}{1910} \\ \hline    \end{tabular}
    \label{tab:Citation table}
\end{table}

\subsection{Discrepancy in evaluating Hirniak's work}
\label{sec:Discrepancy in evaluating Hirniak's work}

The importance and impact of \textsc{Hirniak}'s articles published in 1908 and 1911 have been viewed very differently over the last 100+ years. Initially seen in a very positive way until \textsc{Skrabal} pointed towards the chemical problems in his 1930 article. \textsc{Jost} and \textsc{Hearon} discussed similar issues in their 1947 and 1963 articles, respectively. Since then, only \textsc{Zhabotinsky} and \textsc{Cervellati} raised questions in their history of science papers.

As can be seen in Table~\ref{tab:Citation table}, the large majority of authors cite \textsc{Hirniak}'s articles in a favorable context. But one has to distinguish between \textsc{Hirniak}'s proposed reaction scheme and his belief that periodic chemical reactions in homogeneous systems are possible. There was a reason why \textsc{Bray} didn't publish immediately after the first experimental observation. He wanted to be certain that the effect happens in a homogeneous chemical system. Only after several more years working in the lab, he finally published his seminal paper in 1921. And \textsc{Belousov}, the discover of the Belousov-Zhabotinsky reaction, was still unable to publish his now famous oscillatory system in the 1950s.

Many scientists see \textsc{Hirniak} as an early advocate and, perhaps, rely on the fact that \textsc{Lotka} and \textsc{Bray} themselves cited \textsc{Hirniak}'s work (see Section \ref{sec:Citations by Alfred Lotka and William Bray}). \textsc{Hirniak}'s work might also be cited because authors find the terms `\textit{periodic chemical reaction}' in the English translation of his articles.

As it is common among scientists, references are added because of i) the reputation of the citation, ii) its historical significance, iii) others cited the work, iv) key words in the title, or v) the author(s) read the article and therefore chose to add it. \textsc{Hirniak}'s citations are certainly no exception. But, because of \textsc{Lotka} and \textsc{Bray}'s `publication year mistake', one can look at the used publication year of his 1911 article in bibliographies. It is no surprise that the majority of publications in the first 100 years using the correct year were not written in English. They probably had access to the original German article. Therefore, it's also no surprise that all critical articles until today, except of \textsc{Zhabotinsky}'s 1991 `BZ history article', used the correct year of 1911. This includes all history of science publications, because those authors must have read the original article and therefore knew the real publication year.

Detailed discussion of how \textsc{Lotka} and \textsc{Bray} cited \textsc{Hirniak}'s work can be found in Section \ref{sec:Citations by Alfred Lotka and William Bray}. Citations of all other authors are listed and discussed in Appendix \ref{app:Favorable and critical citations of Hirniak's work}.

\subsection{Citations by Alfred Lotka and William Bray}
\label{sec:Citations by Alfred Lotka and William Bray}

As mentioned earlier, \textsc{Alfred Lotka} was the first to cite \textsc{Hirniak}'s 1911 paper~\cite{Hirniak1911} in the introductory paragraph of his 1912 article in the \textit{\foreignlanguage{German}{Zeitschrift für Physikalische Chemie}}~\cite{Lotka1912}. He wrote:
\begin{quote}
     ``{\small \foreignlanguage{German}{In einer frühern Mitteilung$^{1)}$ wurde ein Fall einer gewissen Reihe von Folgereaktionen mit autokatalytischer Beschleunigung behandelt, wobei sich ergab, dass unter gewissen Um\-stän\-den der Reaktionsverlauf oscillatoriech wird. Für die auftretenden Differentialgleichungen wurde derzeit eine angenäherte Lösung angegeben, welche allerdings nur in der unmittelbaren Nähe des schliesslichen stationären Zustande gültig ist, wie dies schon in der ursprünglichen Mitteilung angedeutet und später noch von J. Hirniak$^{2)}$ besonders hervorgehoben wurde.}}"\\[2mm]
    ``{\small In an earlier communication$^{1)}$ an example of a certain series of subsequent reactions with autocatalytic acceleration was discussed, whereby it emerged that under certain circumstances the course of the reaction becomes oscillatory. At that time, an approximate solution for the differential equations was given which, however, is only valid in the immediate vicinity of the final stationary state, as already indicated in the original communication and later particularly emphasized by J. Hirniak$^{2)}$.}"
\end{quote}
The references $^{1)}$ and $^{2)}$ refer to \textsc{Lotka}'s 1910~\cite{Lotka1910} and \textsc{Hirniak}'s 1911~\cite{Hirniak1911} papers, respectively. But, as mentioned before, \textsc{Lotka} used the year 1910 in the bibliography entry. \textsc{Hirniak} mentioned his 1908 paper in his 1911 paper, but \textsc{Lotka} did not include this publication as a citation in any of his own publications. Unfortunately, we will never know if this was a conscious decision by \textsc{Lotka} to keep his 1910 article the first theoretical description of a homogeneous periodic chemical reaction, just an oversight, if he thought that the article in  Ukrainian wasn't worth citing, or any other reason.

\textsc{Lotka} also cited \textsc{Hirniak}'s 1911 paper in the introductory paragraph of his 1920 \textit{Proceedings of the National Academy of Sciences} paper~\cite{Lotka1920a}. He writes:
\begin{quote}
    ``{\small In chemical reactions rhythmic effects have been observed experimentally, and have also been shown, by the writer$^1$ and others,$^2$ to follow, under certain conditions, from the laws of chemical dynamics.}"
\end{quote}
The references $^1$~\cite{Lotka1910a,Lotka1910,Lotka1912,Lotka1912a,Lotka1920b} and $^2$~\cite{Hirniak1911,Lowry1910} are given as:
\begin{quote}
    $^1 $Lotka, A.~J., \textit{J.\ Phys.\ Chem.}, \textbf{14}, 1910 (271-274); \textit{Zs.\ physik.\ Chem.}, \textbf{72}, 1910 (508-
511); \textbf{80}, 1912 (159-164); \textit{Phys.\ Rev.}, \textbf{24}, 1912 (235-238); \textit{Proc.\ Amer.\ Acad.\ Arts Sci.}, 55, 1920 (137-153).
$^2$ Hirniak, J., \textit{Zs.\ physik.\ Chem.}, \textbf{75}, 1910 (675); compare also Lowry and John, \textit{J.\ Chem.\ Soc.}, \textbf{97}, 1910 (2634-2645).
\end{quote}
\textsc{Lotka}'s reference `\textit{Phys.\ Rev.}, \textbf{24}, 1912 (235-238)'~\cite{Lotka1912a} was a presentation at the sixty-first meeting of the American Physical Society on March, 2, 1912, though the volume number should be \textbf{34} instead of \textbf{24}. The article by \textsc{Lowry} and \textsc{John}~\cite{Lowry1910} describes a chemical reaction with two consecutive unimolecular changes, not resulting in any oscillatory behavior.

In the same year, \textsc{Lotka} published an article in the \textit{Proceedings of the American Academy of Arts and Sciences}~\cite{Lotka1920b} on a related concept and wrote in Section `General Case':
\begin{quote}
    ``{\small In general some or all of these roots may be complex, so that in general a more or less complicated series of reactions may give rise to oscillatory phenomena.$^{16}$}"
\end{quote}
Here, the reference $^{16}$ lists three articles in the \textit{\foreignlanguage{German}{Zeitschrift für Physikalische Chemie}}: \textsc{Lotka}'s 1910~\cite{Lotka1910}, \textsc{Lotka}'s 1912~\cite{Lotka1912},  and \textsc{Hirniak}'s 1911~\cite{Hirniak1911} articles, without further distinction.

The most important reference to \textsc{Hirniak}'s work can probably be found in \textsc{Lotka}'s famous 1920 \textit{JACS} article ``Undamped oscillations derived from the law of mass action''~\cite{Lotka1920c}. This article starts with
\begin{quote}
    ``{\small It was shown by the writer on a former occasion$^1$ that the course of a chemical reaction, as computed from the laws of chemical dynamics, may, in certain circumstances, assume an oscillatory character.
    The case considered led, however, to damped oscillations fading off into equilibrium, not to a continued periodic process;\ldots}"
\end{quote}
Citation $^1$ then lists ``A.~J.~Lotka, \textit{J.\ Phys.\ Chem.}, \textbf{14}, 271 (1910); \textit{Z.\ physik.\ Chem.}, \textbf{72}, 508 (1910); \textbf{80}, 159 (1912); see also Hirniak, \textit{ibid.}, \textbf{75}, 675 (1910); and compare also Lowry and John, \textit{J.\ Chem.\ Soc.}, \textbf{97}, 2634 (1910); Rakowski, \textit{Z.\ physik.\ Chem.}, \textbf{57}, 321 (1906).''
The corresponding articles are \cite{Lotka1910a,Lotka1910,Lotka1912, Hirniak1911,Lowry1910,Rakowski1907}.
The last article by \textsc{Adam Rakowski} also describes a linear reaction scheme but limited to first orders. Similar to \textsc{Hirniak}'s 1911 paper, \textsc{Lotka} also used a wrong publication year for \textsc{Rakowski}'s article. It should be 1907 instead of 1906.

In \textsc{William Bray}'s 1921 seminal publication ``A periodic reaction in homogeneous solution and its relation to catalysis''~\cite{Bray1921}, he presents the first experimental evidence of a periodic reaction in a homogeneous system. \textsc{Bray} discovered an oscillating iodine concentration during the catalytic conversion of hydrogen peroxide to oxygen and water by iodate. After presenting his experimental results, \textsc{Bray} writes
\begin{quote}
    ``{\small \ldots the writer believes the present example to be the first instance of a periodic reaction in homogeneous solution. The possibility of such periodicity had, however, been appreciated. Lotka$^4$ and Hirniak$^5$ independently examined the problem; each assumed a definite mechanism for a hypothetical reaction, set up differential equations for the various intermediate reactions, and by a mathematical analysis setup conditions sufficient to account for periodicity.}"
\end{quote}
In citation $^4$ he lists ``Lotka, \textit{J.\ Phys.\ Chem.}, \textbf{14}, 271 (1910); reprinted \textit{Z.\ physik.\ Chem.}, \textbf{72}, 508 (1910); \textsc{This Journal}, \textbf{42}, 1595 (1920)'' and in $^5$ he refers to ``Hirniak, \textit{Z.\ physik.\ Chem.}, \textbf{75}, 675 (1910)''.

Here, \textsc{Hirniak}'s 1911 article has been reported, again, as published in 1910, the same year as \textsc{Lotka}'s two first papers on damped periodic reactions. Because of this citations and the error by \textsc{Lotka}, \textsc{Hirniak}'s 1911 paper has been, most of the time, cited with the wrong year.

Following the quoted sentences, \textsc{Bray} discusses the autocatalytic reaction within \textsc{Lotka}'s reaction scheme as an essential feature but does not mention \textsc{Hirniak}'s name or reaction scheme at all. Therefore, it is questionable if \textsc{Bray} consciously chose to cite \textsc{Hirniak}'s paper because of its content, or if he simply added the citation because \textsc{Lotka} used the same citation in his 1912 and 1920 papers.

\section{Conclusion}

In 1908, the Ukrainian scientist, \textsc{Julian Hirniak}, published an article in a Ukrainian journal claiming to show that the interconversion of three isomers can, under certain conditions, approach equilibrium by damped oscillations. His article was undoubtedly the first publication to propose the possibility of oscillatory behaviour in a homogeneous chemical solution, and, as such, it has received considerable attention in the scientific literature since 1912. We started this project to resolve some historiographical uncertainties about this publication, and also to re-evaluate \textsc{Hirniak}’s contribution to the theory of oscillatory chemical reactions, in light of his controversy with \textsc{Alfred Lotka} over priority for the idea.

The historical issues involve his first name (\textsc{Julian} or \textsc{Julius}) and the publication date (1910 or 1911) of his more accessible 1911 paper in \textit{\foreignlanguage{German}{Zeitschrift für Physikalische Chemie}}~\cite{Hirniak1911}, which is very similar to his 1908 Ukrainian article~\cite{Hirniak1908a}. \textsc{Hirniak} published his work in Ukrainian, Polish, and German but never in English. The use of \textsc{Julius} as his first name in bibliographies is a result of his \textit{\foreignlanguage{German}{Zeitschrift}} article, where he used \textsc{Julius Hirniak}, although standard translation of his Ukrainian first name \textsc{\foreignlanguage{Ukrainian}{Юліан}} into German, or any other non-Cyrillic language, would have never resulted in \textsc{Julius}. More confusing still has been the practice of citing his 1911 \textit{\foreignlanguage{German}{Zeitschrift}} article as ‘1910’ in bibliographies since the 1920s, as a result of oversights by Alfred Lotka and William Bray. Lotka used the wrong year when citing \textsc{Hirniak} in his famous 1912 article~\cite{Lotka1912} and in three of his 1920 articles ~\cite{Lotka1920a,Lotka1920b,Lotka1920c} and \textsc{Bray} used the wrong year in his 1921 article~\cite{Bray1921}.

In addition to resolving these citation issues, we wanted to address the question of whether \textsc{Julian Hirniak} is a `forgotten pioneer' of oscillatory chemical reactions, overshadowed by \textsc{Alfred Lotka} because \textsc{Hirniak} did not publish in English journals, or is he a `misguided proponent' of chemical oscillations, who made a serious mistake in his analysis of the interconversion of three isomers? To answer this question, we translated \textsc{Hirniak}’s Ukrainian 1908 article~\cite{Hirniak1908a} and his 1911 German article~\cite{Hirniak1911} into English. Besides the 29 favorable references to \textsc{Hirniak}’s work that we found, including those by \textsc{Lotka} and \textsc{Bray}, only six critical reflections have been published between 1930 and 2017. Most of the favorable citations seem to be re-citations of references to \textsc{Hirniak} found in \textsc{Lotka} and \textsc{Bray}’s bibliographies, because they cite the wrong publication year 1910. Another reason might be the original article title ``\foreignlanguage{German}{Zur Frage der periodischen Reaktionen}" (``On the question of periodic reactions"). If an author searches for early work about oscillatory chemical reactions, \textsc{Hirniak}’s paper would seem to be a reasonable citation.

Our analysis of \textsc{Hirniak}’s work shows that he was a visionary scientist who was convinced - early on - that a homogeneous chemical reaction system can exhibit damped oscillatory behavior. Unfortunately, a close examination of his example of three-isomer interconversions reveals that this simple reaction scheme cannot exhibit damped oscillations without violating the Second Law of Thermodynamics, embodied in \textsc{Wegscheider}’s \emph{Principle of Detailed Balance}\footnote{A translations of the important Section IV of \textsc{Rudolf Wegscheider}'s 1901 article~\cite{Wegscheider1901} can be found in Appendix \ref{app:1901Translation} (\textbf{\emph{Translation of Section IV of Wegscheider's 1901 article}}).}. Comparing their papers (see translations in App.~\ref{app:1908Translation} and \ref{app:1911Translation}), it is obvious that \textsc{Hirniak} borrowed much of his analysis of the three-isomer example from \textsc{Wegscheider}’s 1901 German article~\cite{Wegscheider1901}, so he knew about the restrictive conditions introduced by \textsc{Wegscheider}. Nonetheless, \textsc{Hirniak} tried to explain the absence of any known oscillations in homogeneous chemical reaction systems as an experimental problem, and he hoped that the rapid damping of oscillations that he observed in his theoretical model (which rendered periodic behavior impossible to observe) could be resolved later by a better choice of parameters or experimental detection methods.

It turns out that \textsc{Julian Hirniak} was correct that oscillating chemical reactions are possible in homogeneous solution, as would soon be demonstrated by \textsc{Bray} and \textsc{Liebhafsky}~\cite{Bray1921,Bray1931b}, but his reasoning was flawed. \textsc{Alfred Lotka}’s reasoning, on the other hand, was correct in identifying the key requirements for both damped and sustained chemical oscillations: an open reaction system driven out-of-equilibrium by the dissipation of free energy, governed by nonlinear reaction rate laws, and involving negative feedback.

Nonetheless, prompted by \textsc{Hirniak}'s idea of a cyclic set of linear chemical reactions and \textsc{Lotka}'s idea of an open reaction system operating far-from-equilibrium, we re-considered the case of the enzyme-catalyzed phosphorylation of sugar molecules, driven by the spontaneous hydrolysis of ATP to ADP. We showed that this system of chemical reactions, while satisfying \textsc{Wegscheider}'s \emph{Principle of Detailed Balance}, can, indeed, exhibit damped oscillations on its way to its steady-state rate of product formation (i.e., the classic \textsc{Michaelis--Menten} rate law). However, as noted by \textsc{Hirniak} in 1908, the oscillations are so highly damped as to be unobservable, which accounts for the fact that biochemists have never seen such behavior in enzyme-catalyzed reactions. 

\bmhead{Acknowledgements}

YH thanks \textsc{Oleh Petruk} for providing the photo of Julian Hirniak, and most importantly for implementing the project that resulted in publishing the book \cite{Holovatch2021_book-chapter}. YH also acknowledges long-standing collaboration with \textsc{Maksym Dudka}, \textsc{Yulian Honchar}, and \textsc{Mariana Krasnytska} on some of the issues discussed in this paper. 
NM and YH thank \textsc{Kostyantyn Kurylyshyn} for his hospitality at the Stefanyk National Scientific Library in Lviv.
We thank \textsc{Martin Rohde} for useful discussions. 
NM thanks the Shevchenko Scientific Society in New York City (especially \textsc{Marko Slyz}) for providing a copy of Hirniak’s 1908 article, which proved very difficult to obtain, \textsc{Zach Rewinski} for his linguistic help, and the Hamburger Collaborative Project Fund at The College of Wooster for support, which made this project possible.

\newpage

\begin{appendices}


\newpage
\section{Julian Hirniak's publications}
\label{app:Julian Hirniak's publications}
\setcounter{footnote}{0}

Some of \textsc{Hirniak}'s research papers\footnote{Other references, listing some of \textsc{Hirniak}'s publications are \cite{Hryvnak2011,Shevchuk2018,Holovatch2018}.} were published in the Ukrainian journal  
\textit{\foreignlanguage{Ukrainian}{Збiрник Ма\-те\-матично-Природописно-Лїкарської Секциї Наукового Товарист\-ва імени Шевченка}}, which also had the German title 
\textit{\foreignlanguage{German}{Sammelschrift der Mathematisch-Naturwissenschaftlich-Ärztlichen Sektion der \v{S}ev\v{c}enko-Ge\-sell\-schaft der Wissenschaften in Lemberg}} at that time. The English translation is \textit{Proceedings of the Mathematical-Natural Scien\-ces-Medical Section of the Shevchenko Scientific Society in Lviv}. The journal was established in 1897 and published in Lviv until 1939. Though the journal title and the table of content was bilingual, the actual articles were only in Ukrainian. 
This journal will be abbreviated as \foreignlanguage{Ukrainian}{ЗМПЛС} in the publication list. In Fig. \ref{fig:1908_coverpage} we show the cover page of the 1908, Vol.~XII journal issue.

\begin{enumerate}
    \item 
    \foreignlanguage{Ukrainian}{Юлiян Гiрняк},
    ``\foreignlanguage{Ukrainian}{Роля сталої, плинної i ґазової фази в хемiчнiй рiвновазї}"
    (``The role of the solid, liquid and gas phase in chemical equilibrium"),
     \foreignlanguage{Ukrainian}{ЗМПЛС} (1903)~\cite{Hirniak1903}.
    
    \item 
    \foreignlanguage{Ukrainian}{Юлiян Гiрняк},
    ``\foreignlanguage{Ukrainian}{Ненастанна деґрадация енерґiї конечна проява й причина всякого руху i житя в природi}"
    (``Constant degradation of energy is an inevitable manifestation and cause of all movement and of life in nature"),
     \foreignlanguage{Ukrainian}{Літературно-науковий вістник}\footnote{A monthly journal published in 1898–1906 in Lviv, in 1907–1914 and 1917–1919 in Kyiv, and in 1922–1932 again in Lviv. The cover page of volume 21 is shown in Fig.~\ref{fig:1903_coverpage}.} (Literary-Scientific Herald) \textbf{21} (1903)~\cite{Hirniak1903a}. 
    \begin{figure}
        \centering
        \includegraphics[width=0.4\linewidth]{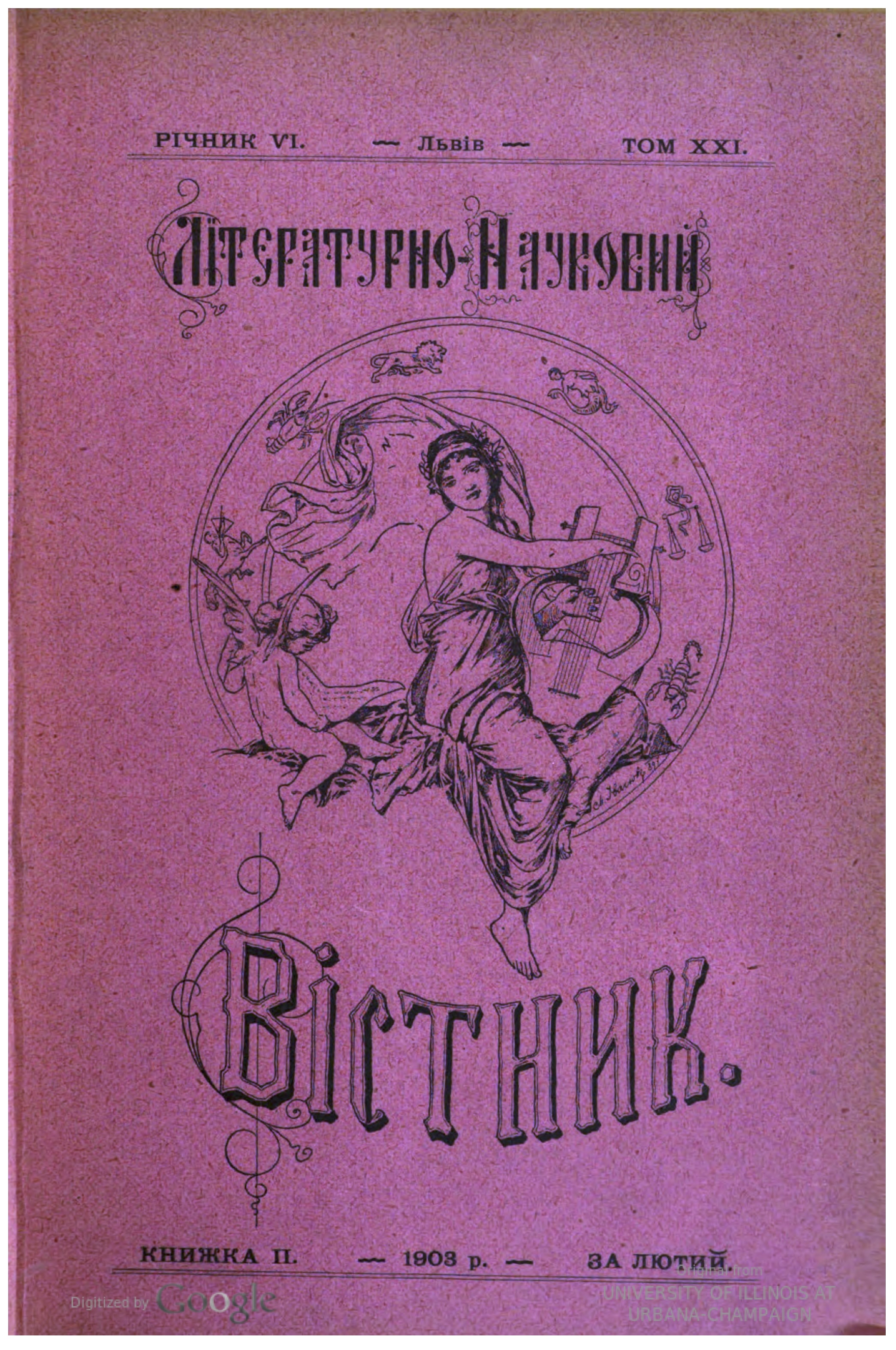}
        \caption{Cover page of \textit{\foreignlanguage{Ukrainian}{Літературно-науковий вістник}}, Vol.~21 from 1903 (Public Domain. Digitized from original document provided by the Stefanyk National Scientific Library in Lviv).
         }
        \label{fig:1903_coverpage}
    \end{figure}

    \item 
    \foreignlanguage{Ukrainian}{Юлiян Гiрняк},
    ``\foreignlanguage{Ukrainian}{О проводї тепла цукру у воднiм розчинї}" (``On the heat conduction of sugar in an aqueous solution"),
    \foreignlanguage{Ukrainian}{ЗМПЛС} (1907)~\cite{Hirniak1907}.
    
    \item 
    \foreignlanguage{Ukrainian}{Юлiян Гiрняк},
    ``\foreignlanguage{Ukrainian}{Про вплив синхронїчної змiни концентрациї на хiд мономолекулярної реакциї}" (``On the effect of a synchronous change in concentration on flow of the monomolecular reaction"),
    \foreignlanguage{Ukrainian}{ЗМПЛС}  (1908)~\cite{Hirniak1908b}.

    \item 
    \foreignlanguage{Ukrainian}{Юлiян Гiрняк},
    ``\foreignlanguage{Ukrainian}{Про перiодичнi хемiчнi реакциї}" (``On periodic chemical reactions"),
    \foreignlanguage{Ukrainian}{ЗМПЛС} (1908)~\cite{Hirniak1908a}. The cover page of this volume is shown in Fig.~\ref{fig:1908_coverpage}.
    \begin{figure}
        \centering
        \includegraphics[width=0.4\linewidth]{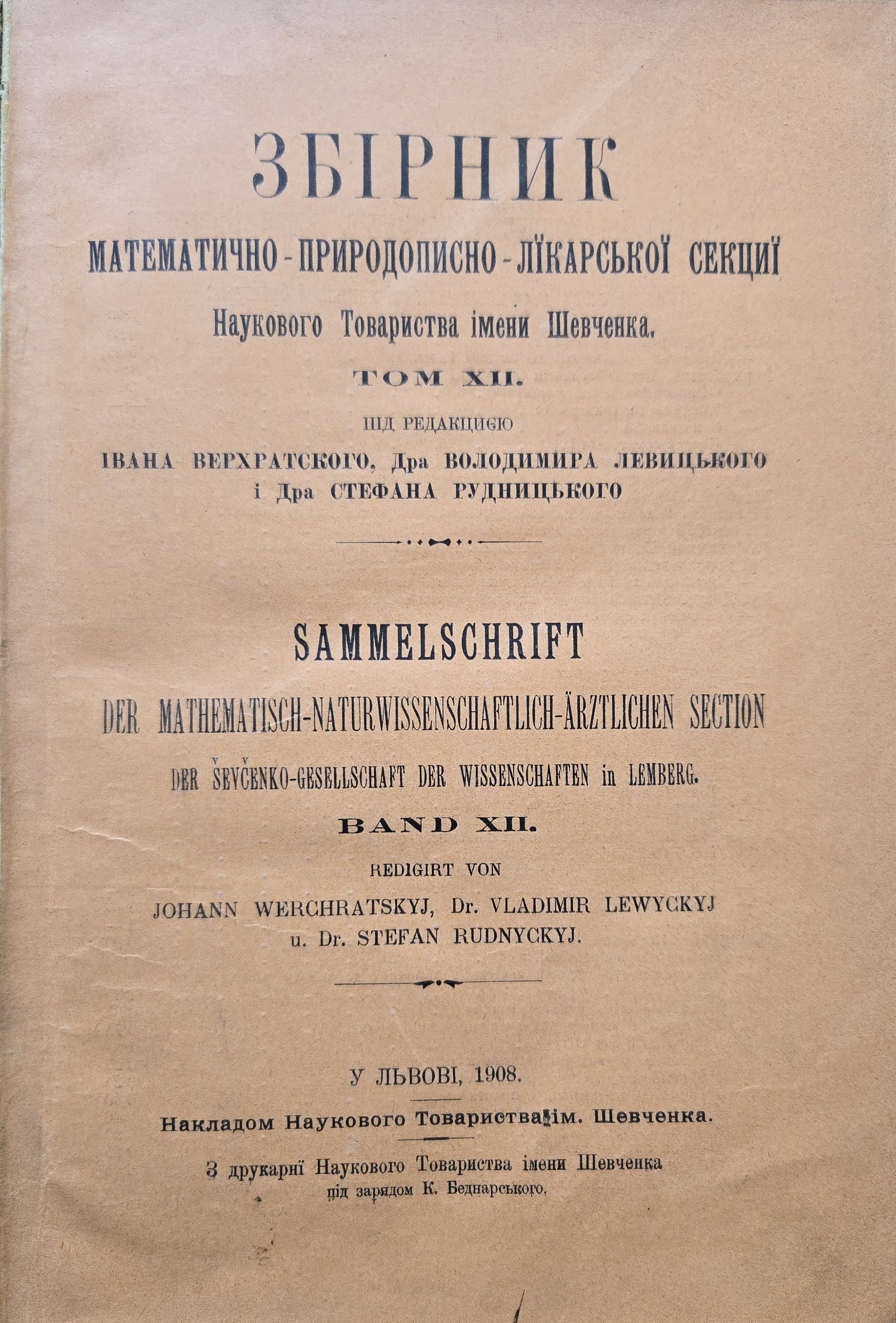}
        \caption{Cover page of \textit{\foreignlanguage{Ukrainian}{ЗМПЛС}}, Vol.~XII from 1908 (Public Domain. Digitized from original document provided by the Stefanyk National Scientific Library in Lviv).
        }
        \label{fig:1908_coverpage}
    \end{figure}
  
    \item 
    \foreignlanguage{Ukrainian}{Юлiян Гiрняк},
    ``\foreignlanguage{Ukrainian}{Замiтки до рiвнань моно- i бi-молекулярної хемiчної кiнетики}" (``Notes on the equations of mono- and bi-molecular chemical kinetics"),
    \foreignlanguage{Ukrainian}{ЗМПЛС} (1909)~\cite{Hirniak1909a}.
    
    \item 
    \foreignlanguage{Ukrainian}{Юлiян Гiрняк},
    ``\foreignlanguage{Ukrainian}{Вплив температури на скорiсть декiлькох хемiчних реакцiй}" (``Effect of temperature on the rate of several chemical reactions"),
    \foreignlanguage{Ukrainian}{ЗМПЛС} (1909)~\cite{Hirniak1909b}.

    \item 
    \foreignlanguage{Ukrainian}{Юлiян Гiрняк},
    ``\foreignlanguage{Ukrainian}{Вплив температури на скорiсть декiлькох хемiчних реакцiй (доповненє)}" (``Effect of temperature on the rate of several chemical reactions (Supplement)"),
    \foreignlanguage{Ukrainian}{ЗМПЛС} (1910)~\cite{Hirniak1910}.

    \item 
    Julius Hirniak,
    ``\foreignlanguage{German}{Zur Frage der periodischen Reaktionen}" (``On the question of periodic reactions"),
    \foreignlanguage{German}{Zeitschrift für Physikalische Chemie} (1911)~\cite{Hirniak1911}.

    \item 
    Julius Hirniak,
    ``\foreignlanguage{German}{Beiträge zur chemische Kinetik. I. (In zwei Teilen)}" (``Contributions to chemical kinetics. I (In two parts)"),
    {\v S}ev{\v c}enko-Gesellschaft der Wissenschaften in Lemberg, pp.~101 (1911)~\cite{Hirniak1911_book}. \\
    $[$Note: This work was intended for his habilitation (from \cite{Hirniak1941})$]$.
    
    \item 
    \foreignlanguage{Ukrainian}{Юлiян Гiрняк},
    ``\foreignlanguage{Ukrainian}{Мінеральоґия і хемія: Начерк мінеральоґиї і хемії для середних шкіл}" 
    (``Mineralogy and Chemistry: Outline of Mineralogy and Chemistry for Secondary Schools"),
    \foreignlanguage{Ukrainian}{Руське Товариство Педагогiчне} (Ruthenian Pedagogical Society)\footnote{The Ruthenian Pedagogical Society [Rusjke Tovarysto Pedahohichne], later Ukrainian Pedagogical Society [Ukrajinsjke Pedahohichne Tovarystvo] (popular name - Ridna Shkola), was established in Lviv in 1881 to promote Ukrainian-language education.}(1912)~\cite{Hirniak1912_book}.

    \item 
    \foreignlanguage{Ukrainian}{Юлiян Гiрняк},
    ``\foreignlanguage{Ukrainian}{Дещо про теоретичне i методичне значiнє сочинника температури скоростий процесiв для хемiчної кiнетики}" (``About the theoretical and methodical value of the temperature component of the rates of processes for chemical kinetics"),
    \foreignlanguage{Ukrainian}{ЗМПЛС} (1913)~\cite{Hirniak1913}.

    \item 
    \foreignlanguage{Ukrainian}{Юлiян Гiрняк},
    ``\foreignlanguage{Ukrainian}{Основи хемії для висших кляс гімназияльних}" (``Basics of chemistry for upper classes in a gymnasium"),
    \foreignlanguage{Ukrainian}{Українськe Товариство Педагогічнe} (Ukrainian Pedagogical Society) (1914)~\cite{Hirniak1914_book}.

   \item 
    \foreignlanguage{Ukrainian}{Юлiян Гiрняк},
    ``\foreignlanguage{Ukrainian}{Кілька слів про наукову діяльність нашого земляка Володимира Бачинського}" (``A few words about the scientific activity of our compatriot Volodymyr Bachynskyi"),
    \foreignlanguage{Ukrainian}{ЗМПЛС} (1919)~\cite{Hirniak1919}.

    \item 
    \foreignlanguage{Ukrainian}{Юлiян Гiрняк},
    ``\foreignlanguage{Ukrainian}{Організуймо великий рідний капітал для українських кооператив!}" (``Let's organize a large native capital for Ukrainian cooperatives!"),
    \foreignlanguage{Ukrainian}{Краєве товариство «Ощадність»} (Regional Society «Oshchadnistj» («Frugality»))  (1927)~\cite{Hirniak1927_book}.
    
    \item Juljan Hirniak,
    ``\foreignlanguage{polish}{O zasadniczych poj{\c e}ciach, nekt{\'o}rych nowszych kierunkach i ogolnych paradoksach chemicznej kinetyki}" (``About fundamental concepts, some new trends and general paradoxes of chemical kinetics"),
    Towarzystwo Aptekarskie we Lwowie (Pharmacy Society in Lviv)  (1936)~\cite{Hirniak1936_book}. \\
    $[$Note: \foreignlanguage{polish}{Odczyt wygloszony dnia 28 maja 1936 na I. posiedzeniu Oddzialu Lwowskiego Zwi{\c a}zku Chemik{\'o}w Polskich} (Lecture delivered on May 28, 1936 at the first meeting of the Lviv Association Branch of Polish Chemists).$]$

    \item Julian Hirniak,
    ``\foreignlanguage{German}{Zur Kinetik der Folgereaktionen zweiter Ordnung}" (``About the kinetics of second order subsequent reactions"),
    Acta Physicochimica U.R.S.S. (1941)~\cite{Hirniak1941}.

\end{enumerate}

In 1909 \textsc{Hirniak} also reviewed seven German publications for the \foreignlanguage{Ukrainian}{Бібліографія} (Bibliography) section of \foreignlanguage{Ukrainian}{ЗМПЛС}, the same volume he published his two 1909 articles in~\cite{Hirniak1909a,Hirniak1909b}. These reviews appeared on pages 26 -- 34 and included one book by \textsc{Walter Nernst}~\cite{Nernst1909_book}, three articles in \foreignlanguage{German}{\textit{Annalen der Physik}} by \textsc{Grüneisen}~\cite{Grueneisen1908a,Grueneisen1908b} and \textsc{Engler}~\cite{Engler1908}, two articles in the \foreignlanguage{German}{\textit{Physikalische Zeitschrift}} by \textsc{Schlesinger}~\cite{Schlesinger1909} and \textsc{Goldschmidt}~\cite{Goldschmidt1909}, and one article by \textsc{Krüger} in the \foreignlanguage{German}{\textit{Nachrichten von der Königlichen Gesellschaft der Wissenschaften zu Göttingen}}~\cite{Krueger1908}.
\begin{enumerate}
    \item On page 26: \foreignlanguage{German}{W.~Nernst – Theoretische Chemie vom Standpunkte der Avogadroschen Regel und der Thermodynamik. (VI. Auflage. Stuttgart, 1909).} 
    \item On page 27: \foreignlanguage{German}{H.~Schlesinger – Die spezifischen Wärmen von Lösungen. I., Physikalische Zeitschrift. \textbf{10}(6), 210-215 (1909).}
    \item On page 27: \foreignlanguage{German}{E.~Grüneisen – Über die thermische Ausdehnung und die spezifische Wärme der Metalle, Annalen der Physik, \textbf{331}(7) 211-216 (1908).}
    \item On pages 28--30: \foreignlanguage{German}{E.~Grüneisen – Zusammenhang zwischen Kompressibilität, thermischer Ausdehnung, Atomvolumen und Atomwärme der Metalle. Annalen der Physik, \textbf{331}(7) 393--402 (1908).}
    \item On pages 30--31: \foreignlanguage{German}{H.~Goldschmidt – Über die Abhängigkeit der Reaktionsgeschwindigkeit von der Temperatur in homogenen gasförmigen Systemen, Phy\-si\-ka\-lische Zeitschrift, \textbf{10}(6), 206--210 (1909).}
    \item On pages 31--33: \foreignlanguage{German}{F.~Krüger – Zur Kinetik des Dissotiationsgleichgewichtes und der Reaktiongeschwindigkeit, Göttinger-Nachrichten, (4), 318-336 (1908).}
    \item On page 34: \foreignlanguage{German}{W.~Engler – Über den Einfluß der Temperatur auf radioaktive Umwandlungen, Annalen der Physik, \textbf{331}(8), 483--520 (1908).}
\end{enumerate}

\newpage
\section{Favorable and critical citations of Hirniak's work}
\label{app:Favorable and critical citations of Hirniak's work}
\setcounter{footnote}{0}

\subsection{Favorable citations}
\label{app:Favorable citations of Hirniak's work}

The first decades viewed \textsc{Hirniak}'s work very positive. \textsc{Lotka} and \textsc{Bray} cited \textsc{Hirniak}'s publications when discussing the theoretical possibility of a periodic chemical reaction. Since then, many other authors followed their perception.\\

\noindent $\bullet$ The second article citing \textsc{Hirniak} had been published by \textsc{Max Trautz} in October 1912~\cite{Trautz1912} (seven months after \textsc{Lotka}'s 1912 paper). He discusses the chemical kinetics and reaction velocities of reversible successive reaction steps and first presents the purely mathematical result published by \textsc{Jüttner} in 1911 that the equilibrium point will be approached aperiodically~\cite{Juettner1911}. He continues describing a system of two simultaneous monomolecular reactions with an autocatalytic second step and states that the intermediate isomer concentration can't go through a maximum, citing \textsc{Lowry} and \textsc{John}'s 1910 article~\cite{Lowry1910}. But under certain conditions, the systems can show periodic behavior, as outlined by \textsc{Lotka} in 1910~\cite{Lotka1910,Lotka1910a}. He then describes the scientific discussion between \textsc{Hirniak} and \textsc{Lotka} in
 \begin{quote}
     ``{\small \foreignlanguage{German}{Lotkas Rechnungen werden von Hirniak zum Teil nicht für ganz einwurfsfrei gehalten. Er behandelt den Fall gegenseitiger Umwandlung dreier Isomeren nach erster Ordnung, für den zurzeit noch kein experimentelles Beispiel vorliegt und findet, daß der Wert von ,,Schwingungsdauer" und ,,Dämpfung" von der Anfangskonzentration unabhaängig sein muß$^4$), worauf Lotka wieder in sehr allgemeiner Weise die allgemeine Lösung für eine Reihe von Folgereaktionen mit autokatalytischem Verlauf gibt$^5$).}}"\\[2mm]
     ``{\small In some cases, Hirniak does not consider Lotka's calculations to be completely free of objections. He deals with the case of mutual conversion of three isomers to the first order, for which there is currently no experimental example, and finds that the value of "oscillation period" and "damping" must be independent of the initial concentration$^4$), whereupon Lotka again presents, in a very general manner, the general solution for a series of subsequent reactions with an autocatalytic process$^5$).}"
 \end{quote}
while citing \textsc{Hirniak}'s 1911~\cite{Hirniak1911} and \textsc{Lotka}'s 1912 article~\cite{Lotka1912} in the \textit{\foreignlanguage{German}{Zeitschrift für Physikalische Chemie}}. As a `funny' typo one can find that he used the name \textit{\foreignlanguage{German}{Journal für Physikalische Chemie}} instead of \textit{\foreignlanguage{German}{Zeitschrift für Physikalische Chemie}} for \textsc{Hirniak}'s paper.

\noindent $\bullet$ 
In 1913, \textsc{Robert Kremann} published the book ``Die periodischen Erscheinungen in der Chemie'' (``The periodic phenomena in chemistry")~\cite{Kremann1913_book} and discussed the work of \textsc{Hirniak} and \textsc{Lotka} in section ``\textit{Theoretische Diskussionen über periodische Erscheinungen in der Chemie von Lotka und Hirniak}" (``\textit{Theoretical discussions about periodic phenomena in chemistry by Lotka and Hirniak}"). In his first paragraph, he writes
\begin{quote}
    ``{\small \ldots gibt J.\ Hirniak$^1$) ein Beispiel für die Möglichkeit einer periodischen Reaktion in homogenen System, eines Falles, der bis jetzt noch völlig unbekannt ist.}"\\[2mm]
    ``{\small \ldots J.\ Hirniak$^1$) gives an example of the possibility of a periodic reaction in a homogeneous system, a case that is still completely unknown.}"
\end{quote}
and cites both of \textsc{Hirniak}'s articles, the 1908 and the 1911 article with the correct year. This is the first citation of \textsc{Hirniak}'s 1908 article. It will take about 80 years until this publications will be cited again in \textsc{Krug} and \textsc{Kuhnert}'s 1985 article~\cite{Krug1985}. But it should also be noted that \textsc{Kremann} followed the same idea \textsc{Hirniak} claimed already in his 1911 article: \textsc{Lotka}'s theoretical model considers a heterogeneous system whereas \textsc{Hirniak}'s reaction scheme describes a homogeneous system, which has later been shown to be not true.

\textsc{Kremann} reproduces the triangular reaction scheme from \textsc{Hirniak}'s 1911 article (see Fig.~\ref{fig:HirniakReactionSchemes}) and discusses the problem of necessary positive rate constants, even uses several of the same equations as in \textsc{Hirniak}'s 1911 article (see App.~\ref{app:1911Translation}). He concludes subsection ``\textit{II. Reaktionen im homogenen System}" (``\textit{II. Reactions in a homogeneous system}") with
\begin{quote}
    ``{\small \foreignlanguage{German}{Diese Ausführungen Hirniaks deuten also deutlich die Möglichkeit einer periodischen Reaktion auch im homogenen System an.}}"\\[2mm]
    ``{\small These statements by Hirniak clearly indicate the possibility of a periodic reaction even in a homogeneous system.}"
\end{quote}

\noindent $\bullet$
In 1929 \textsc{Nicolas von Raschevsky} published the article ``\foreignlanguage{German}{Beitrag zur Theorie der physikalisch-chemischen Periodizität}" (``Contribution to the theory of physical-chemical periodicity"), discussing the possibility of periodic and rhythmic processes.~\cite{Raschevsky1929} Before introducing the first differential equations, he writes
\begin{quote}
    ``{\small \foreignlanguage{German}{Im Gebiet der homogenen Reaktionen ist das Auftreten von zeitlicher Periodlzit\"at theoretisch sehr eingehend, besonders von A. Lotka untersucht worden *. Man kann wohl sagen, dass bei gekoppelten Reaktionen oszillierende L\"osungen der Differentialgleichungen allgemein vorkommen. Jedoch sind diese Oszillationen allgemein positiv oder negativ ged\"ampft. Eine echte Periodizit\"at tritt nur dann auf, wenn die Reaktionen von sehr spezieller Art sind, und daneben noch die Reaktionsgeschwindigkeiten ganz speziellen Bedingungen unterworfen sind.}}"\\[2mm]
    ``{\small In the field of homogeneous reactions, the occurrence of temporal periodicity has been studied theoretically in great detail, especially by A. Lotka *. One can probably say that oscillating solutions of the differential equations generally occur in coupled reactions. However, these oscillations are generally positively or negatively damped. Real periodicity only occurs when the reactions are of a very specific type and the reaction rates are also subject to very specific conditions.}"
\end{quote}
and lists seven of \textsc{Lotka}'s articles~\cite{Lotka1910,Lotka1910a,Lotka1912,Lotka1912a,Lotka1920c,Lotka1920,Lotka1920b} and his \textit{Elements of Physical Biology} book, \textsc{Hirniak}'s 1911 article with the wrong publication year of 1910, and an article by \textsc{Rakowsky}~\cite{Rakowski1907}, also with the stated publication year of 1906 to be one year before the actual year of 1907. This error indicates that he `copied' this citation from \textsc{Lotka}'s 1920 \textit{JACS} article.

After emigrating to the USA, he used the name \textsc{Nicolas Rashevsky} and became an influential proponent of the nascent field of mathematical biology.

\noindent $\bullet$
\textsc{Ernest Hedges} and \textsc{James Myers} published their book ``\textit{The Problem of Physico-Chemical Periodicity}" in 1926 and briefly mentioned \textsc{Hirniak}'s name in their `Theoretical Discussions'. They distinguish clearly between all earlier heterogeneous reactions and `true periodic chemical reactions' and start Chapter IV with
\begin{quote}
    ``{\small The most important type of periodicity in many respects is a rhythmic time change and to this type belong all true periodic chemical reactions.}"
\end{quote}
They discuss the difference between damped and undamped `vibrations', introduce the need for an autocatalytic reaction step, and present a formula for the `period of vibration'. But then, they write
\begin{quote}
    ``{\small Other mathematical expositions of how an autocatalytic system may become periodic are given by Lotka and by Hirniak.}"
\end{quote}
and citing \textsc{Lotka}'s 1912 \textit{Physical Review} paper~\cite{Lotka1912a} and \textsc{Hirniak}'s 1911 paper~\cite{Hirniak1911}, using the correct year. Although they used the correct publication year, it seems unclear if they actually read \textsc{Hirniak}'s 1911 article because \textsc{Hirniak}'s cyclic reaction scheme does not include any autocatalytic step.

\noindent $\bullet$
\textsc{Maurice Copisarow} published his 1931 article ``\textit{\foreignlanguage{German}{Periodizität und ihre Grundlagen}}" (``\textit{Periodicity and its foundations}") in the journal \foreignlanguage{German}{Kolloid-Zeitschrift}~\cite{Copisarow1931} and discussed various phenomena in physical systems. He only touches shortly chemistry when writing
\begin{quote}
    ``{\small \foreignlanguage{German}{\ldots die verschiedenen Beispiele periodischer Strukturen aus dem Gebiete der Chemie$^{11-14}$\ldots}}"\\[2mm]
    ``{\small \ldots the various examples of periodic structures from the field of chemistry$^{11-14}$\ldots}"
\end{quote}
In $^{11}$, he cites four of \textsc{Lotka}'s publications~\cite{Lotka1910a,Lotka1912,Lotka1912a,Lotka1920c}, \textsc{Hirniak}'s 1911 article~\cite{Hirniak1911} in $^{12}$ (using the correct year), two papers about rhythmic behavior in heterogeneous systems by \textsc{Okaya}~\cite{Okaya1918,Okaya1919} in $^{13}$, and another heterogeneous chemical system by \textsc{Hughes}~\cite{Hughes1925} in $^{14}$.

\noindent $\bullet$
\textsc{Suzanne Veil} published a small book in 1934 with the title ``\textit{\foreignlanguage{french}{Les phénomènes périodiques de la chimie. II. Les  périodicités cinétiques}}" (``\textit{Periodic phenomena in chemistry. II. Kinetic periodicities}")~\cite{Veil1934_book2}, in which she discusses various temporal oscillatory chemcial systems. The first volume has the extension ``\textit{\foreignlanguage{french}{I. Les  périodicités de structure}}" (``\textit{I. Structural periodicities}")~\cite{Veil1934_book1} and discusses \textsc{Liesegang} rings and similar precipitation pattern. In Chapter \textit{\foreignlanguage{french}{Autres périodicités suggestions d'ordre mathématique}} (Other mathematically suggested periodicities) of Vol.~I, she starts Section \textit{\foreignlanguage{french}{Suggestions d'ordre mathématique}} (Mathematical  suggestions) with 
\begin{quote}
    ``{\small \foreignlanguage{french}{Par la seule voie théorique, certain auteurs Lotka (93), Hirniak (94), Okaya (13) ont cherché si, par de hypothèses convenables sur la marche et le jeu de réactions engagées, il état possible de prévoir une allure rythmique des effets observés.}}"\\[2mm]
    ``{\small Through the theoretical approach alone, certain authors Lotka (93), Hirniak (94), Okaya (13) have sought whether, through suitable hypotheses on the progress and interaction of the reactions involved, it was possible to predict a rhythmic behavior of the observed effects.}"
\end{quote}
She cited \textsc{Lotka}'s 1910~\cite{Lotka1910,Lotka1910a} and 1912~\cite{Lotka1912,Lotka1912a} papers for the damped oscillatory system, and only one 1920 paper~\cite{Lotka1920c} for the undamped system. In \textsc{Hirniak}'s case, she cited his 1911 article~\cite{Hirniak1911} (with the correct year), and for \textsc{Okaya}, his 1918 and 1919 articles~\cite{Okaya1918,Okaya1919}.

In the next paragraph, she presents \textsc{Lotka}'s autocatalytic reaction scheme and finishes with ``\ldots the oscillations to be expected are, depending on the case, either damped or maintained." But, dispite the fact that she discussed \textsc{Bray}'s result at the beginning of the book, she finishes with
\begin{quote}
    ``{\small \foreignlanguage{french}{Malheureusement ces spéculations, pour intéressantes qu'elles puissent être, se prêtent assez peu au contrôle de l'expérience, et n'ont pas encore reçu, jusqu'ici de confirmation directe.}}"\\[2mm]
    ``{\small Unfortunately these speculations, however interesting they may be, do not lend themselves well to experimental testing, and have not yet received direct confirmation.}"
\end{quote}
which is interesting to read as \textsc{Bray} published the experimental proof more than a decade earlier.

\noindent $\bullet$
In 1938, \textsc{Shemiakin} and \textsc{Mikhalev} published the book ``\textit{\foreignlanguage{russian}{Физико-химические периодические процессы}}" (``\textit{Physico-chemical periodic processes}")~\cite{Shemiakin1938_book} with more than 900 citations. In section ``Chemical kinetics and periodic processes" on page 112, they write
\begin{quote}
    ``\foreignlanguage{russian}{Математическое доказательство возможности существования химических реакций, протекающих периодически во времени, было впервые дано А.\ Лотка [616].}"\\[2mm]
    ``The mathematical proof of the possibility of the existence of chemical reactions occurring periodically in time was first given by A.\ Lotka [616]."
\end{quote}
citing four of \textsc{Lotka}'s articles between 1910 and 1920 and added `\foreignlanguage{russian}{Теория периодических реакций}' (`Theory of periodic reactions') after the references.

They also presented \textsc{Hirniak}'s work in an interesting way. After showing the reaction scheme (very similar to Fig.~\ref{fig:HirniakReactionSchemes} but with the arrows of $k_3$ to $k_6$ in the wrong directions), they reproduced the first two equations from \textsc{Hirniak}'s 1911 article before stating ``For a periodic solution it is necessary that", followed by the forth equation in \textsc{Hirniak}'s 1911 article. And based on the different rate constant numbering in their reaction scheme, this equation can't be correct. To distinguish \textsc{Hirniak}'s work from \textsc{Lotka}, they added `\foreignlanguage{russian}{Теория периодической реакции изомеризации}' ('Theory of a periodic isomerization reaction') after the reference.

\noindent $\bullet$
In 1939, \textsc{Angélique Arvanitaki} wrote an article about the squid giant axon~\cite{Arvanitaki1939} and discussed various oscillatory systems including chemical periodic reactions by simply citing \textsc{Kremann}~\cite{Kremann1913_book}, Lowry and John~\cite{Lowry1910}, Lotka~\cite{Lotka1910a,Lotka1912,Lotka1920c}, \textsc{Hirniak}~\cite{Hirniak1911}, and \textsc{Hedges} and \textsc{Meyers}~\cite{Hedges1926_book}. As often, all citations, except the more mathematically focused article by \textsc{Lowry}, contained the word `periodicity'. Two pages later, they write
\begin{quote}
    ``{\small \foreignlanguage{french}{En suivant la voie de LOTKA (33, 34, 25), HIRNIAC (28), etc.\ on peut partir de deux réactions chimiques consécutives qui seraient ici, comme nous le dictent nos observations, électroactives, et qui rempliraient certaines conditions d’autocatalyse et de réversibilité.}}"\\[2mm]
    ``{\small Following the path of LOTKA (33, 34, 25), HIRNIAC (28), etc.\ we can start from two consecutive chemical reactions which would be here, as our observations dictate, electroactive, and which would meet certain conditions of autocatalysis and reversibility.}" 
\end{quote}
citing the same publications. But it seems that the reference (25) should actually be (35), as earlier, because the number 25 in the article refers to a 1933 physiology paper, not of interest there.

\noindent $\bullet$
In the 1939 article ``\textit{\foreignlanguage{russian}{Периодические процессы в кинетике окислительных реакций}}" (``\textit{Periodic processes in the kinetics of oxidation reactions}") \textsc{David Frank-Kamenetskii}~\cite{Frank-Kamenetskii1939} writes
\begin{quote}
    ``{\small \foreignlanguage{russian}{Возможность колебательных процессов в кинетике окислительных реакций представляет общий интерес [см. ссылки на литературу ($^\text{2-8}$)]}}"\\[2mm]
    ``{\small The possibility of oscillatory processes in oxidizing reaction kinetics is of common interest [see references to the literature ($^\text{2-8}$)]}"
\end{quote}
The citations list includes $^2$ \textsc{Quincke}, Wied.\ Ann., \textbf{35}, 614 (1888).~\cite{Quincke1888} [Note: The citation uses page 614 but it should be page 580, as correctly used in \cite{Bernstein1900}.] $^3$ Bernstein, Pflüg.\ Zrch., \textbf{80}, 628 (1900)~\cite{Bernstein1900}. $^4$ Ostwald, Vorlesungen über Naturphilosophie, Leipzig.\ S.\ 274, 315, 352 (1902).~\cite{Ostwald1902_book} $^5$ Lotka, ZS.\ phys.\ Chemie, \textbf{72}, 508 (1910); \textbf{80}, 159 (1912).~\cite{Lotka1910,Lotka1912} $^6$ Hirniak, ZS.\ phys.\ Chemie, \textbf{75}, 675 (1910)~\cite{Hirniak1911} [Note: Wrong publication year] $^7$ Skrabal, ZS.\ phys.\ Chemie, \textbf{6}, 382 (1930).~\cite{Skrabal1930} $^8$ Kremann, \foreignlanguage{German}{Die periodischen Erscheinungen in der Chemie, Sammlung chemischer Vorträge}, \textbf{19}, 289, Stuttgart (1913)~\cite{Kremann1913_book}.

In this list of citations, \textsc{Frank-Kamenetskii} mixes mechanical oscillations (page 274 in~\cite{Ostwald1902_book}), electrochemical oscillations~\cite{Bernstein1900}, periodic phenomena in biological systems (page 315 in \cite{Ostwald1902_book}) with the proposed periodic behavior in chemical systems by the last four authors.

\noindent $\bullet$
\textsc{Herbert Pohl} published two articles in 1981 and citing \textsc{Hirniak}'s 1911 article, though as published in 1910. In both cases, it was one citation in a longer list with \textsc{Bray}, \textsc{Lotka}, and others when stating
\begin{quote}
    ``{\small Early reports of even chemical reactions of periodic nature were largely ignored (Bray, 1921; Hirniak, 1910; Lotka, 1910, 1920).}"
\end{quote}
in~\cite{Pohl1981a} or 
\begin{quote}
    ``{\small The study of oscillating reactions is presently in a growingly active state.}"
\end{quote}
in~\cite{Pohl1981b}. After this statement, he lists articles from \textsc{Lotka}, \textsc{Hirniak}, and \textsc{Bray} but also includes 14 other publications from, for example, \textsc{Belousov}, \textsc{Prigogine}.

\noindent $\bullet$
In 1985, \textsc{Hans-Jürgen Krug} and \textsc{Lothar Kuhnert} published an article about oscillating systems with an autocatalytic step in \textit{\foreignlanguage{German}{Zeitschrift für Physikalische Chemie}}~\cite{Krug1985}. In the introduction they write
\begin{quote}
    ``{\small \foreignlanguage{German}{Dazu wurde bereits 1908 von Hirniak [1] und 1910 von Lotka [2] auf die Möglichkeit des Ablaufs periodischer chemischer Reaktionen in homogener Phase hingewiesen.
    In den didaktischen Reaktionsschemata von Hirniak und Lotka sowie in weiteren Modellen (z.B.\ Lotka-II-Schema [3], Brüsselator [4], Oregonator [5], Lotka-Selkov- Schema [6]) wurde gezeigt, daß oszillatorisches Verhalten eng an die Beteiligung von Rückkopplungen in Form von zyklischer Kopplung, Autokatalyse, Autoinhibition oder Kreuzkatalyse gebunden ist.}}"\\[2mm]
    ``{\small The possibility of periodic chemical reactions taking place in a homogeneous phase was already pointed out by Hirniak [1] in 1908 and by Lotka [2] in 1910.
    In the didactic reaction schemes of Hirniak and Lotka as well as in other models (e.g., Lotka-II scheme [3], Brusselator [4], Oregonator [5], Lotka-Selkov scheme [6]) it was shown that oscillatory behavior is closely linked to the participation of feed backs in the form of cyclic coupling, autocatalysis, autoinhibition, or cross-catalysis.}"
\end{quote}
This is the second publication citing \textsc{Hirniak}'s 1908 paper~\cite{Hirniak1908a} in reference [1]. But they also added \textsc{Hirniak}'s 1911 article~\cite{Hirniak1911}, though still using the wrong year of 1910. In \textsc{Lotka}'s reference [2], they only cited the German article in \textit{\foreignlanguage{German}{Zeitschrift für Physikalische Chemie}}~\cite{Lotka1910} but not the English version in \textit{The Journal of Physical Chemistry} of the same year~\cite{Lotka1910a}. This might have been a conscious decision and is the earliest finding of placing \textsc{Hirniak} and \textsc{Lotka} at the same `scientific' level while indicating that \textsc{Hirniak} published one article before \textsc{Lotka} and the other at the same year. The references [3-6] are~\cite{Lotka1920c,Prigogine1968,Field1974a,Feistel1978}.

\noindent $\bullet$
\textsc{P.E.~Rapp} cited \textsc{Lotka} and \textsc{Hirniak} equally in his 1987 article ``\textit{Why are so many biological systems periodic?}"~\cite{Rapp1987}. But he also highlights the initially difficult start 
\begin{quote}
    ``{\small Periodic chemical reactions were anticipated theoretically by Lotka and Hirniak in 1910 (Lotka, 1910, 1920; Hirniak, 1910). Accounts of an experimental example published in 1921 (Bray, 1921) seemingly attracted limited attention.}"
\end{quote}
As can be seen, they cited \textsc{Lotka}'s damped~\cite{Lotka1910a} and undamped~\cite{Lotka1920c} systems and \textsc{Hirniak}'s 1911 article~\cite{Hirniak1911} with the year 1910 as used in \textsc{Lotka} and \textsc{Bray}'s work~\cite{Bray1921}.

\noindent $\bullet$
\textsc{Jason Gallas} cited \textsc{Hirniak}'s work in his 2015 article~\cite{Gallas2015} while talking about 
\begin{quote}
    ``{\small The origins of the oscillator known as Brusselator can be traced to a work in 1956 by Prigogine and Balescu$^5$ showing that undamped oscillations$^{6-15}$ are supported far from thermodynamic equilibrium in open chemical systems governed by nonlinear kinetic laws.}"
\end{quote} 
The references are \cite{Prigogine1956} for $^5$ and a list of nine references with work from \textsc{Prigogine}~\cite{Nicolis1977_book,Prigogine1968}, \textsc{Scott}~\cite{Scott1993_book}, and seven early publications between 1910 and 1926 (books by \textsc{Kremann} and \textsc{Hedges}~\cite{Kremann1913_book,Hedges1926_book} articles by \textsc{Lotka}~\cite{Lotka1910,Lotka1920b,Lotka1912,Lotka1920c}  and \textsc{Hirniak}~\cite{Hirniak1911} (using the 1910 publication year).

\noindent $\bullet$
 \textsc{Eberhard Bruckner}'s 1987  historical review article ``\textit{\foreignlanguage{German}{Die ersten theoretischen Modelle für oszillatorisches Verhalten in der Chemie von J.\ Hirniak und A.\ J.\ Lotka - zwei Publikationen in der Zeitschrift für Physikalische Chemie aus den Jahren 1910/11}}" (``The first theoretical models for oscillatory behavior in chemistry by J.\ Hirniak and A.\ J.\ Lotka - two publications in the Journal of Physical Chemistry from 1910/11")~\cite{Bruckner1987} was published in conjunction with the 100$^\text{th}$ anniversary of the journal \textit{\foreignlanguage{German}{Zeitschrift für Physikalische Chemie}}. He presents the submission dates as January 31, 1910 for \textsc{Lotka}'s  and November 22, 1910 for \textsc{Hirniak}'s articles in the same journal and uses 1910 as the publications year for both in the publication list. He quotes \textsc{Hirniak}'s initial sentence with the reference to the 1908 article but does not provide any bibliographic information.
 
 \textsc{Bruckner} discusses the acceptance problems for periodic chemical reactions in the early 20$^\text{th}$ century with examples from \textsc{Ostwald}, \textsc{Bray}, \textsc{Lotka}, and \textsc{Hirniak}. He always refers to \textsc{Lotka} and \textsc{Hirniak} together as in, for example, 
 \begin{quote}
    ``{\small \foreignlanguage{German}{\ldots die in den theoretischen Modellen von Lotka und Hirniak ausgepr\"agt ist. Beiden gelang die Entdeckung der reaktionskinetischen Bedingungen für oszillatorisches Verhalten \ldots}}"\\[2mm]
 	``{\small \ldots  which can  be found in the theoretical models by Lotka and Hirniak. Both discovered the reaction-kinetic conditions for oscillatory behavior \ldots}"
 \end{quote}
 
\noindent $\bullet$
In a 2016 article by \textsc{Alexander Pechenkin}~\cite{Pechenkin2016},  with the title ``\foreignlanguage{russian}{\textit{Две истории периодических процессов в химии} }" (``Two stories of periodic processes in chemistry"), \textsc{Hirniak} is again cited while \textsc{Pechenkin} introduces \textsc{Lotka}'s reaction scheme with its autocatalytic step, creating periodic behavior in Section 2. While talking about \textsc{Lotka}'s 1912 article, he writes
\begin{quote}
    ``{\small \foreignlanguage{russian}{В 1912 г. А. Лотка опубликовал статью, где сопроводил свой анализ химических колебаний ссылкой на статью Ю.\ Хирниака, опубликованную в 1910 г.}}"\\[2mm]
    ``{\small In 1912, A. Lotka published an article in which he accompanied his analysis of chemical oscillations with a link to an article by J.\ Hirniak, published in 1910.}"
\end{quote}
He continues in the next paragraph with
\begin{quote}
    ``{\small \foreignlanguage{russian}{Интересно, что сам Хирниак в своей статье 1910 г. сослался на свою статью, появившуюся раньше, чем первая статья Лотки. Однако эта первая статья Хирниака вышла на русинском языке и прошла незамеченной.}}"\\[2mm]
    ``{\small Interestingly, Hirniak himself, in his 1910 article, referred to his article appearing earlier than Lotka's first article. However, this first article by Hirniak was published in Rusyn and went unnoticed.}"
\end{quote}

With the second quote, \textsc{Pechenkin} points out that Lotka `ignored' \textsc{Hirniak}'s original 1908 paper, adding to the controversy about who published the first paper on the theoretical description of a periodic homogeneous chemical reaction. A more detailed discussion on the term \foreignlanguage{russian}{русинском} (Rusyn) can be found in the article's Section \textbf{\emph{Ruthenian Language}}.

But, as many other scientists, \textsc{Pechenkin} uses the wrong publication year of \textsc{Hirniak}'s 1911 paper. In addition, he used the year 2010 for \textsc{Lotka}'s 1910 article in the \textit{Journal of Physical Chemistry} and his `back transliteration' of \textsc{Hirniak}'s last name from English to Cyrillic resulted in \foreignlanguage{Ukrainian}{Хирниак}. Here, the original name \foreignlanguage{Ukrainian}{Гірняк}, as used in all of \textsc{Hirniak}'s Ukrainian articles, is barely recognizable.

\noindent $\bullet$
In another 2016 article by \textsc{Pechenkin}~\cite{Pechenkin2016b}, he discusses the 1938 book ``\textit{\foreignlanguage{russian}{Физико-химические периодические процессы}}" (``\textit{Physico-chemical periodic processes}") by \textsc{Shemiakin} and \textsc{Mikhalev} and starts Section `Mathematical equations' with
\begin{quote}
    ``{\small Already R. Kremann (1913, 411) describes A. Lotka’s mathematical technique (1910, 1912) which shows that under certain conditions some of the complicated heterogeneous autocatalytic systems may show oscillatory behavior. As an example of the homogeneous oscillatory system Kremann refers to Hirniak’s theoretical construction of a monomolecular interconversion of three isomers (Hirniak 1911).}"
\end{quote}
and citing the three German publications by \textsc{Kremann}~\cite{Kremann1913_book}, \textsc{Lotka}~\cite{Lotka1910}, and \textsc{Hirniak}~\cite{Hirniak1911}, though missing the citations for \textsc{Lotka}'s 1912 article. Here, \textsc{Pechenkin} used the correct year for \textsc{Hirniak}'s 1911, probably because \textsc{Kremann} used it in his 1913 book.

This quote is another example of scientist referring to \textsc{Lotka}'s 1910 model as heterogeneous whereas \textsc{Hirniak}'s is considered homogeneous.

\noindent $\bullet$
In 2017, \textsc{Joana Freire}, \textsc{Marcia Gallas}, and \textsc{Jason Gallas} cited \textsc{Hirniak}'s 1911 paper, as citation [8], in a paragraph when talking about the long history of periodic chemical reactions`~\cite{Freire2017}. On page 1989, they write 
\begin{quote}
    ``{\small In chemistry, periodic oscillations already have an almost two-centuries long history, dating back at least to the beginnings of the 19th century as evidenced by the many experimental works collected and described in almost forgotten books [5,6] and papers [7–9], a history that deserves to be better known.}"
\end{quote}
The two cited books are from \textsc{Kremann} in 1913~\cite{Kremann1913_book} and from \textsc{Hedges} and \textsc{Myers} from 1926~\cite{Hedges1926_book}. The three articles are German publications in the \textit{\foreignlanguage{German}{Zeitschrift für Physikalische Chemie}} by \textsc{Lotka}~\cite{Lotka1910,Lotka1912} and \textsc{Hirniak}~\cite{Hirniak1911}, with the publication year 1910.

\noindent $\bullet$
\label{Pechenkin2018book}
 Some information about \textsc{Hirniak}'s life and scientific work can be found in Section ``Julian Hirniak" in \textsc{Alexander Pechenkin}'s 2018 book ``\textit{The History of Research on Chemical Periodic Processes}~\cite{Pechenkin2018_book}. This section gives a short overview about \textsc{Hirniak}'s live but contains several factual errors. For example, he discusses \textsc{Hirniak}'s claim for priority presenting the first theoretical model for periodic chemical reactions and how it was related to \textsc{Lotka}'s work. But it is not clear how he comes to the conclusion when starting the section with ``In 1910, Hirniak in ``\foreignlanguage{German}{Zeitschrift für Physikalische Chemie}" proposed that cyclic reactions can be oscillatory'', using the year 1910, which, too often, had been used as the actual publication year. It's also confusing because he finishes the introductory Abstract of that chapter with 
 \begin{quote}
     {\small Besides Lotka, Hirniak (Lvov) contributed to the early mathematical studies of chemical periodical processes (1908, 1911)}
 \end{quote}
 using the correct publication year. But after \textsc{Hirniak}'s name, he uses the transliterated Russian spelling \foreignlanguage{russian}{Львов} [Lvov], not the correct English version of this Ukrainian city.
 
 \textsc{Pechenkin} also mentions \textsc{Hirniak}'s 1908 publication but does not provide a reference. He only gives the German name of the journal, with a few typos, and  presents the journal cover page of the 1908 edition, but stating that it's the issue of \textsc{Hirniak}'s 1911 paper.
 
When discussing \textsc{Lotka} and \textsc{Hirniak}'s papers, he points out that \textsc{Lotka} probably didn't know about \textsc{Hirniak}'s 1908 article and suggests that \textsc{Hirniak} thought, while writing his 1911 article, that this was the same reason why \textsc{Lotka} didn't cite his own work.  Unfortunately, \textsc{Pechenkin} falsely states
\begin{quote}
    ``{\small It should be noted that Lotka in his 1912 and 1920 papers referred to Hirniak’s 1908 paper.}"
\end{quote}
but no citation or reference can be found in any of \textsc{Lotka}'s 1912~\cite{Lotka1912,Lotka1912a} or 1920~\cite{Lotka1920,Lotka1920b,Lotka1920c,Lotka1920a} publications, except to \textsc{Hirniak}'s 1911 (with the wrong year 1910 as mentioned earlier).

Notably, \textsc{Pechenkin} also writes that \textsc{Hirniak} published his 1908 article in the ``Rusyn language" and ``Hirniak believed that Lotka did not refer to this paper because it was published in a rare periodical in the obscure language." We explain the differences between \textit{Rusyn} and \textit{Ruthenian} (the term \textsc{Hirniak} used in his 1911 German article) and \textit{Ukrainian} in Section~\ref{sec:Ruthenian Language}.
Perhaps due to the language confusion, \textsc{Pechenkin} also wrote ``Probably, Hirniak belonged to Rusyn (Ruthenians), a small ethnic group which belongs neither to Russians, nor to Ukrainians.", which is not correct. He also used three differently wrong versions for the long English and German names of the Shevchenko Scientific Society. The correct spelling can be found in App~\ref{app:Julian Hirniak's publications}.

\noindent $\bullet$
\textsc{Christophe Letellier} cites \textsc{Hirniak} in Chapter 5 ``\textit{Chemical Reactions}" of his 2019 book ``\textit{Chaos in Nature}~\cite{Letellier2019_book-chapter1}. He quotes a section from \textsc{Lotka}'s 1910 article and includes \textsc{Hirniak}'s article
\begin{quote}
    ``{\small The possibility of decaying oscillations had been announced after a study in 1904 of a hypothetical autocatalytic reaction$^3$ --- since at the time ``\textit{no reaction is known which follows the} [mentioned]\textit{law} [\ldots] \textit{of damped oscillations}" --- by Alfred Lotka (1880-1949) and by Julius Hirniak.$^4$ Both studied the differential equations for the intermediary reactions and, after a mathematical analysis, defined the conditions under which an oscillating chemical reaction could be seen.}"
\end{quote}
Citation $^3$ lists \textsc{Lotka}'s 1910 article~\cite{Lotka1910a} and $^4$ refers to \textsc{Hirniak}'s 1911 article~\cite{Hirniak1911}, using the correct year.

\noindent $\bullet$ 
In a 2021 article by \textsc{Bernold Fiedler}~\cite{Fiedler2021} one can find
\begin{quote}
	 ``{\small It is a lasting merit of [Hir1911] to point at the possibility of (damped) oscillations, once the Wegscheider constraints \ldots are strongly violated.}"
\end{quote}
This is an interesting approach to \textsc{Hirniak}'s paper. \textsc{Fiedler} clearly realized that  \textsc{Hirniak}'s reaction does not fulfill the \textsc{Wegscheider} constraints but rephrases it in a positive way.

\noindent $\bullet$
In a 2022 article by  \textsc{Roman Cherniha} and \textsc{Vasyl' Davydovych}~\cite{Cherniha2022}, they write in the introduction
\begin{quote}
	``{\small Following some earlier papers, in which linear ODEs were used for mathematical modeling of chemical reactions (in particular, see [3–5]), Lotka has shown that the densities in periodic reactions can be adequately described by a model involving ODEs with quadratic nonlinearities.}"
\end{quote}
while citing both of \textsc{Hirniak}'s articles~\cite{Hirniak1908a,Hirniak1911} and \textsc{Lotka}'s 1910 German article~\cite{Lotka1910}. No further discussion happens and it seems simply to be a citation with the first three articles including ODEs and having `periodic reactions' in the title.

\noindent $\bullet$
\textsc{Jason Gallas} cited \textsc{Hirniak}'s 1911 article in his 2022 article~\cite{Gallas2022}, but it is completely 'hidden' within the five citations of the first sentence 
\begin{quote}
    ``{\small In pioneering works done in 1910–1920, Lotka [27–31] considered theoretically the problem of chemical oscillations. He was able to derive equations which produced sustained oscillations [31, 32].}"
\end{quote}
The references [27-31] list four of \textsc{Lotka}'s papers~\cite{Lotka1910,Lotka1910a,Lotka1912,Lotka1920c} and \textsc{Hirniak}'s 1911 paper~\cite{Hirniak1911} as [28], though using the wrong year. The additional citation [32] refers to a 2018 article by \textsc{Erneux}~\cite{Erneux2018}. 

\subsection{Critical citations}
\label{app:Critical citations of Hirniak's work}

\noindent $\bullet$
\textsc{Anton Skrabal} wrote in 1930 a 42-page article about the theory of periodic reactions in homogeneous systems. He starts with
\begin{quote}
    ``{\small \foreignlanguage{German}{Im allgemeinen nähern sich die chemischen Reaktionen aperiodisch und asymptotisch ihrem Gleichgewicht$^1$), doch kennt man auch periodisch verlaufende Vorgänge$^2$). Unter den letzteren sind die periodischen Reaktionen im homogenen System am rätselhaftesten und daher von besonderm theoretischem Interesse.
    Die Möglichkeit solcher Reaktionen haben vor längerer Zeit Jul.\ Hirniak$^3$) und A.\ Lotka$^4$) theoretisch gezeigt. Beiden theoretischen Betrachtungen liegt ein System von Simultanreaktionen zugrunde.}}"\\[2mm]
    (`` {\small In general, chemical reactions approach their equilibrium aperiodically and asymptotically$^1$), but periodic processes are also known$^2$). Of the latter, the periodic reactions in the homogeneous system are the most puzzling and therefore of particular theoretical interest.
    The possibility of such reactions was shown theoretically a long time ago by Jul.\ Hirniak$^3$) and A.\ Lotka$^4$). Both theoretical considerations are based on a system of simultaneous reactions.}"
\end{quote}
using the following citations: $^1$) Siehe W.\ Nernst, Theoretische Chemie, 11.\ bis 15.\ Aufl., S.\ 634 und 766, Stuttgart 1926~\cite{Nernst1909_book}. $^2$) Vgl.\ R.\ Kremann. \foreignlanguage{German}{Die periodischen Erscheinungen in der Chemie in Sammlung chemischer Vorträege \textbf{19}, 289, Stuttgart 1913.}~\cite{Kremann1913_book}, $^3$) J.\ Hirniak, \foreignlanguage{German}{Beiträge zur chemischen Kinetik}, Lemberg 1911. Z.\ physikal.\ Chem.\ \textbf{75}, 675, 1911.~\cite{Hirniak1911} $^4$) A.\ Lotka, Z.\ physikal.\ Chem. \textbf{72}, 508, 1910. \textbf{80};, 159, 1912~\cite{Lotka1910a,Lotka1912}.

But \textsc{Skrabal} realized already the `problems' with \textsc{Hirniak}'s math and chemistry. On page 388 he writes
\begin{quote}
    ``{\small \foreignlanguage{German}{In seiner Arbeit erwähnt J.\ Hirniak eine etwaige ``unbekannte Gesetzmässigkeit", die das Negativwerden des Ausdrucks unter dem Quadratewurzelzeichen verhindert. Eine solche Gesetzmässigkeit wäre die Beziehung (22), falls die Zukunft lehren sollte, dass sie eine notwendige Beziehung ist.
    Im stärkeren Widerspruch zu dieser Beziehung stehen die sogenannten ``Zirkularreaktionen", bei welchen die eine Seite der Beziehung (22) endlich, die andere Null ist.}}"\\[2mm]
    (`{\small In his work, J.\ Hirniak mentions a possible ``unknown law" that prevents the expression under the square root sign from becoming negative. Such a law would be the relation (22), if the future should teach that it is a necessary relationship.
  In greater contradiction to this relationship are the so-called "circular reactions", in which one side of the relationship (22) is finite and the other is zero.}")
\end{quote}
Equation (22) is given as $k_2 k_4 k_6 = k_1 k_3 k_5$ because he used the same reaction scheme as \textsc{Hirniak} in his 1911 article (see Eq.~\eqref{eq:PDB equation} and right sketch in Fig.~\ref{fig:HirniakReactionSchemes}). He continued discussing \textsc{Wegscheider}'s \emph{Principle of Detailed Balance} and closes his article with
\begin{quote}
    ``{\small \foreignlanguage{German}{Periodische Reaktionen sind aus thermodynamischen und kinetischen Gründen möglich, aber aus denselben Gründen nicht wahrscheinlich. Damit im Einklang stehen die experimentellen Erfahrungen.}}"\\[2mm]
    ``{\small Periodic reactions are possible for thermodynamic and kinetic reasons, but not probable for the same reasons. The experimental experiences are consistent with this.}"
\end{quote}
This quote is another indication about the difficult start of periodic chemical reactions in homogeneous systems. Even \textsc{Skrabel} was still very skeptical, nine years after \textsc{William Bray}'s experimental proof of oscillatory chemical reactions in 1921~\cite{Bray1921}.

\noindent $\bullet$
In 1947, \textsc{Wilhelm Jost} published his article ``\foreignlanguage{German}{Über den Ablauf zusammengesetzter chemischer Reaktionen - Systeme von Reaktionen I. Ordnung}" (``About the process of compound chemical reactions - systems of first order reactions") and discussed a reaction scheme using first-order kinetics between three isomers, identical to \textsc{Hirniak}'s system.

He starts his article with
\begin{quote}
	``{\small \foreignlanguage{German}{Es ist bekannt, daß man bei der Berechnung der wechselseitigen Umlagerung dreier Stoffe nachnebenstehendem Schema von Reaktionen I. Ordnung Lösungen konstruieren kann, die einem Einstellen des Gleichgewichts durch gedämpfte Schwingungen entsprechen, sofern man beliebige Verhältnisse zwischen den sechs Geschwindigkeitskonstanten zuläßt$^1$. Die Periodizität geht verloren, wenn man die Gültigkeit des Prinzips der mikroskopischen Reversibilität zugibt, wie ebenfalls seit langem bekannt, aber erst von L.\ Onsager$^2$ in allgemeinerem Zusammenhang diskutiert.}}"\\[2mm]
    ``{\small It is known that when calculating the mutual rearrangement of three substances according to the diagram of first-order reactions shown opposite, solutions can be constructed which correspond to establishing the equilibrium through damped oscillations, provided that arbitrary ratios between the six rate constants are allowed$^1$. Periodicity is lost when one admits the validity of the principle of microscopic reversibility, which has also long been known but was only discussed in a more general context by L.\ Onsager$^2$.}"
\end{quote}
In reference $^1$, \textsc{Jost} cited articles by \textsc{Hirniak}~\cite{Hirniak1911} and \textsc{Lotka}~\cite{Lotka1910,Lotka1912,Lotka1910a,Lotka1932} and refers to \textsc{Skrabal}'s work~\cite{Skrabal1930,Skrabal1941_book} for numerical calculations and general overview about the topic.
In reference $^2$, \textsc{Jost} cited \textsc{Onsager}'s two famous articles from 1931~\cite{Onsager1931a,Onsager1931b}.
He then gives the balanced rate constants equation 
and concludes that the system's eigenvalues are real, while, again, citing \textsc{Skrabal}'s work. He continues with two exemplary four-component reaction schemes, generalizes his mathematical proof to $n$-component systems, and summarizes (which is actually his abstract) with
\begin{quote}
	``{\small \foreignlanguage{German}{1.\ Es kann gezeigt werden, daß in beliebig kompliziert zusammengesetzten chemischen Umwandlungen nach der I. Ordnung keine Periodizität auftreten kann, sofern das Prinzip der mikroskopischen Reversibilität erfüllt ist.
    2.\ Darüber hinaus gilt die schärfere Aussage: die Konzentration jeder Komponente dieses Systems kann als Funktion der Zeit höchstens ($n-2$) Extrema durchlaufen, wenn $n$ die Zahl der Komponenten ist.}}"\\[2mm]
    ``{\small 1.\ It can be shown that no periodicity can occur in first-order chemical transformations of any complexity, provided the principle of microscopic reversibility is fulfilled.
    2.\ Furthermore, the stricter statement applies: the concentration of each component of this system can pass through at most ($n-2$) extremes as a function of time if $n$ is the number of components.}"
\end{quote}
These statements shows clearly that \textsc{Hirniak}'s proposed periodic behavior wasn't possible in \textsc{Jost}'s analysis.

\noindent $\bullet$
\textsc{John Hearon} included \textsc{Hirniak}'s 1911 article (with the correct year) in his 1953 article ``\textit{The kinetics of linear systems with special reference to periodic reactions}"~\cite{Hearon1953} in the introduction with 
\begin{quote}
    ``{\small The possibility of periodic reactions was early considered (Lotka, 1910; Hirniak, 1911; Lotka, 1920) and an experimental case was shortly reported (Bray, 1921).}"
\end{quote}
After introducing \textsc{Lotka}'s system, he continues with
\begin{quote}
    ``{\small In all of the above cases, except that of Hirniak (\textit{loc.\ cit.}), the differential equations which describe the system are \textit{non-linear}, whether or not the system is open.}"
\end{quote}
His mathematical analysis of different chemical reaction schemes is similar, but more general than our examination of \textsc{Hirniak}'s work. The summary of his paper can be found in the first sentence of his abstract
\begin{quote}
    ``{\small It is shown on the basis of (1) conservation of mass, (2) positive concentrations, and (3) the principle of detail balancing that periodic reactions cannot occur in a closed system described by linear differential equations.}"
\end{quote}

\noindent $\bullet$
10 years later, in 1963, \textsc{Hearon} again cited \textsc{Hirniak}'s 1911 article in his  article ``\textit{Theorems on Linear Systems}"~\cite{Hearon1963}. There, he wrote
\begin{quote}
    ``{\small Hypothetical (but assertedly realizable), chemical systems, obeying \textit{linear} kinetics, have been discussed (as early as 1911, Hirniak$^{38}$ and late as 1955, Landahl$^{39}$; other references are given in Hearon$^8$) as exhibiting periodic behavior. In all cases either the \emph{Principle of Detailed Balance} has been assumed out of existence, or the equations used show that conservation of mass is infringed.}"
\end{quote}
with the two citations \cite{Hirniak1911,Landahl1955} and his own article~\cite{Hearon1953}, clearly indicating the impossibility of the periodic behavior.

\noindent $\bullet$ 
In the 1973 article ``\textit{Oscillations, multiple steady states, and instabilities in illuminated systems}" by \textsc{Abraham Nitzan} and \textsc{John Ross}~\cite{Nitzan1973}, the authors state 
\begin{quote}
    ``{\small The possibility of oscillatory phenomena in a cyclic reaction scheme was long ago demonstrated by Hirniak.$^{10}$}"
\end{quote}
while citing \textsc{Hirniak}'s 1911 paper correctly. They start similarly as in Section \textbf{\emph{Analysis of Hirniak's work}} but assume specific conditions for the rate constants,
\begin{equation}
    k_1 \gg k_4,\;\; k_5 \gg k_2,\; \text{and } k_3 \gg k_6
    \label{eq:Nitzan condition}
\end{equation}
which lead to a damped oscillation frequency of 
\begin{equation}
    \omega = \frac{1}{2}\left[4 k_1 k_5 - \left(k_1 + k_5 - k_3\right)^2\right]^{1/2}
\end{equation}
and a damping constant of
\begin{equation}
    \gamma = \frac{1}{2}\left(k_1 + k_5 + k_3\right)~.
\end{equation}
Because $\gamma > \omega$ they conclude that ``in most situations the oscillatory behavior will not be amendable to experimental observation due to the large damping rate.'' 
Although they realize that the
 \emph{Principle of Detailed Balance} 
rules out damped oscillations in a closed system on its way to chemical equilibrium, they write
\begin{quote}
    ``{\small However, in an open system in a nonequilibrium steady state [the conditions proposed in Eq.~\eqref{eq:Nitzan condition}] may be achieved and one way of doing so is by optical excitation.}"
\end{quote}
Though this might be possible, it is certainly not what \textsc{Hirniak} had in mind when proposing his reaction scheme.

\noindent $\bullet$
\textsc{Fong} \textit{et al.} wrote in their 1982 photochemistry paper~\cite{Fong1982}
\begin{quote}
    ``{\small The feedback mechanism responsible for the oscillatory delayed fluorescence could conceivably be given by a cyclic scheme generically related to that discussed by Hirniak$^{51}$ and Nitzan and Ross,$^{52}$ although the large amplitudes of the oscillations are not easily accounted for by the description of an incoherent kinetic system.}"
\end{quote}
while citing \cite{Hirniak1911} and \cite{Nitzan1973}. Here it is interesting that they cited \textsc{Hirniak} for a system the original work was certainly not intended for.

\noindent $\bullet$
\textsc{Anatol Zhabotinsky} discussed \textsc{Hirniak}'s ``cyclic interconversion of three isomers'' in his 1991 short historical review article ``\textit{A history of chemical oscillations and waves}"~\cite{Zhabotinsky1991a} and explains why periodic behavior seems to be possible but it thermodynamically impossible. He writes
\begin{quote}
    ``{\small In 1910, Hirniak$^4$ proposed that cyclic reactions can be
oscillatory. He used the simplest example: the cyclic interconversion of three isomers.}"
\end{quote}
citing \textsc{Hirniak}'s 1911 article, but using the wrong year here and in the bibliography. He discusses that thermodynamics puts ``strong restrictions on the rate constants in this system'' and adds 
\begin{quote}
    ``{\small This condition immediately forbids any oscillations in the system. Later, it was shown that it is impossible to have any concentration oscillations in the vicinity of the thermodynamic equilibrium state.$^5$ This thermodynamic analysis made a very strong impression on the majority of chemists, who interpreted it as being valid for all homogeneous closed chemical systems.}"
\end{quote}
citing \textsc{Jost} (1947)~\cite{Jost1947}, \textsc{Hearon} (1953)~\cite{Hearon1953}, and \text{Gray} (1970)~\cite{Gray1970} in reference $^5$. With the last statement, \textsc{Zhabotinsky} gives an explanation for the continued diverse view of \textsc{Hirniak}'s work. Many scientists see \textsc{Hirniak} as an early advocate for chemical oscillatory systems, similar to \textsc{Lotka}, whereas others highlight the thermodynamic errors in his reaction scheme.

\noindent $\bullet$
 In their 2017  article  ``\textit{Periodic Reactions: The Early Works of William C. Bray and Alfred J. Lotka}, \textsc{Rinaldo Cervellati} and \textsc{Emanuela Greco}~\cite{Cervellati2017} compared shortly \textsc{Hirniak} and \textsc{Lotka}'s work and  \textsc{Hirniak}'s claim of priority based on his 1908 article, though they dated both articles, in their references (31) and (33), as published in 1911.
 
 They specifically addressed \textsc{Hirniak}'s misunderstanding of \textsc{Lotka}'s work. They write
\begin{quote}
    ``Hirniak distinguished his work from that of Lotka by insisting$^{34}$ particularly on the homogeneous nature of his model."
\end{quote}
with the reference text as
\begin{quote}
   ``Hirniak did not want to recognize that even the example of Lotka was hypothetical and concerned a homogeneous system. The presence of a condensed phase was perhaps misleading for Hirniak; however, Lotka’s hypothetical reaction then proceeded in a homogeneous phase, gas or solution." 
\end{quote}

\newpage
\section{Translation of Hirniak's 1908 article}
\label{app:1908Translation}
\setcounter{footnote}{0}

 \begin{quote}
    Translation from Ukrainian of the article ``\foreignlanguage{Ukrainian}{Про перiодичнi хемiчнi реакциї}", \textit{\foreignlanguage{Ukrainian}{Збiрник Математично-Природописно-Лїкарської Секциї Наукового Товариства імени Шевченка}} \textbf{XII} (1908) \cite{Hirniak1908a} by \foreignlanguage{Ukrainian}{Юрій Головач} (Yurij Holovatch).
 \end{quote}

\noindent Note: In this translation, we reproduce scans of the equations from the original publication. The articles have been provided by the Stefanyk National Scientific Library in Lviv, Ukraine, where originals of the journals are stored.

\begin{center}
    \textbf{About periodic chemical reactions.}\\
    written by\\
    \textbf{Dr. Julian Hirniak.}\\
    (reported at the meeting of the Mathematical-Natural Sciences-Medical Section on September 10, 1908)
\end{center}

In this study I want to attract attention to one consequence that is very easy to derive from the equations of chemical kinetics.
Making a closer look on the great interest to the problems in this field of physical chemistry, one has to conclude that the research is pursued exclusively in the experimental direction and that the studies
with the help of mathematical analysis\footnote{R. Wegscheider -- Chemische Kinetik homogener Systeme -- Monatshefte für Chemie. B. XXI. 1900. page 693.} appear in the literature almost as exceptions. If one adds to this that the very value of mathematical analysis in chemistry is very often directly or indirectly questioned, it is of no surprise that on some issues  completely wrong views can be spread, although a simple and easy calculation can clarify the matter indisputably. I mean here the fact that in one, recently published place, the periodicity of a chemical process is considered a characteristic sign of catalysis. As I will show by the calculations below, periodicity can appear in processes in which there is no catalysis at all. And therefore, it is not in a causal relationship with catalysis, although such a phenomenon can be observed with the latter.

So far, I have very limited and fragmentary information about the extent to which periodic reactions have been studied experimentally, and I cannot now make it easier for myself to access the relevant literature sources. The work of Prof.\ Dr.\ W.\ Ostwald about the dissolution of chromium by mineral acids, which gave a periodic rate of this rather physical process, does not fit the analysis below. 

Without resorting to premature assumptions about the significance of such reactions, either for chemical dynamics itself, or for application to the explanation of some subtler process in nature, I will try to use one theoretical example to show that in some cases they not only can, but must occur. Moreover, the possibility of such cases is not limited at all.

Let's take one well-known example. Prof.\ R.\ Wegscheider studied the relationship between reaction rate constants and chemical equilibrium constants. In his work,\footnote{\foreignlanguage{German}{Über simultane Gleichgewichte etc. -- Monatshefte für Chemie. -- B. XXII. 1901. p. 860.}} he discusses, among others, the problem of three substances and presents the reaction equation between the three isomers. In this case, he considers unimolecular reactions, which essentially facilitates the calculations. There, the reaction occurs reversibly in all possible combinations. Substance $M_1$ turns to $M_2$, and $M_2$ to $M_1$, $M_2$ turns to $M_3$, and $M_3$ to $M_2$, $M_3$ turns to $M_1$, and $M_1$ to $M_3$.
Thus, the total change
\begin{center}
    \includegraphics[width=0.6\linewidth]{figures/Hirniak1908_Eq01.png}
\end{center}
consists of as many as six reaction phases, or six separate component processes. Each of them has its own specific and separate rate, which is determined by the corresponding factors $k_1$, $k_2$, $k_3$, $k_4$, $k_5$, $k_6$.

Let's denote these phases with Roman numerals as follows:
\begin{center}
    \includegraphics[width=0.7\linewidth]{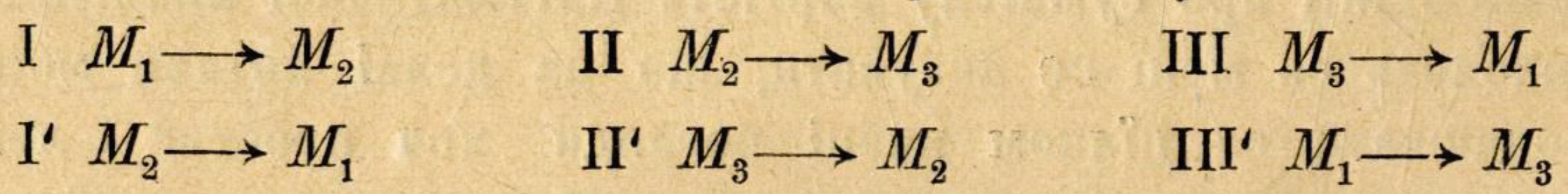}
\end{center}

The initial concentrations of the three isomers $M_1$, $M_2$, $M_3$ will be denoted as $A_1$, $A_2$, $A_3$. Let the decrease of these concentrations during time $t$ be $\xi_1$, $\xi_2$, $\xi_3$. We define individual reaction rates in a sequential order, and by subscribing individual factors for clarity, we obtain:
\begin{center}
    \includegraphics[width=0.5\linewidth]{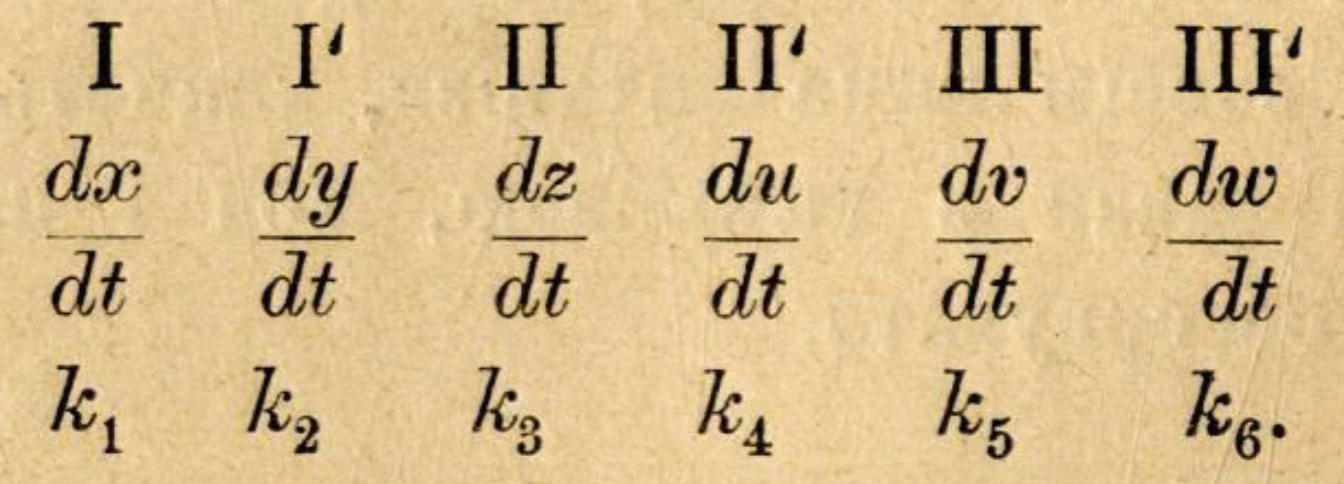}
\end{center}

Due to the mass action law the rates equal to:
\begin{center}
    \includegraphics[width=0.8\linewidth]{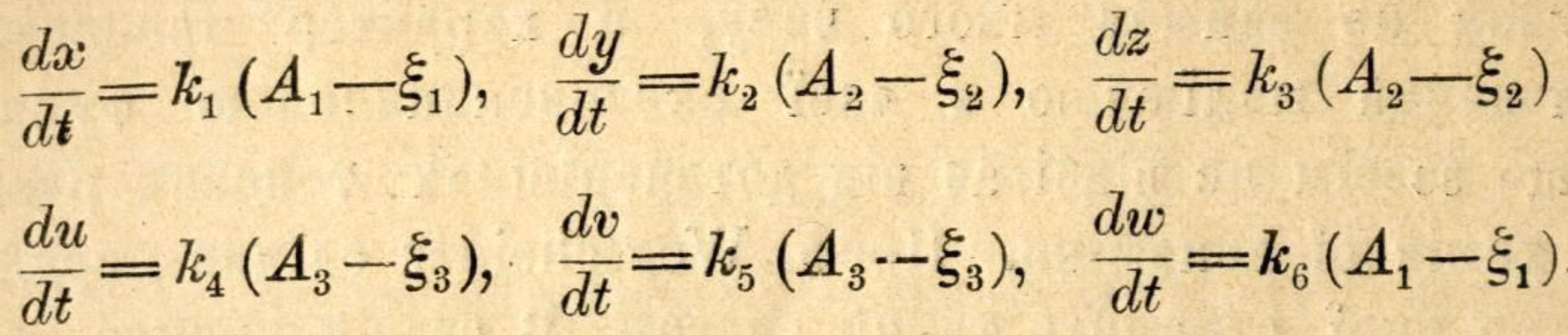}
\end{center}

Considering these equations in more detail, it is easy to obtain the following decrease rates of individual concentrations:
\begin{center}
    \includegraphics[width=1.0\linewidth]{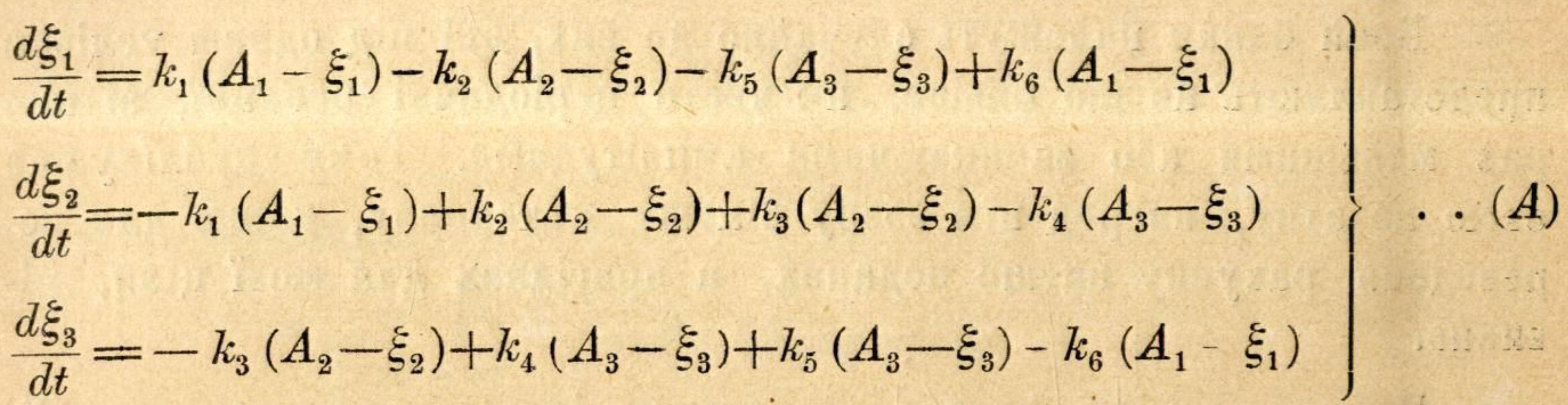}
\end{center}

However, the following equations are obtained from the equations of hylotropic groups, in which reactions can be formulated according to Ostwald.\footnote{See Wegscheider -- Ibid.} The mass conservation law is expressed here by the relation: $\xi_1+\xi_2+\xi_3=0$, from which it also follows that $\dfrac{d\xi_1}{dt}+\dfrac{d\xi_2}{dt}+\dfrac{d\xi_3}{dt}=0$. The last connection indeed follows from the system of equations ($A$). The integration of this system gives the following result:
\begin{center}
    \includegraphics[width=1.0\linewidth]{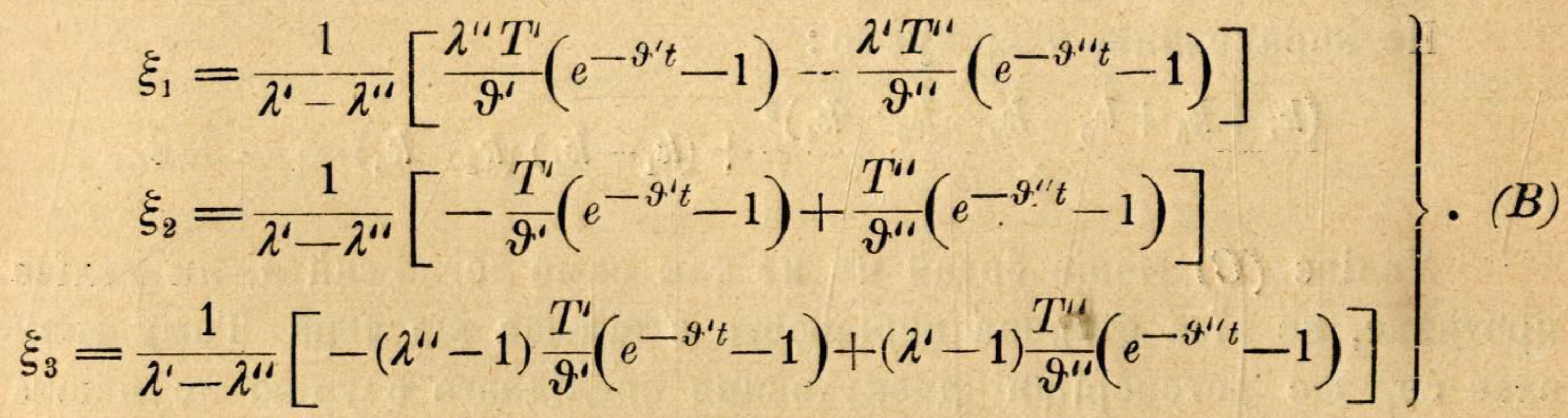}
    \label{eq:Hirniak1908_Eq06}
\end{center}

where the following notations have been introduced:

\begin{center}
    \includegraphics[width=1.0\linewidth]{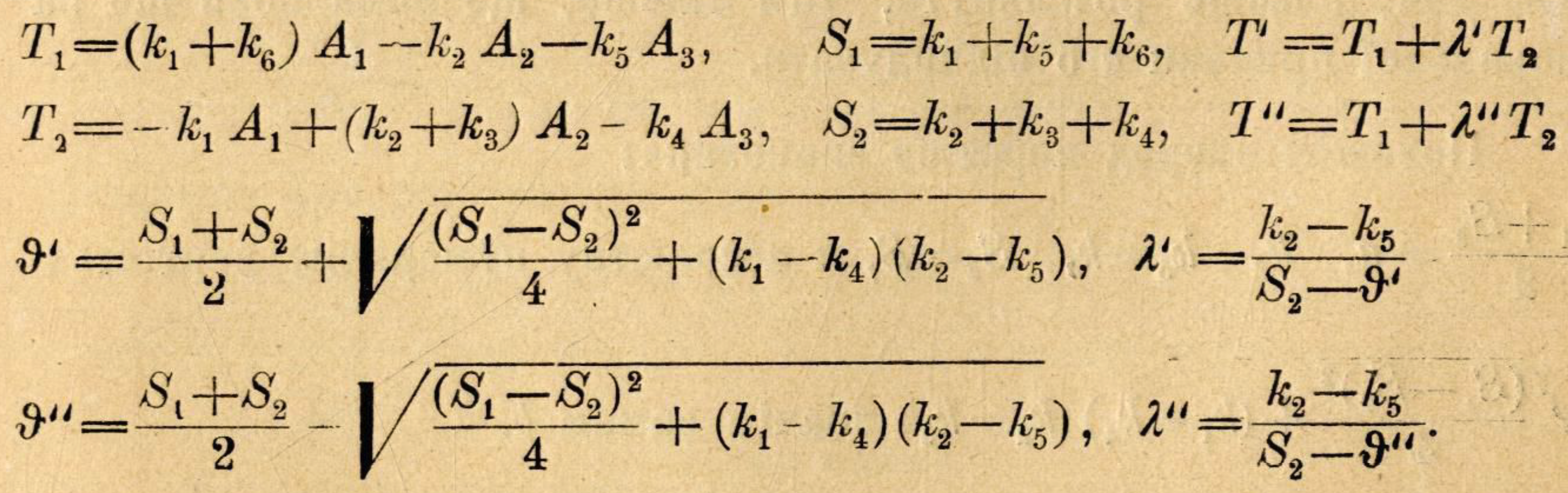}
    \label{eq:Hirniak1908_Eq07}
\end{center}
Although the equations ($B$) look quite complicated, the calculations do not take much time, and the nature of the process, which is described by an exponentially decaying in time function, is such that it does not deviate at all from the current typical solutions to the problems of chemical dynamics. Prof.\ Wegscheider studied first of all the equilibrium to which such a system leads. He had no reason to delve further into the process itself, that is, into the way the state of equilibrium is approached. Therefore, he did not conduct a more detailed discussion of his equations ($B$).

However, it so happened that these equations belong to those which, under one condition, are nothing else but purely periodic reactions with increasingly smaller or decaying amplitudes. Therefore, in publishing this note before collecting more material, I will use the above equations for the calculations that are suitable for my purpose.

So let's see if the system of integral equations ($B$) can be physically interpreted in the case when the inequality holds:
\begin{center}
    \includegraphics[width=1.0\linewidth]{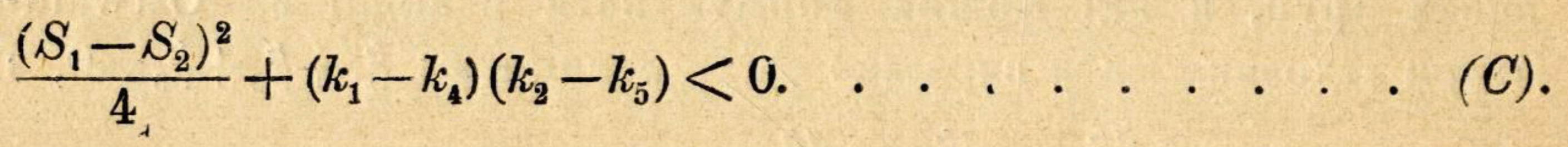}
\end{center}

Let's put, for example, the following values for individual $k$:
\begin{center}
    \includegraphics[width=0.7\linewidth]{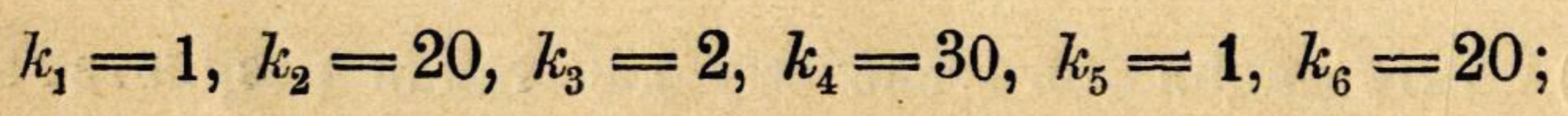}
    \label{eq:first set of rate constants}
\end{center}

After calculations, we get:
\begin{center}
    \includegraphics[width=0.8\linewidth]{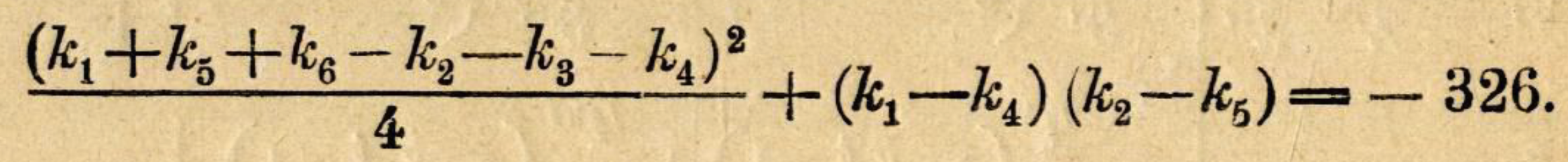}
    \label{eq:Hirniak1908_Eq10}
\end{center}

Condition ($C$) can be fulfilled as easily as the opposite condition, when the left side of the inequality is greater than zero. Therefore, it seems to me that the discussions of the last case until now have no advantage over the previous one ($C$), and it is this one that I want to take a closer look at, especially since there is no rational reason to reject it yet. \\

We will use shorter notations in the future: 
\begin{center}
    \includegraphics[width=0.5\linewidth]{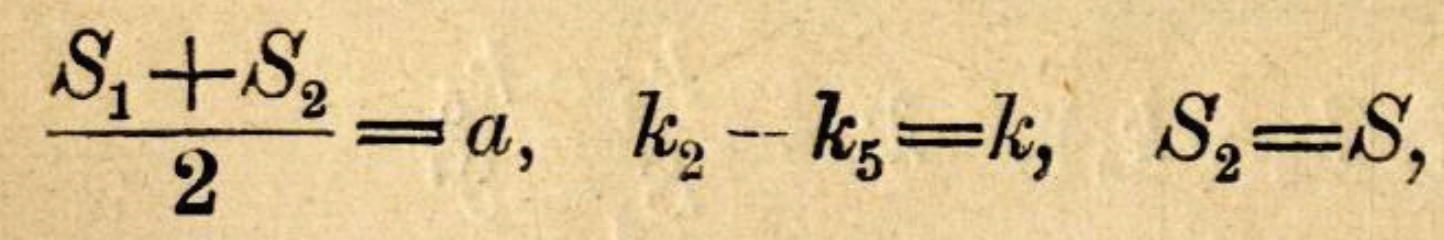}
\end{center}
and the absolute numerical value
\begin{center}
    \includegraphics[width=0.4\linewidth]{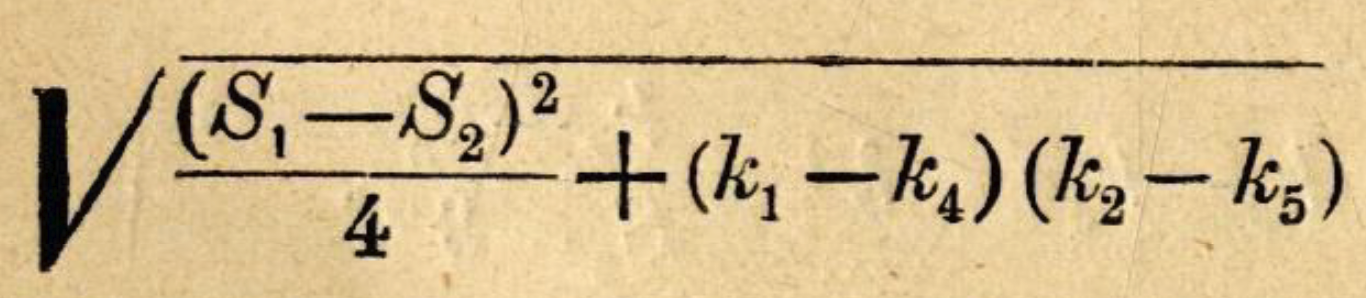}
\end{center}
will be denoted as $b$. \\

Substituting these notations into the abbreviations given above 
(p. 3), we get:
\begin{center}
    \includegraphics[width=0.8\linewidth]{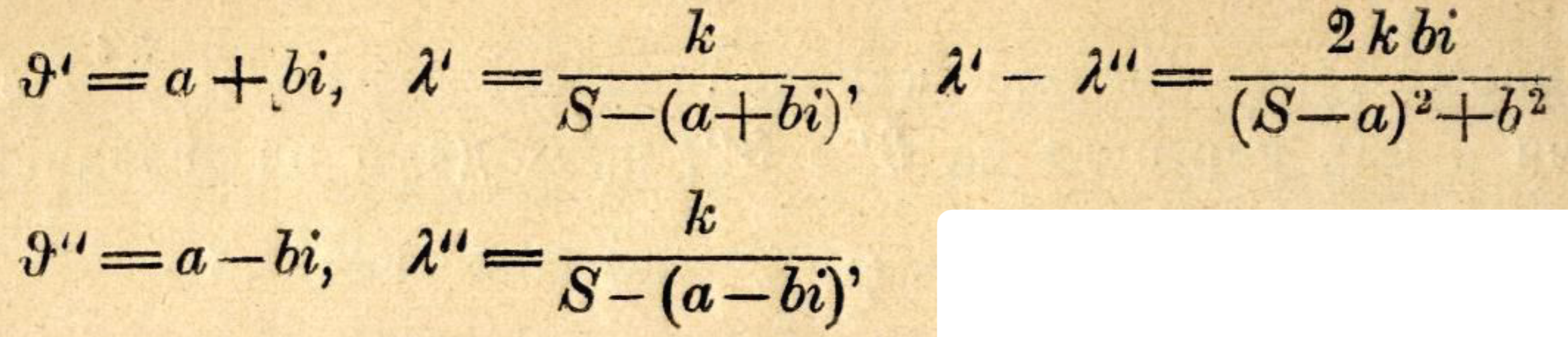}
 \hspace{-30mm}where $i=\sqrt{-1}$.
\end{center}

Let's proceed to the calculation of integrals $\xi_1$, $\xi_2$, $\xi_3$.
We obviously have here:
\begin{center}
    \includegraphics[width=0.8\linewidth]{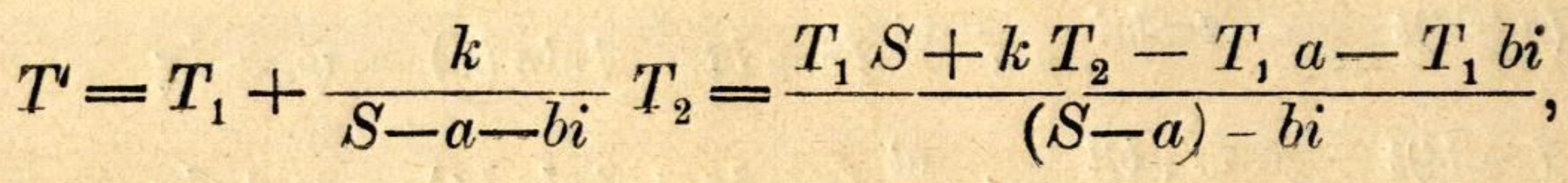}
\end{center}

or, with further shortenings
\begin{center}
    \includegraphics[width=0.8\linewidth]{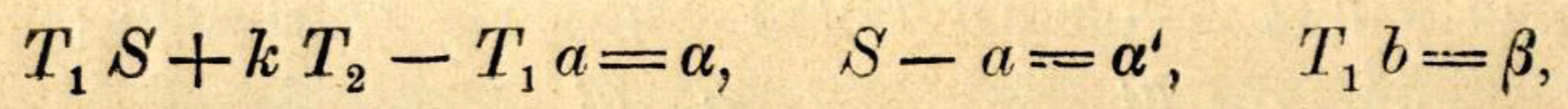}
\end{center}

instead of the former expression
\begin{center}
    \includegraphics[width=0.2\linewidth]{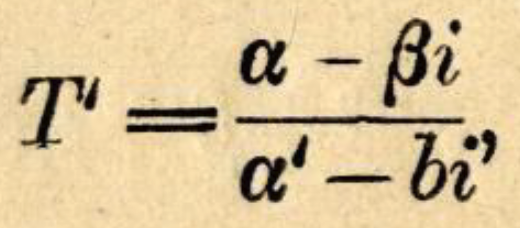}
\end{center}
and quite similarly
\begin{center}
    \includegraphics[width=0.2\linewidth]{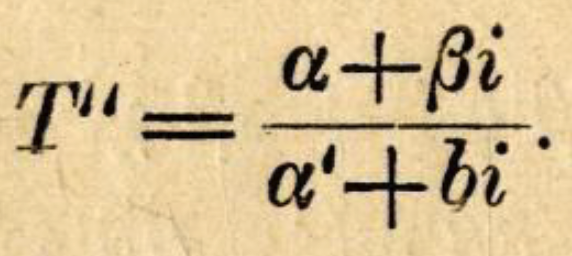}
\end{center}

Let's calculate the values of $\dfrac{T'}{\vartheta'}$ and $\dfrac{T''}{\vartheta''}$. The first one:
\begin{center}
    \includegraphics[width=0.9\linewidth]{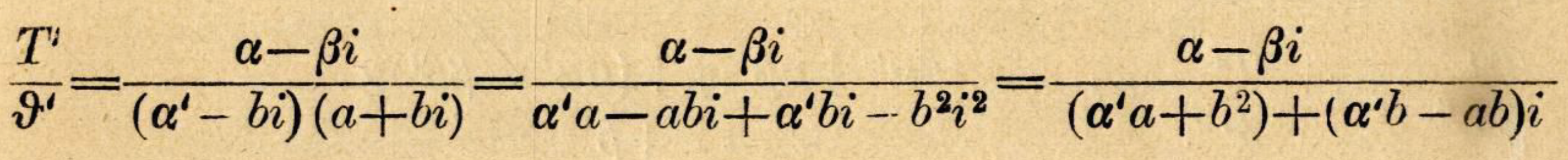}
\end{center}

gives by definition
$(\mathbf{\alpha'} a + b^2) = \mathbf{\alpha''},\;\; \mathbf{\alpha'} b - ab = \beta'$, the value $\dfrac{\mathbf{\alpha} - \beta i}{\mathbf{\alpha''} + \beta' i}$,
and the second one in the same way $\dfrac{T''}{\vartheta''}=\dfrac{\mathbf{\alpha} + \beta i}{\mathbf{\alpha''} - \beta' i}~.$
From which it is easy to calculate again
\begin{center}
    \includegraphics[width=0.5\linewidth]{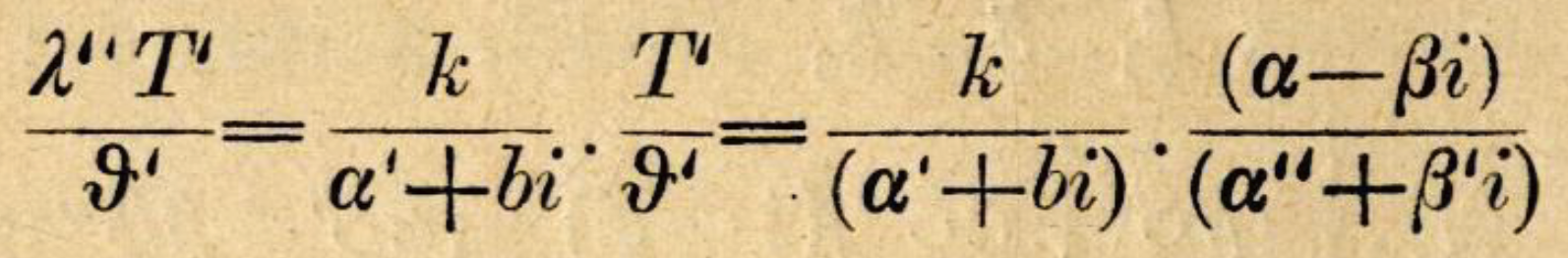}
\end{center}

Proceeding like this further
\begin{center}
    \includegraphics[width=0.6\linewidth]{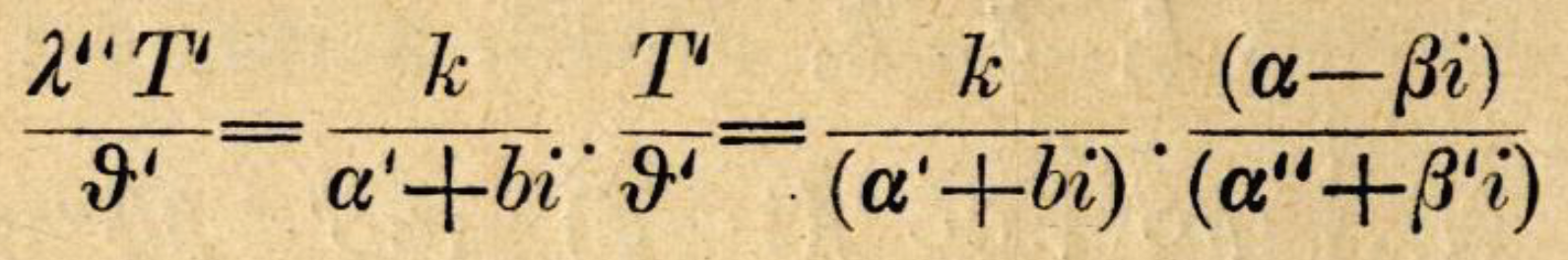}
\end{center}
we get with simultaneous reduction
\begin{center}
    \includegraphics[width=0.5\linewidth]{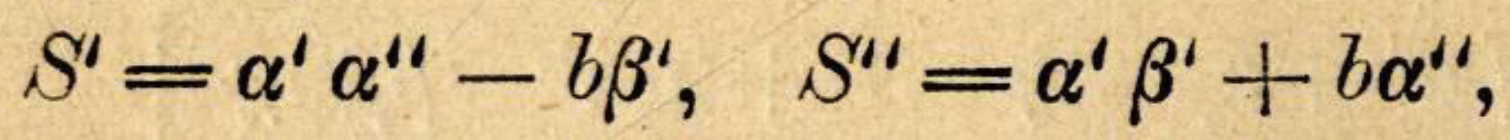}\\
    \includegraphics[width=0.2\linewidth]{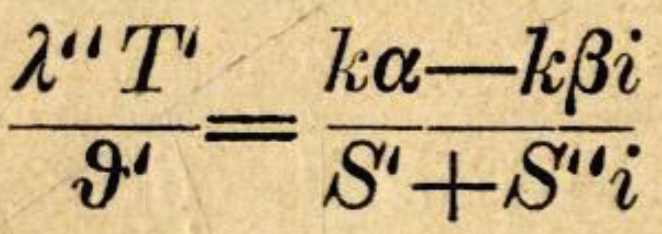}
\end{center}

and is the same way:
\begin{center}
    \includegraphics[width=0.2\linewidth]{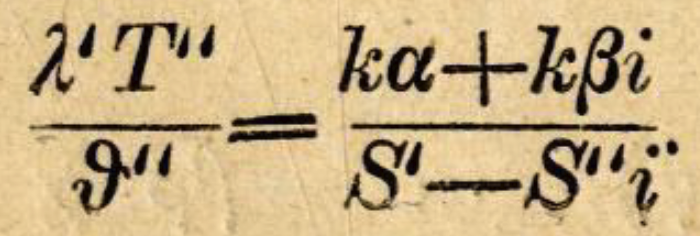}
\end{center}

Now let's take a closer look at exponential expressions $e^{-\varphi't}$ and $e^{-\varphi''t}$.
By substituting there the  values of $\varphi'$ and $\varphi''$, we arrive due to the definition
\begin{center}
    \includegraphics[width=0.2\linewidth]{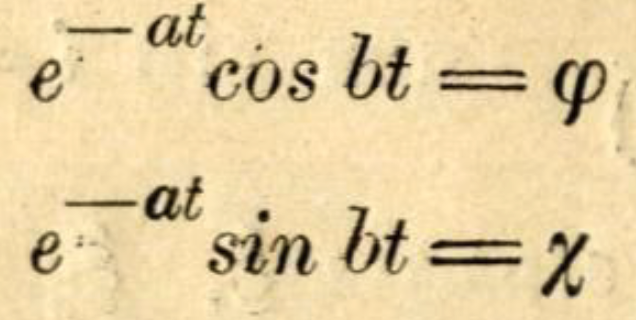}
\end{center}

at the entirely new forms:
\begin{center}
    \includegraphics[width=01.0\linewidth]{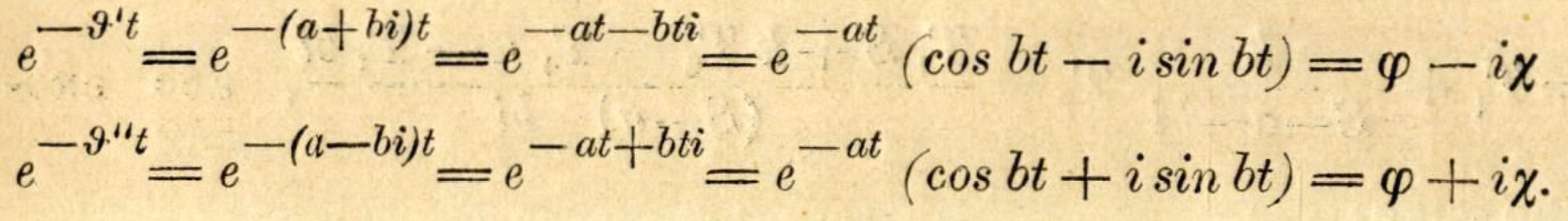}
\end{center}

Using them, let us now calculate the following expression:
\begin{center}
    \includegraphics[width=0.8\linewidth]{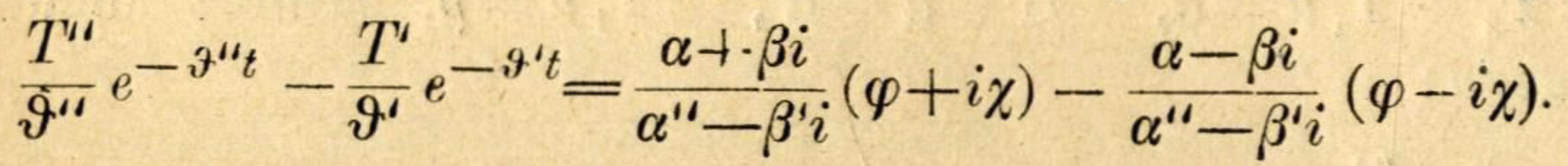}
\end{center}

Bringing to the common denominator $\alpha''^{2} + \beta'^{2}$ we get the following value for the numerator:
\begin{center}
    \includegraphics[width=0.9\linewidth]{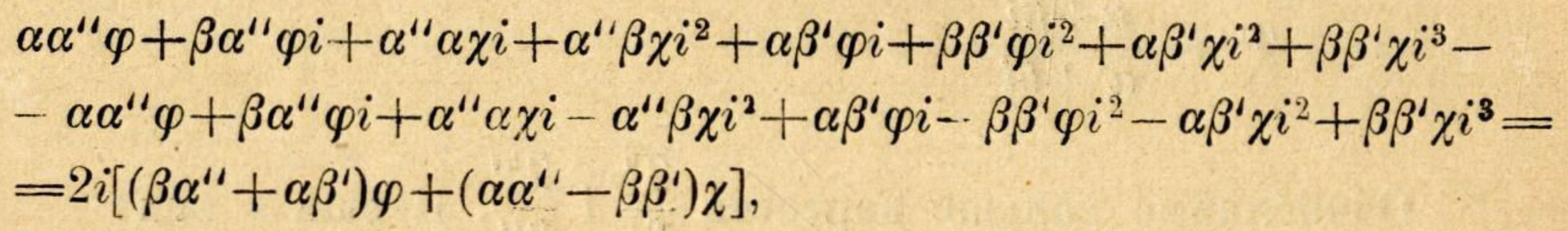}
\end{center}
resulting in
\begin{center}
    \includegraphics[width=0.8\linewidth]{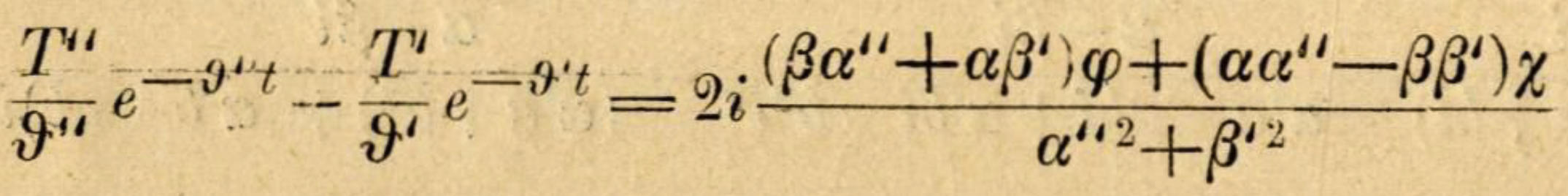}
\end{center}
Proceeding similarly, we can calculate easily as well
\begin{center}
    \includegraphics[width=0.9\linewidth]{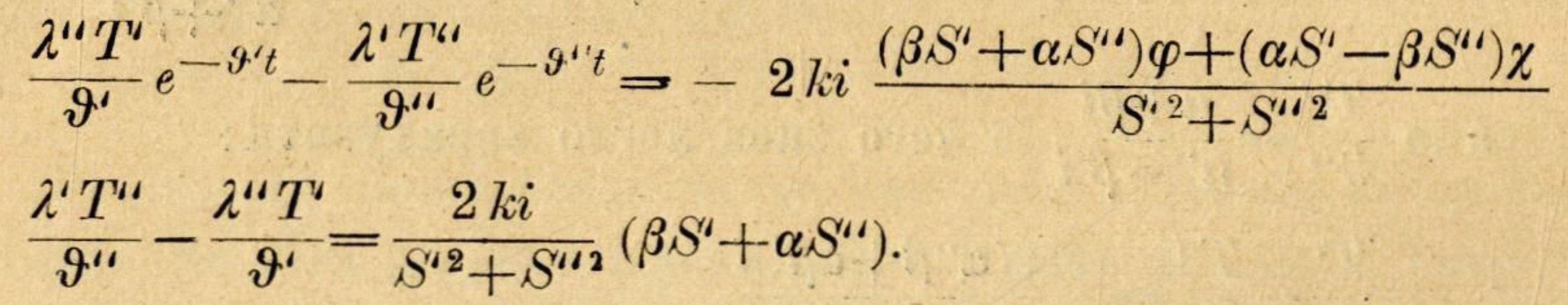}
\end{center}

Having calculated in this way all the constituent terms of the system of integral equations (\textit{B}), we can now easily write the equations describing the changes of individual $\xi_1$, $\xi_2$, $\xi_3$,  over time, namely:
\begin{center}
    \includegraphics[width=1.0\linewidth]{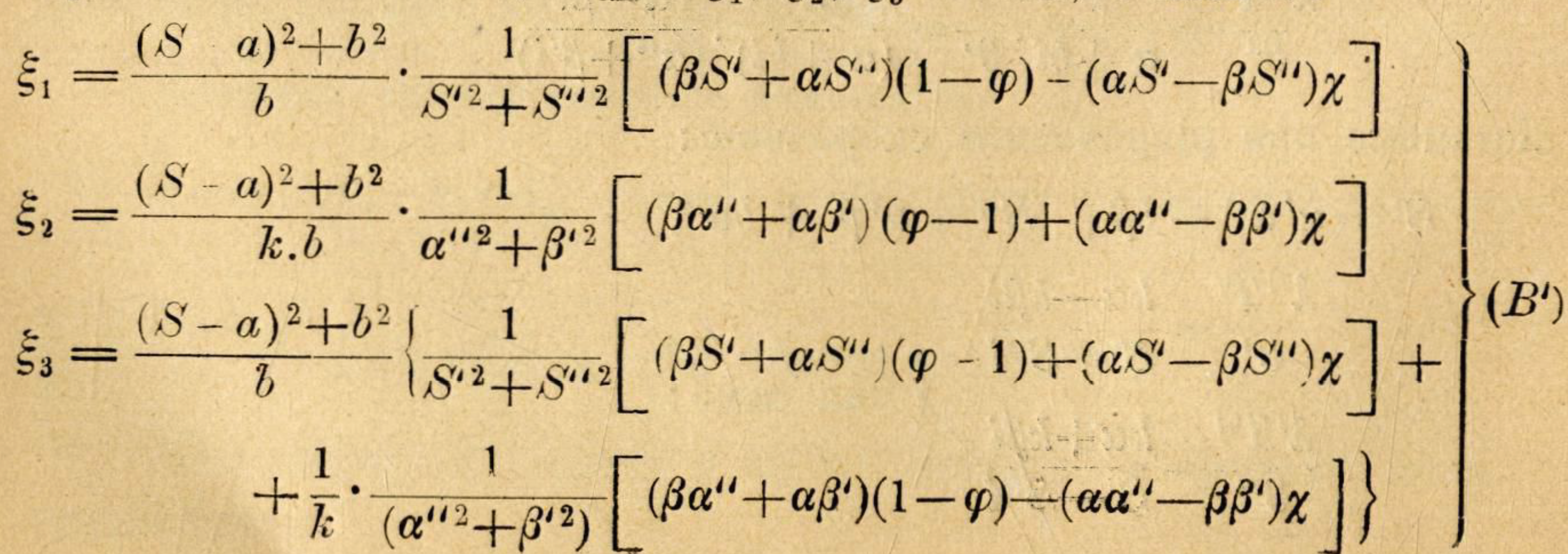}
\end{center}

Comparing the new form ($B'$) of the system of integral equations with the form ($B$), we immediately see that all complex (imaginary) numbers marked in ($B$) with letters $\lambda',\, \lambda'',\, T',\, T'',\, \varphi',\, \varphi''$ fell out, and only real factors remained:
\begin{center}
    \includegraphics[width=0.8\linewidth]{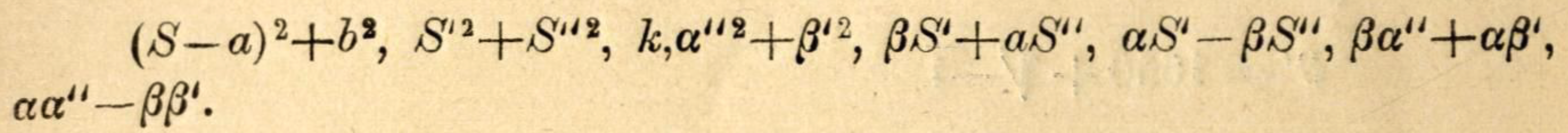}
\end{center}

Exponential terms $e^{-\vartheta't}$ and $e^{-\vartheta''t}$ have become periodic functions $e^{-a t} \cos bt$ and $e^{-at} \sin bt$. Thanks to this, the integrals $\xi_1$, $\xi_2$, and $\xi_3$ became suitable for physical interpretation also under condition ($C$), and the integrals ($B'$) in this case are nothing else than the dependence of changes in individual concentrations on time, expressed by one of the periodic functions, i.e., the goniometric one. \\

Let us consider a numerical example for a more convenient illustration. Let us put, for example, $k_1=1,\, k_2=601,\, k_3=1,\, k_4=501,\, k_5=1,\, k_6= 2196.426$. 
\label{eq:second set of rate constants}
Then $a=1650.713$ and $b=1$. 
Let us choose as well $A_1=1.3684. \, A_2=5,\, A_3=0.0002$, having in mind here such units that the solution comprising $A_1+A_2+A_3$ remains in a liquid state. After performing rather long calculation, we finally get:
\begin{center}
    \includegraphics[width=0.8\linewidth]{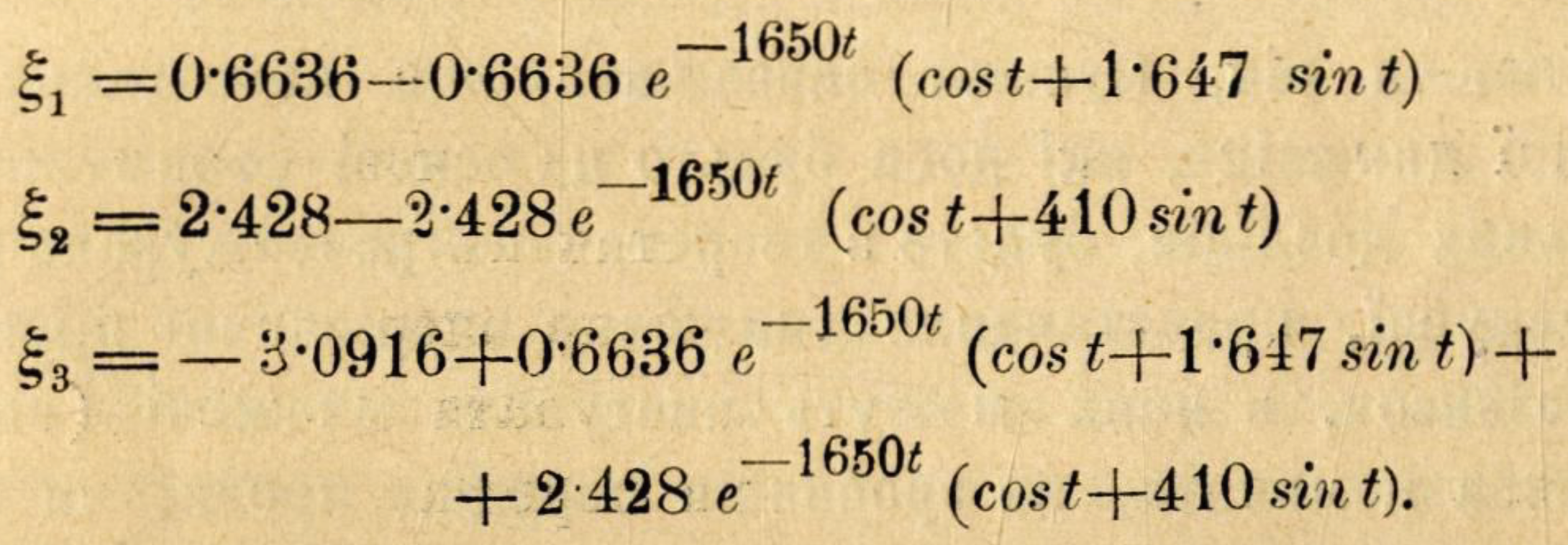}
\end{center}

From the form of these equations it follows first of all that
\begin{itemize}
\item[a)] $\xi_1+\xi_2+\xi_3=0$, that should be kept.
\item[b)] at time $t=0$ we have $\xi_1=\xi_2=\xi3=0$, i.e., the concentrations do not decrease by any amount, but only rest at the initial levels $A_1,\, A_2,\, A_3$. 
\item[c)] in time $t=\infty$ the substance $M_1$ falls from the initial level $1.3684$ by the amount $0.7048$, $M_2$ from $5$ by $2.572$, and $M_3$ rises from $0.0001$ to the height $3.0917$.  
\end{itemize}

This numerical example gives only a significant concentration shift, but due to the extremely high value of the exponent (1650) of the exponential function $e^{-\alpha t}$, deviation or periodic fluctuations of the initial composition of the entire system are so strongly damped from the first period that, for example, they could not be observed experimentally. As it is very easy to further calculate, the first deviation, which is also the maximal one, immediately has a value close to the composition at $t=\infty$. This does not surprise us at all, considering that there are relations as
\begin{center}
    \includegraphics[width=0.25\linewidth]{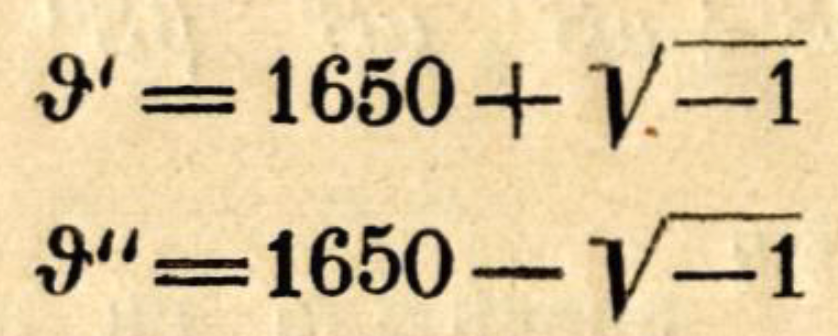}
\end{center}
in which the value of $b$ differs from zero by only one. The more the value of $b$ increases, the more pronounced will be the periods and the longer they will be kept within the limits available for analytical observation. \\

We also immediately see that because of the relation 
$$a=\dfrac{S_1 + S_2}{2}= \dfrac{k_1+k_2+k_3+k_4+k_5+k_6}{2}$$ 
harmonic, that is, continuous periodic changes in the composition of the system are absolutely excluded, because $a$ would be $=0$ only if all $k$ were $=0$, which would be identical to the absence of any reaction. \\

Capitulating the matter, we must note that the most general equations of chemical dynamics, which have been discussed so far either on the basis of the formulation of experiments or for theoretical reasons, represented exclusively aperiodic processes. As we can see, one has not bypass solutions that may contain complex (imaginary) terms. It should first be checked whether they will not fall out of the equations due to the  appropriate computational transformation. \\

It seems to me that there is little prospect and there is no great need to engage in general searches to predict in advance the possibility of periodic processes on the basis of appropriately combined reactions. Therefore, for now I will limit myself to the discussion of one arbitrarily chosen example given above.

\newpage
\section{Translation of Hirniak's 1911 article}
\label{app:1911Translation}
\setcounter{footnote}{0}

 \begin{quote}
    Translation from German of the article ``\foreignlanguage{German}{Zur Frage der periodischen Reaktionen}", \textit{\foreignlanguage{German}{Zeitschrift für Physikalische Chemie}} \textbf{75U} (1911) \cite{Hirniak1911} by Niklas Manz.
 \end{quote}

\begin{center}
    \textbf{On the question of periodic reactions.}\\
    by\\
    \textbf{Julius Hirniak.}\\
    (Received on 22.11.10.)
\end{center}
Mister A.\ Lotka published in the April issue of the $72^\text{nd}$ volume of this magazine an article entitled: "On the theory of periodic reactions". I just recently read this publication and hence, I am only now getting around to pointing out that I dealt with the same question some time ago;
my first note in this regard is in the ``Proceedings of the mathematical-natural Sciences-medical Section"\footnote{Zbirnyk mat.-pryr.-lik Sekcyji.}), Vol. XII, 1908, of the Shevchenko Society of sciences in Lviv in Ruthenian language with the title: ``The periodic chemical reactions", published. In Vol.~XIII of the same publication, I addressed the same question again in the article ``Remarks on the equations of mono- and bimolecular chemical kinetics".

Since my work in other languages has not yet been reviewed, they probably remained generally unknown. But it shows that the question is tackled independently from different sides, which may have something to do with its actuality.

I now consider it appropriate to present my first mentioned publication in more detail before adding a few words about the work of Mr.\ A.\ Lotka.
I pointed out the possibility (even necessity) of a periodic chemical reaction using the example of a monomolecular mutual transformation of three isomers.
\begin{center}
    \includegraphics[width=0.25\linewidth]{figures/Hirniak1911_Eq01.png}
\end{center}
The kinetic system of differential equations was already integrated by Prof.\ Wegscheider for a long time\footnote{``About simultaneous equilibria, etc.", Journal of phys.\ Chemistry \textbf{39}, 268 (1902)}).

\newpage
If we define the concentration decreases $\xi_1$, $\xi_2$, $\xi_3$ of the components $M_1$, $M_2$, $M_3$, the rate constants $k_1$, $k_2$, \ldots $k_6$, and the initial concentrations $A_1$, $A_2$, $A_3$, we find the kinetic system of differential equations of the reaction to be:
\begin{center}
    \includegraphics[width=1.0\linewidth]{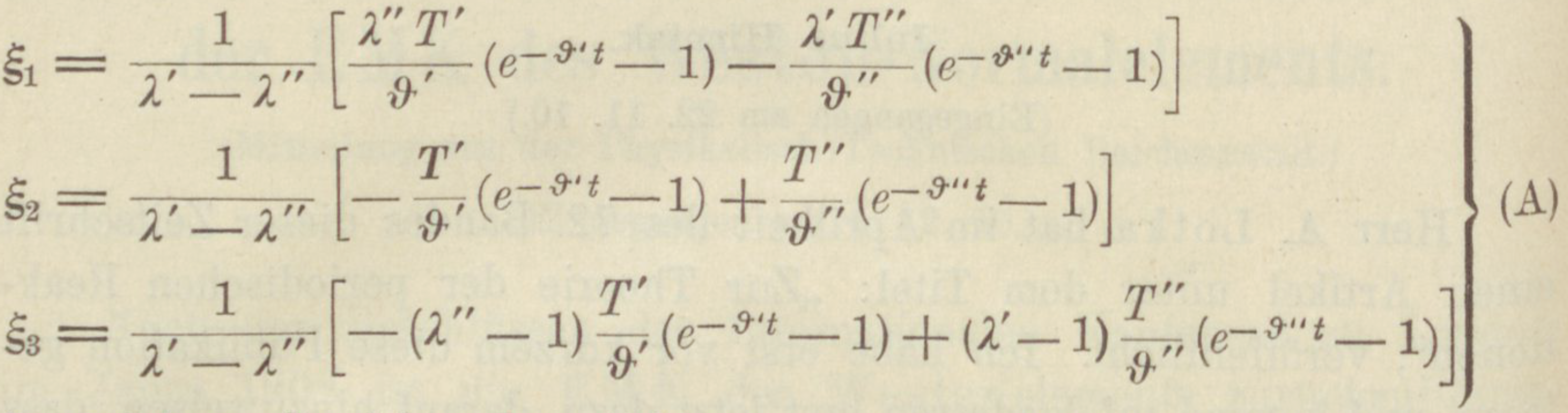}
\end{center}
with the introduction of the following abbreviations:
\begin{center}
    \includegraphics[width=0.8\linewidth]{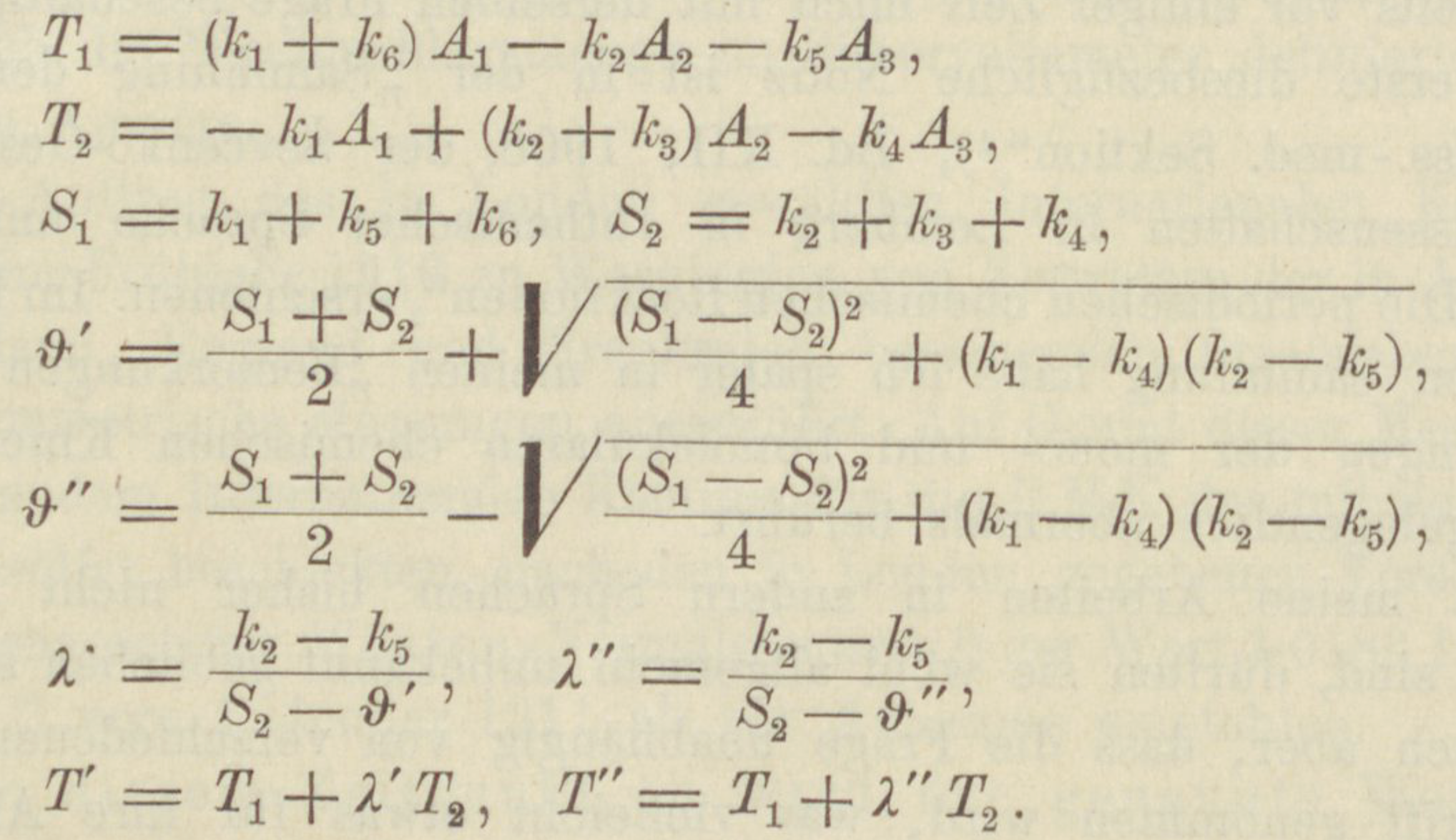}
\end{center}

All six rate constants $k_1$, $k_2$, \ldots $k_6$ must have positive values if the reaction is to retain its character.

It is now also easy to see that the values $\vartheta'$ and $\vartheta''$ can become complex values in these integrals. Given the complete arbitrariness of the values $k_1$, $k_2$ \ldots etc.\ (within the limits of 0 to $\infty$), it can often happen that the expression under the square root sign:
\begin{center}
    \includegraphics[width=0.5\linewidth]{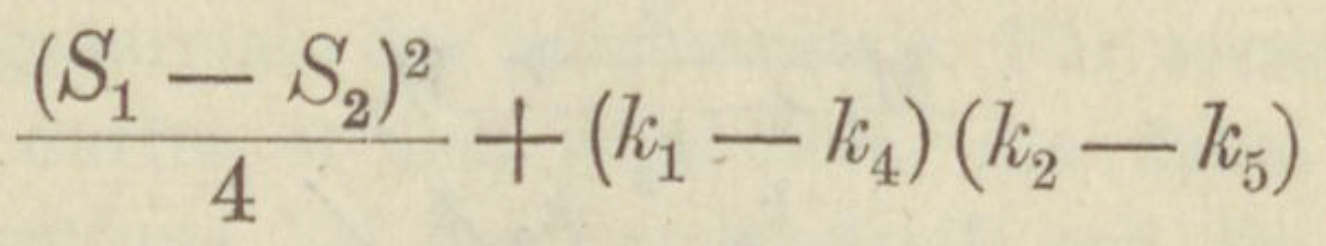}
\end{center}
becomes negative. A numerical example is easy to give. If we insert $k_1 = k_5 = 1$, $k_3 = 2$, $k_2 = k_6 = 20$, $k_4 = 30$ \label{eq:first set of rate constants2}, the value for the discussed expression is $-326$. Due to the connection between the symbols $\vartheta'$ and $\vartheta''$ and the other expressions $\lambda'$, $\lambda''$, $T'$, and $T''$, all expressions of the integral system (A) may become complex, if necessary. \\

\newpage
Without using probability calculus, we can assume that the only conditional equation between the six rate constants 
\begin{center}
    \includegraphics[width=0.8\linewidth]{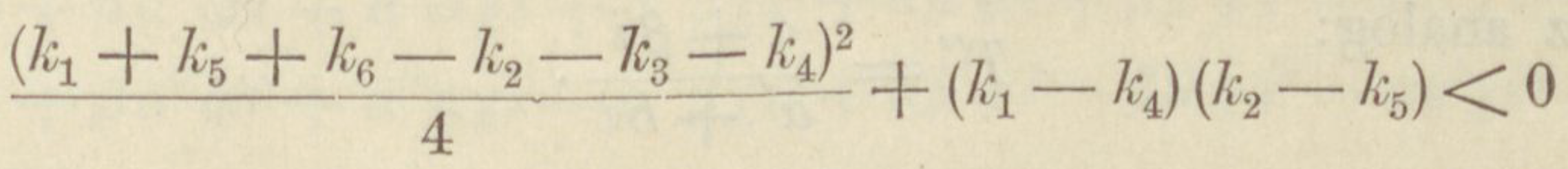}
\end{center}
has a relatively large range of values. This means that the changes of the occurrence of the critical singular case, where everything within the integrals becomes imaginary, are also quite large.

We would come across an imaginary process.

However, it is clear that the case here is completely analogous to that of an electrical discharge, where the process sometimes runs aperiodically, depending on the values of the relevant factors, and at other times resolves into damped sinusoidal oscillations. Unless some yet unknown law prevents the expression under the square root sign from becoming negative, it is difficult to see how one could argue against this conclusion.

This is perhaps so self-evident that further proof is perhaps superfluous. But apart from the fundamental necessity, I also want to show here the simple transformation, on the basis of which the integral system (A) can be freed from all complex values and brought into the form of sine functions.

This should probably make the question a little more tangible.

Let us define:
\begin{center}
    \includegraphics[width=0.6\linewidth]{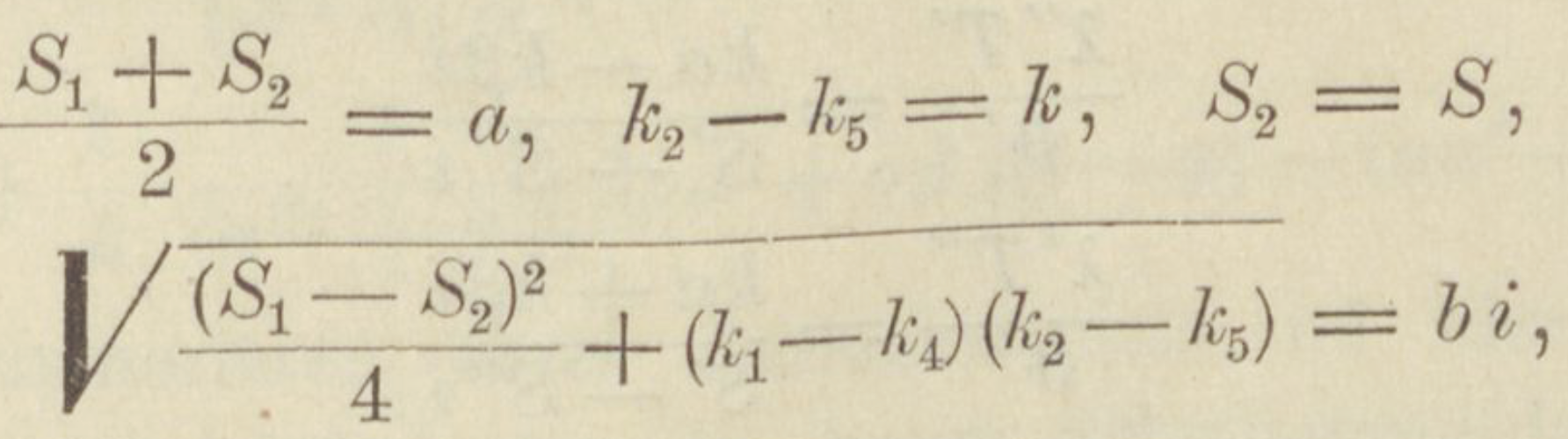}
\end{center}
so follows first:
\begin{center}
    \includegraphics[width=1.0\linewidth]{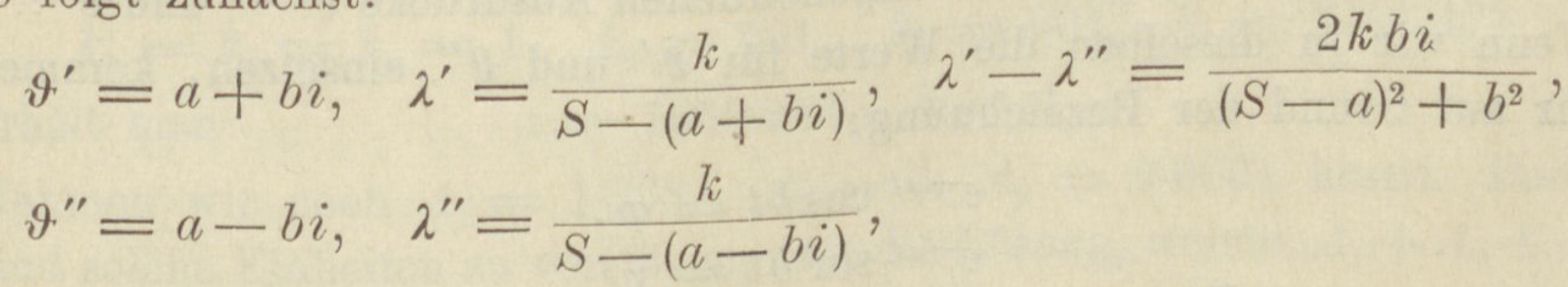}
\end{center}
with $i = \sqrt{-1}$.

Further is:
\begin{center}
    \includegraphics[width=0.9\linewidth]{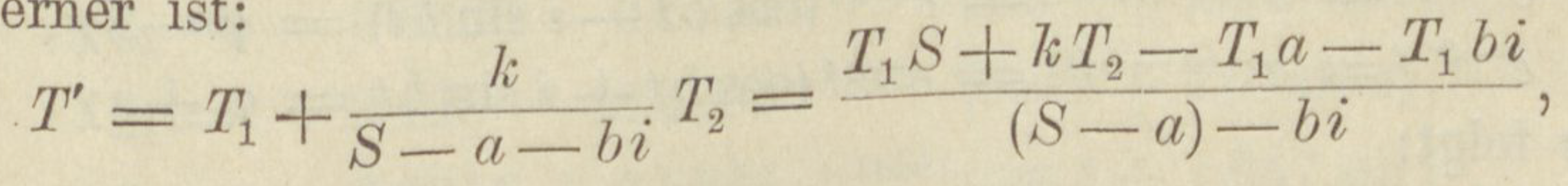}
\end{center}
or, with the introduction of 
\begin{center}
    \includegraphics[width=0.9\linewidth]{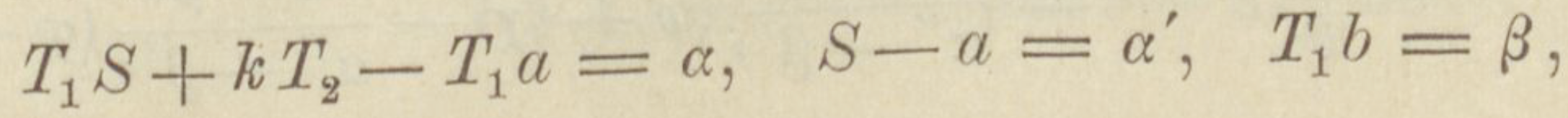}
\end{center}

\noindent also:
\begin{center}
    \includegraphics[width=0.2\linewidth]{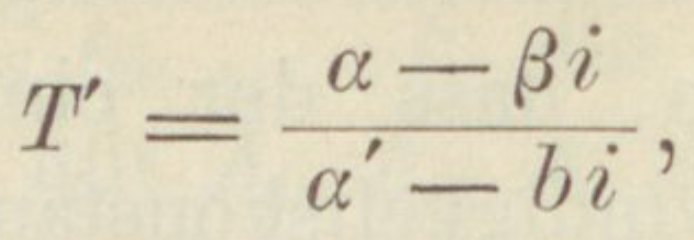}
\end{center}
and similarly:
\begin{center}
    \includegraphics[width=0.2\linewidth]{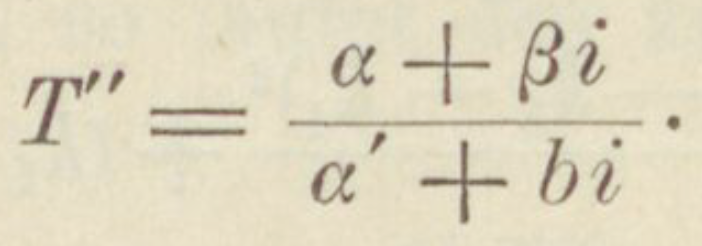}
\end{center}
Let us now calculate the expressions $\dfrac{T'}{\vartheta'}$ and $\dfrac{T''}{\vartheta''}$. The first one:
\begin{center}
    \includegraphics[width=0.8\linewidth]{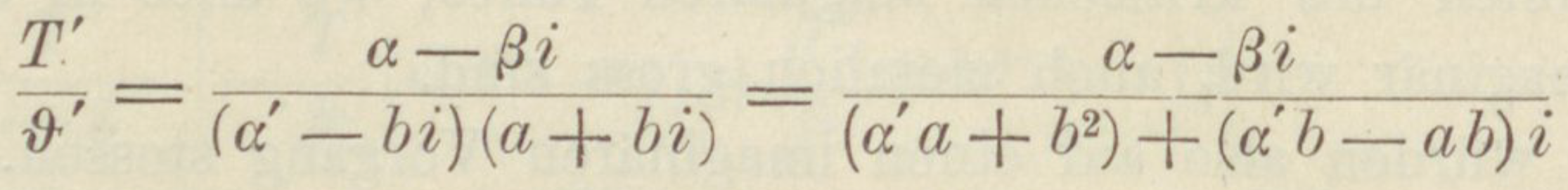}
\end{center}
gives with the following substitutions:
\begin{center}
    \includegraphics[width=0.5\linewidth]{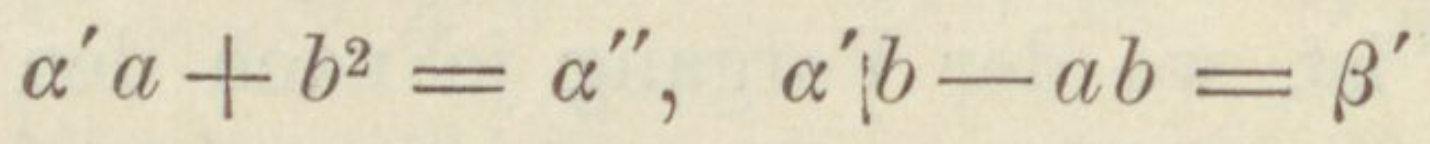}
\end{center}
the value\vspace{-5mm}
\begin{center}
    \includegraphics[width=0.15\linewidth]{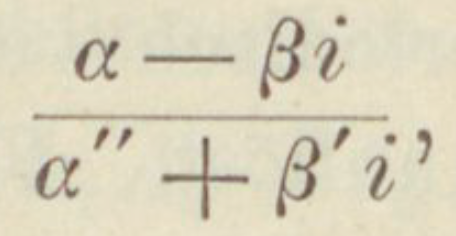}
\end{center}
and the second similarly:\vspace{-5mm}
\begin{center}
    \includegraphics[width=0.25\linewidth]{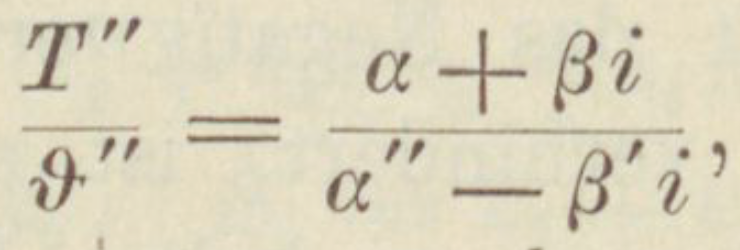}
\end{center}
from which again follows:
\begin{center}
    \includegraphics[width=0.5\linewidth]{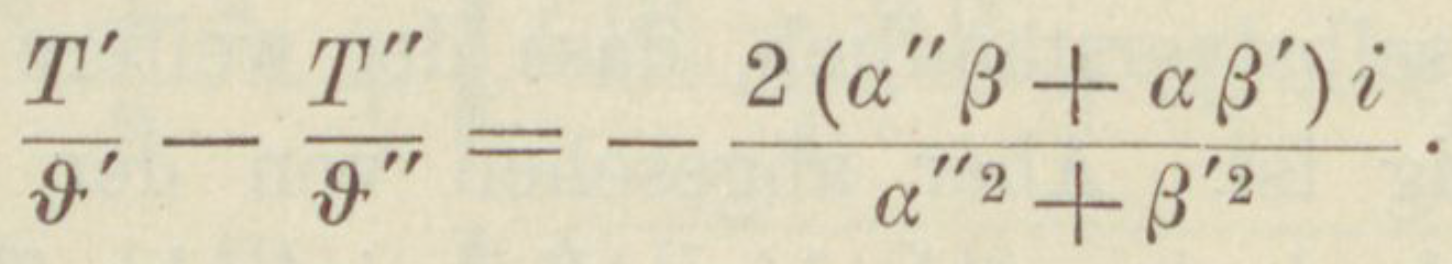}
\end{center}
If we proceed with:
\begin{center}
    \includegraphics[width=0.7\linewidth]{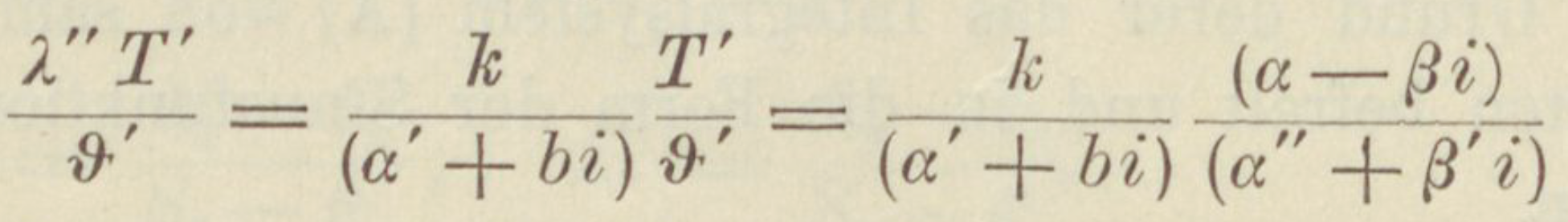}
\end{center}
and use immediately the abbreviations:
\begin{center}
    \includegraphics[width=0.6\linewidth]{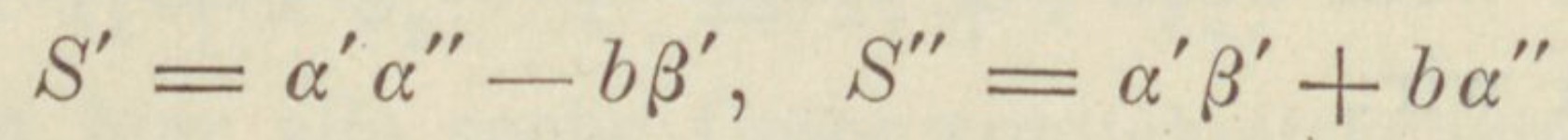}
\end{center}
we obtain:
\begin{center}
    \includegraphics[width=0.3\linewidth]{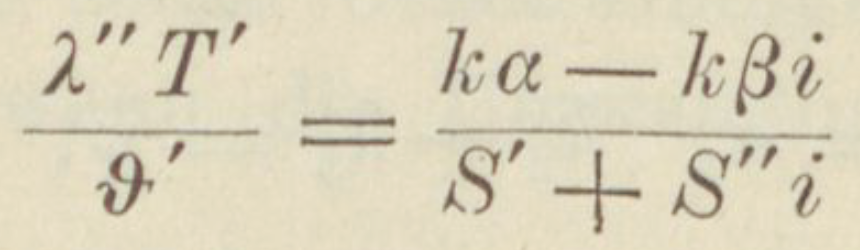}
\end{center}
and also:
\begin{center}
    \includegraphics[width=0.3\linewidth]{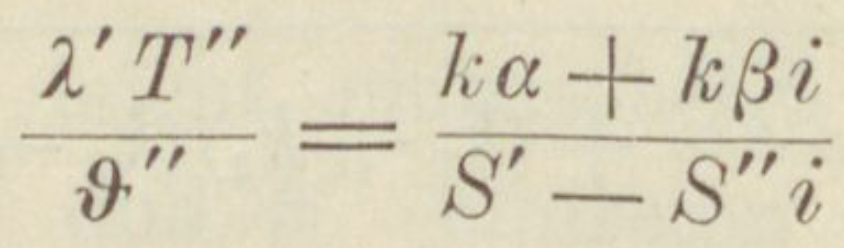}
\end{center}

Let's now take a look at the exponential expressions $e^{-\vartheta't}$ and $e^{-\vartheta''t}$. If we use the values $\vartheta'$ and $\vartheta''$ in those, we come, due to the relationship:
\begin{center}
    \includegraphics[width=0.25\linewidth]{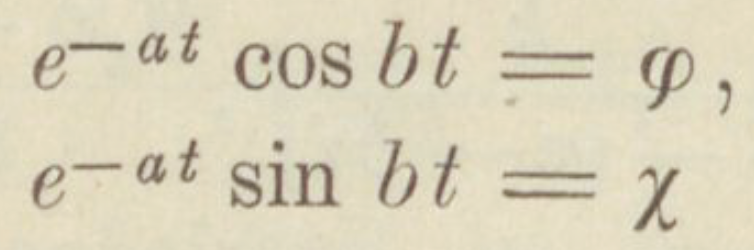}
\end{center}
to completely new forms:
\begin{center}
    \includegraphics[width=0.8\linewidth]{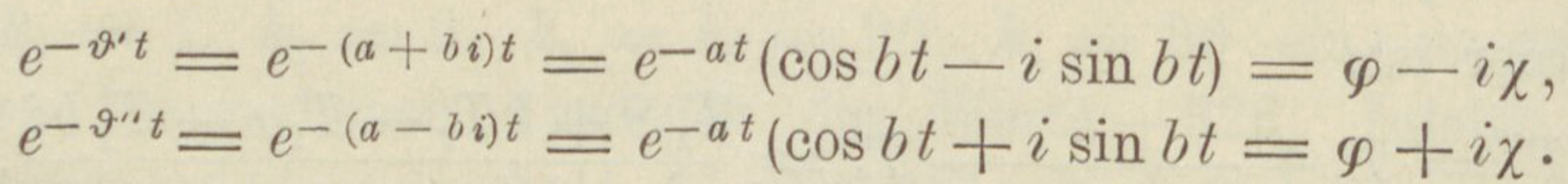}
\end{center}
Therefore, we obtain:
\begin{center}
    \includegraphics[width=0.9\linewidth]{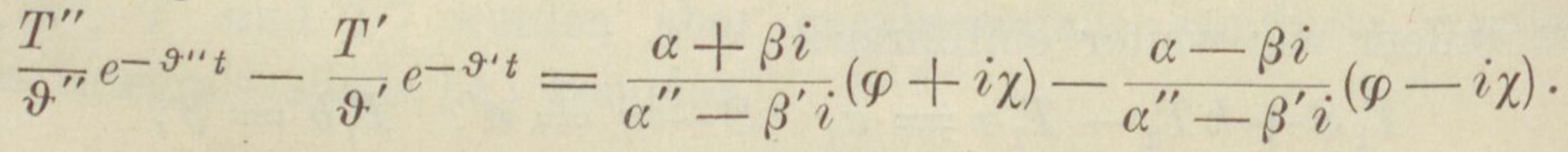}
\end{center}
If we use the same denominator $\alpha''^2+\beta'^2$ for the right side of the last equation, we obtain for the nominator the symmetrical polynomial:
\begin{center}
    \includegraphics[width=1.0\linewidth]{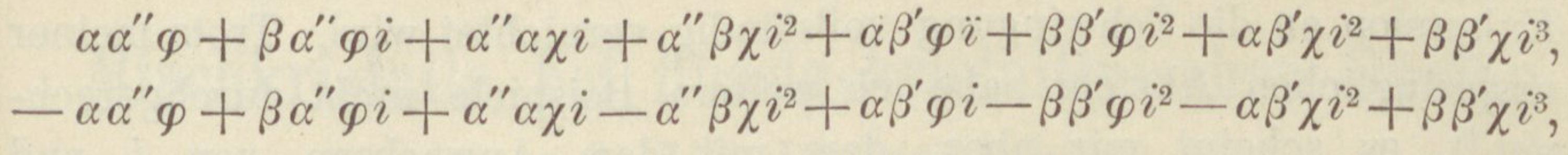}
\end{center}
or immediately:
\begin{center}
    \includegraphics[width=0.5\linewidth]{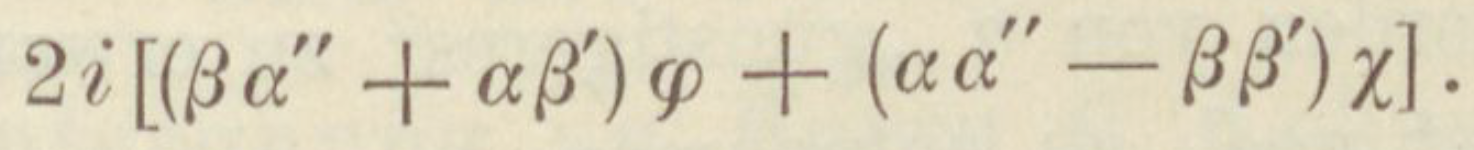}
\end{center}
Therefore, the result is:
\begin{center}
    \includegraphics[width=0.9\linewidth]{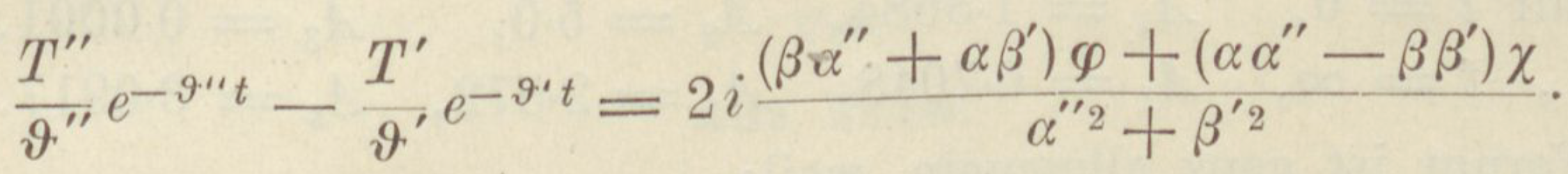}
\end{center}
Using the same path, we obtain:
\begin{center}
    \includegraphics[width=0.9\linewidth]{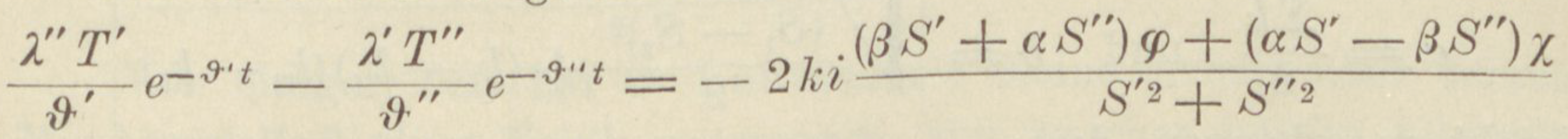}
\end{center}
and:
\begin{center}
    \includegraphics[width=0.7\linewidth]{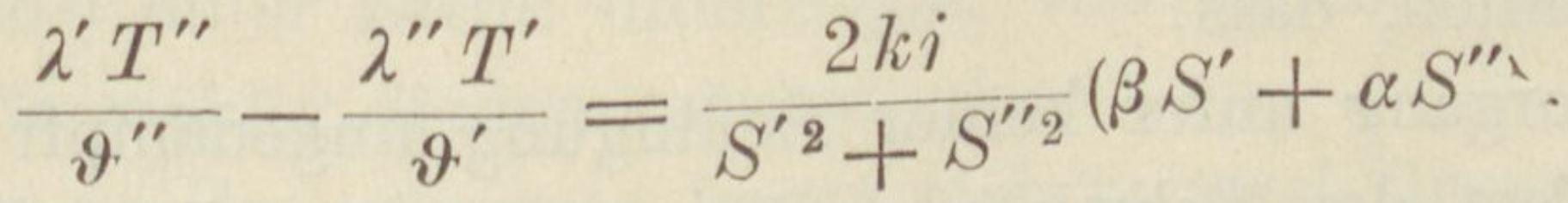}
\end{center}

We now have all the necessary converted terms of the integral system (A), which we can simply write as:
\begin{center}
    \includegraphics[width=1.0\linewidth]{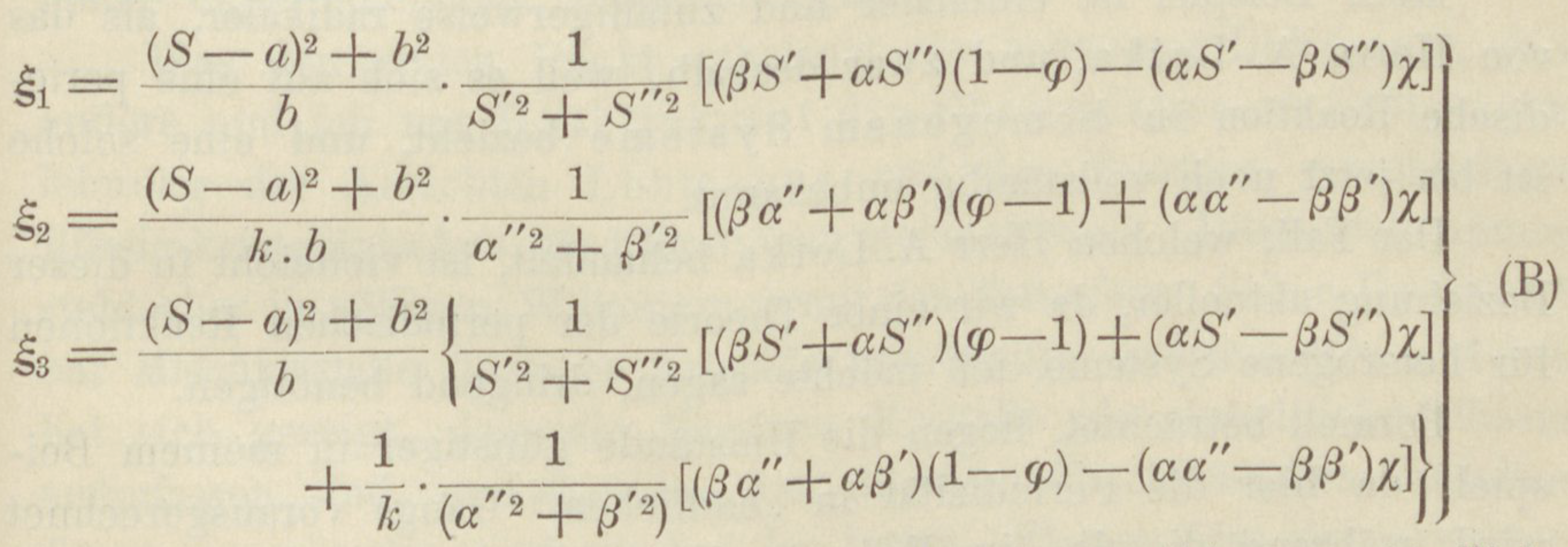}
\end{center}

Calculating a numerical example is complicated.
With the values (intentionally chosen for the short oscillation period):
\begin{center}
    \includegraphics[width=0.9\linewidth]{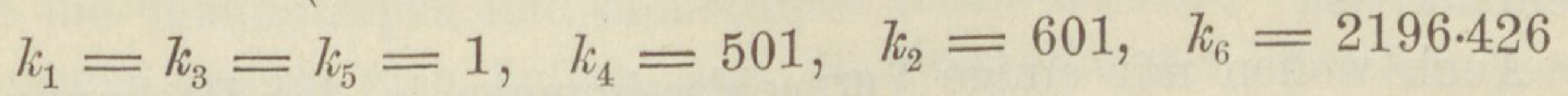}
\end{center}
\label{eq:second set of rate constants2}
we obtain: 
\begin{center}
    \includegraphics[width=0.35\linewidth]{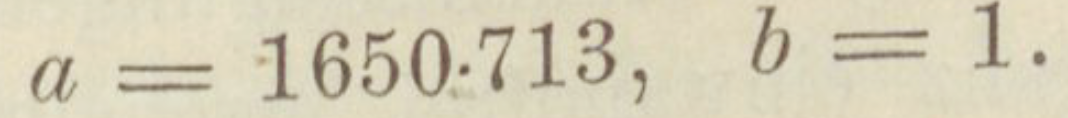}
\end{center}
Let's add $A_1 = 1.3684,\;\; A_2 = 5,\;\; A_3 = 0.0001$. We need to consider units in a way to keep the solution, which includes $A_1 + A_2 + A_3$, diluted. After the insertion follows:
\begin{center}
    \includegraphics[width=0.8\linewidth]{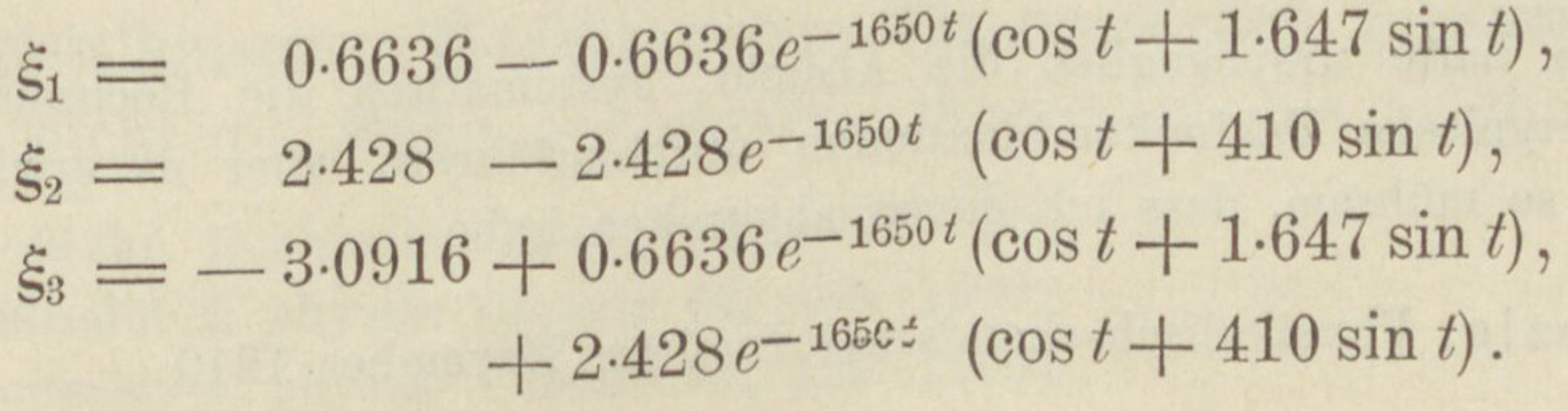}
\end{center}

\setcounter{footnote}{0}
Because of $b = 1$, the oscillation period is short and, because of $a=1650$, the amplitudes are extraordinarily damped. The process here is practically completely aperiodic, because the oscillation is completely destroyed after the first oscillations. Despite my original intention, I did not calculate any further examples\footnote{Originally, I intended to systematically carry out the calculation for various typical number combinations, but the work turned out to be so tedious that I decided against it.}), but it seems to me that as $b$ grows and $a$ becomes smaller, we can expect ``favorable” results. From the given example it can also be seen:
\begin{center}
    \includegraphics[width=0.8\linewidth]{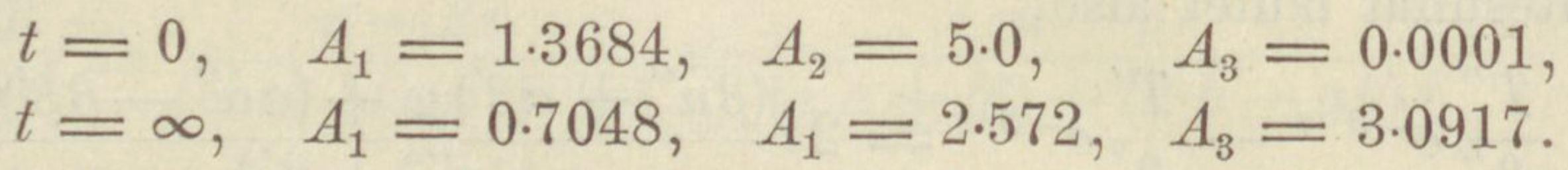}
\end{center}
Furthermore, because of:
\begin{center}
    \includegraphics[width=0.15\linewidth]{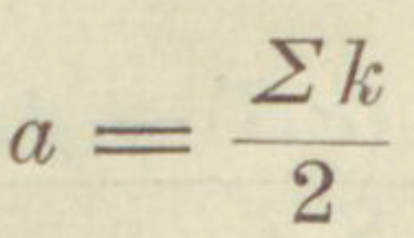}
    \hspace{5mm}\text{and}\hspace{5mm}
    \includegraphics[width=0.6\linewidth]{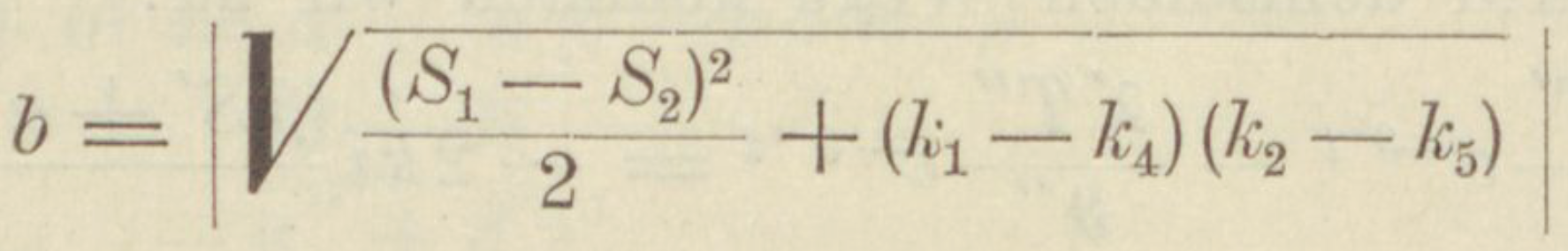}
\end{center}
it is very generally to see that:
\begin{enumerate}
\itemsep0em
    \item the process cannot be undamped under any condition,
    \item the value of the oscillation period and the damping is independent of the initial concentrations.
\end{enumerate}
Finally, I have to make the following comment:

My scheme is simpler and coincidentally more radical than that of Mr.\ A.\ Lotka, because it refers to a periodic reaction in a homogeneous system, and such a reaction is still completely unknown.

The case that Mr.\ A.\ Lotka is dealing with is perhaps more relevant, since we urgently need, I would say, a theory of periodic reactions for heterogeneous systems.

From a formal point of view, the circumstances in my example are more favorable since here the periodicity is calculated in a ``finite" way, while in Mr.\ Lotka's case it is limited to the infinitely small concentration area directly at the equilibrium limit.

Also, some of the neglects/simplifications in Mr.\ Lotka's calculation do not seem entirely correct to me. But since their discussion would be too much, I rather forego it until further research in this area appears to make their considerations important. \\

\hspace{10mm}Leipzig, Physical-Chemistry Institute, November 1910

\newpage
\section{Translation of Section IV of Wegscheider's 1901 article}
\label{app:1901Translation}
\setcounter{footnote}{0}

 \begin{quote}
    Translation from German of Section `\textit{IV. \foreignlanguage{German}{Die gegenseitige Umwandlung von drei Isomeren}}' in the article ``\foreignlanguage{German}{Über simultane {G}leichgewichte und die {B}eziehungen zwischen {T}hermodynamik und {R}eactionskinetik homogener {S}ysteme}" (About simultaneous equilibria and the relationship between thermodynamics and reaction kinetics of homogeneous systems) by \textsc{Rudolf Wegscheider}, \textit{\foreignlanguage{German}{Monatshefte für Chemie}} \textbf{22}(8) (1901) \cite{Wegscheider1901} by Niklas Manz.\footnote{ This paper was identically republished in \textit{\foreignlanguage{German}{Zeitschrift für Physikalische Chemie}} in 1902~\cite{Wegscheider1902} and in the initial journal 10 years later again in 1911~\cite{Wegscheider1911}.}
 \end{quote}

\begin{center}
    \textbf{IV. The interconversion of three isomers.}
\end{center}

If there are three isomeric types of molecules $M_1$, $M_2$, and $M_3$ in a homogeneous system that can convert into one another, then the following six reactions are possible:
\begin{equation}
 \begin{aligned}
    \text{I.}~~~ M_1 \rightarrow M_2, \hspace{5mm}&
    \text{II.} & M_2 \rightarrow M_3, \hspace{5mm}&
    \text{III.} &M_3 \rightarrow M_1. \\
    \text{I'.}~~ M_2 \rightarrow M_1, \hspace{5mm}& 
    \text{II'.} &M_3 \rightarrow M_2, \hspace{5mm}& 
    \text{III'.} &M_1 \rightarrow M_3. \\  
\end{aligned}
\end{equation}
If the concentrations of the three substances are in  equilibrium $C_1$, $C_2$, and $C_3$, then thermodynamics gives the equilibrium conditions
\begin{equation}
    \frac{C_2}{C_1}= K_1,\hspace{5mm}
    \frac{C_3}{C_2}= K_2,\hspace{5mm}
    \frac{C_1}{C_3}= K_3.
\end{equation}\
with the well known relationship
\begin{equation}
    K_3 = \frac{1}{K_1K_2}.
\end{equation}
In this case, the kinetics of the transformation can easily be developed, since we are dealing with simultaneous linear differential equations with constant coefficients.

If the rate coefficients of the six reactions I, I', II, II', III, and III' are denoted in sequence by $k_1$ to $k_6$, their rates are given by
\begin{equation}
    \frac{dx}{dt}, \hspace{3mm} \frac{dy}{dt}, \hspace{3mm}\frac{dz}{dt},\hspace{3mm}
    \frac{du}{dt}, \hspace{3mm}\frac{dv}{dt},\hspace{3mm} \frac{dw}{dt},
\end{equation}
and with the initial concentrations $A_1$, $A_2$, and $A_3$ of the three molecules and the concentration decreases $\xi_1$, $\xi_2$, and $\xi_3$ at time $t$ one can find the six reaction equations as:
\begin{equation}
 \begin{aligned}
    \frac{dx}{dt}=k_1(A_1 -\xi_1), \hspace{3mm} &
    \frac{dz}{dt}=k_3(A_2 -\xi_2),  &
    \frac{dv}{dt}=k_5(A_3 -\xi_3), \\
    \frac{dy}{dt}=k_2(A_2 -\xi_2), \hspace{3mm} &
    \frac{du}{dt}=k_4(A_3 -\xi_3), &
    \frac{dw}{dt}=k_6(A_1 -\xi_1). 
\end{aligned}
\end{equation}
The rate of decrease of the concentrations of the molecules $M_1$, $M_2$, and $M_3$ are 
\begin{equation}
 \begin{aligned}
    \frac{d\xi_1}{dt} & = k_1(A_1 -\xi_1)-k_2(A_2 -\xi_2)-k_5(A_3 -\xi_3)+k_6(A_1 -\xi_1), \\
    \frac{d\xi_2}{dt} & = k_1(A_1 -\xi_1)+k_2(A_2 -\xi_2)+k_3(A_2 -\xi_2)-k_4(A_3 -\xi_3), \\
    \frac{d\xi_3}{dt} & = k_3(A_2 -\xi_2)+k_4(A_3 -\xi_3)+k_5(A_3 -\xi_3)-k_6(A_1 -\xi_1).
\end{aligned}
\end{equation}
The sum of the three $\dfrac{d\xi}{dt}$ is zero, since the law of mass conservation gives $\xi_1+\xi_2+\xi_3=0$. By introducing this relationship, the number of differential equations is reduced to two.

The result of the integration is the following. Introducing these abbreviations:
\begin{equation}
 \begin{aligned}
  T_1 & = (k_1 + k_6)A_1 - k_2 A_2 - k_5 A_3,\\
  T_2 & = -k_1 A_1 + (k_2+k_3) A_2 - k_4 A_3,\\
  S_1 & = k_1 +k_5 + k_6,\\
  S_2 &= k_2 +k+3 + k_4,\\
  \vartheta' &= \frac{S_1 + S_2}{2}+\sqrt{\frac{(S_1 + S_2)^2}{4}+(k_1-k_4)(k_2-k_5)},\\
   \vartheta'' &= \frac{S_1 + S_2}{2}-\sqrt{\frac{(S_1 + S_2)^2}{4}+(k_1-k_4)(k_2-k_5)},\\
   \lambda' & = \frac{k_2-k_5}{S_2 - \vartheta'},\\
   \lambda'' & = \frac{k_2-k_5}{S_2 - \vartheta''},\\
   T' & = T_1 + \lambda' T_2,\\
   T'' & = T_1 + \lambda'' T_2,
\end{aligned}
\end{equation}
one can write
%
 \begin{align}
   \xi_1 & = \frac{1}{\lambda' - \lambda''}
            \left[\frac{\lambda'' T'}{\vartheta'}\left(e^{-\vartheta' t}-1\right)-\frac{\lambda' T''}{\vartheta''}\left(e^{-\vartheta'' t} - 1\right)\right],\nonumber\\
   \xi_2 & = \frac{1}{\lambda' - \lambda''}
            \left[-\frac{T'}{\vartheta'}\left(e^{-\vartheta' t}-1\right)+\frac{T''}{\vartheta''}\left(e^{-\vartheta'' t} - 1\right)\right],\\
   \xi_3 & = \frac{1}{\lambda' - \lambda''}
            \left[-(\lambda'' -1)\frac{T'}{\vartheta'}\left(e^{-\vartheta' t}-1\right)+(\lambda' -1)\frac{T''}{\vartheta''}\left(e^{-\vartheta'' t} - 1\right)\right].\nonumber
\end{align}
%
The equilibrium conditions follow from these formulas by setting $t = \infty$. The equilibrium concentrations $C_1$, $C_2$, and $C_3$ are the concentrations of the three molecules $A_1 - \xi_1$, $A_2 - \xi_2$ and $A_3 - \xi_3$ at $t=\infty$.

Therefore one obtains
\begin{equation}
 \begin{aligned}
C_1 & = [A_1 - \xi_1]_{t=\infty} = \frac{1}{\vartheta' \vartheta''} (k_2k_4 + k_2k_5 + k_3k_5)(A_1 + A_2 + A_3),\\
C_2 & = [A_2 - \xi_2]_{t=\infty} = \frac{1}{\vartheta' \vartheta''} (k_1k_4 + k_1k_5 + k_4k_6)(A_1 + A_2 + A_3),\\
C_3 & = [A_3 - \xi_3]_{t=\infty} = \frac{1}{\vartheta' \vartheta''} (k_1k_3 + k_2k_6 + k_3k_6)(A_1 + A_2 + A_3).
\end{aligned}
\end{equation}
The relationships between equilibrium coefficients and rate constants are as follows:
\begin{equation}
 \begin{aligned}
K_1 & = \frac{C_2}{C_1}= \frac{k_1k_4+k_1k_5+k_4k_6}{k_2k_4+k_2k_5+k_3k_5},\\
K_2 & = \frac{C_3}{C_2}= \frac{k_1k_3+k_2k_6+k_3k_6}{k_1k_4+k_1k_5+k_4k_6},\\
K_3 & = \frac{C_1}{C_3}= \frac{k_2k_4+k_2k_5+k_3k_5}{k_1k_3+k_2k_6+k_3k_6}=\frac{1}{K_1K_2}.
\end{aligned}
\end{equation}
The concentration ratios, which are, according to thermodynamics, independent of the total concentration, have a value independent of the initial concentrations according to the kinetic derivation.

The same equilibrium conditions can be obtained much easier as outlined in the second option of the previous section. Inserting the equilibrium concentrations $C_1$, $C_2$, and $C_3$ into the equations for the $\dfrac{d\xi}{dt}$, the $\dfrac{d\xi}{dt}$ must be zero. This gives the following three equations, which define the conditions for the constant concentration values of the three molecules in accordance with equation 18) of the previously mentioned article:
\begin{equation}
 \begin{aligned}
0 & = k_1C_1 - k_2C_2 - k_5C_3 + k_6C_1,\\
0 & = -k_1C_1 + k_2C_2 + k_3C_2 - k_4C_3,\\
0 & = -k_3C_2 + k_4C_3 + k_5C_3 - k_6C_1.
\end{aligned}
\end{equation}
The third equations is a result of adding the first two after a multiplication with $-1$.

From these equations one finds the same values for the equilibrium concentration conditions as before. For example, if one eliminates $C_1$ from two equations, one obtains a relationship between $C_2$ and $C_3$, which is identical to the one above.

The kinetically derived equilibrium conditions agree with the thermodynamic ones. All six rate constants can have arbitrary values. On the other hand, one can see that the relationships $K_1 = \dfrac{k_1}{k_2}$, $K_2 = \dfrac{k_2}{k_3}$, $K_3 = \dfrac{k_3}{k_1}$, which would hold at each individual equilibrium, would not need to be satisfied if there were simultaneous equilibria. The validity of the equation $K_1 = \dfrac{k_1}{k_2}$ would mean that at equilibrium, the reactions I and I' are also in equilibrium on their own, or that $\dfrac{dx}{dt} - \dfrac {dy}{dt}= 0$. The kinetic equilibrium condition derived above also includes the case discussed in Section II, that the simultaneous pairs of counteractions are not in equilibrium on their own, but that they only reach equilibrium through their interaction; then $\dfrac{dx}{dt} - \dfrac{dy}{dt}$ is not zero even at equilibrium.

In order for the individual pairs of counteractions to be in equilibrium on their own in simultaneous equilibrium, certain relationships between the rate constants must be followed. In equilibrium, one obtains the following values for the speed differences of the pairs of opposing reactions:
\begin{equation}
 \begin{aligned}
\left[\dfrac{dx}{dt} - \dfrac{dy}{dt}\right] & = 
\frac{A_1 + A_2 + A_3}{\vartheta' \vartheta''}(k_1k_3k_5 - k_2k_4k_6),\\
\left[\dfrac{dz}{dt} - \dfrac{du}{dt}\right] & = 
\frac{A_1 + A_2 + A_3}{\vartheta' \vartheta''}(k_1k_3k_5 - k_2k_4k_6),\\
\left[\dfrac{dv}{dt} - \dfrac{dw}{dt}\right] & = 
\frac{A_1 + A_2 + A_3}{\vartheta' \vartheta''}(k_1k_3k_5 - k_2k_4k_6).
\end{aligned}
\end{equation}
The three speed differences are therefore identical at equilibrium, as must be the case from Section II. But only for $k_1k_3k_5 = k_2k_4k_6$ they do become zero. Only then will the three reaction pairs be in equilibrium on their own and only then will the relationships $K_1 = \dfrac{k_1}{k_2}$ etc. apply. One can easily convince oneself that these simpler relationships arise from the previously derived general one if one sets $k_1k_3k_5 = k_2k_4k_6$.

The simpler form of the relationship between equilibrium constants and rate constants always applies, among other things, if one of the three reaction pairs is omitted, e.g., if $k_1 = k_2 =0$.

Simple relationships also occur when equilibrium is reached without any counteracting effects. For example, if only reactions I', II', and III' take place, then $k_1$, $k_3$ and $k_5$ are zero and one obtains $K_1 = \dfrac{k_6}{k_2}$, $K_2 = \dfrac {k_2}{k_4}$, $K_3 = \dfrac{k_4}{k_6}$.

\begin{center}
    \uline{\hspace{20mm}}
    \vspace{5mm}
\end{center}

[Note: We are omitting translating the remaining page because it solely relates to Section I of this paper about the ester formation from acid and alcohol.]

\end{appendices}

\bibliography{2024_Hirniak.bib}

\end{document}